\documentclass[aps,prd,amsmath,amsfonts,amssymb,eqsecnum,nofootinbib,
  floatfix,secnumarabic,preprintnumbers,nobalancelastpage,
 onecolumn,notitlepage]{revtex4-1}
\usepackage[utf8x]{inputenc}
\pdfoutput=1
\usepackage{cancel}
\usepackage{subfigure,color,graphicx,hyperref,dsfont,slashed,multirow}
\usepackage{color}
\usepackage{hyperref}
\usepackage{fontenc}
\usepackage{pbox}
\usepackage{feynmp}

\usepackage{mathtools,nccmath}%
\usepackage{ etoolbox, xparse} 

\usepackage{array}
\newcolumntype{P}[1]{>{\centering\arraybackslash}p{#1}}

\DeclarePairedDelimiterX{\abs}[1]\lvert\rvert{\ifblank{#1}{\,\cdot\,}{#1}}

\let\oldabs\abs
\def\abs{\futurelet\testchar\MaybeOptArgAbs}
\def\MaybeOptArgAbs{\ifx[\testchar\let\next\OptArgAbs
	\else \let\next\NoOptArgAbs\fi \next}
\def\OptArgAbs[#1]#2{\oldabs[#1]{#2}}
\def\NoOptArgAbs#1{\ifblank{#1}{\oldabs{}}{\oldabs[\big]{#1}}}

\DeclarePairedDelimiterX{\set}[1]\{\}{\setargs{#1}}
\NewDocumentCommand{\setargs}{>{\SplitArgument{1}{;}}m}
{\setargsaux#1}
\NewDocumentCommand{\setargsaux}{mm}
{\IfNoValueTF{#2}{#1}{\nonscript\,#1\nonscript\;\delimsize\vert\nonscript\:\allowbreak #2\nonscript\,}}
\let\oldset\set
\def\set{\futurelet\testchar\MaybeOptArgSet}
\def\MaybeOptArgSet{\ifx[\testchar \let\next\OptArgSet
	\else \let\next\NoOptArgSet \fi \next}
\def\OptArgSet[#1]#2{\oldset[#1]{#2}}
\def\NoOptArgSet#1{\OptArgSet[\big]{#1}}
\def\lsim{\raise0.3ex\hbox{$\;<$\kern-0.75em\raise-1.1ex\hbox{$\sim\;$}}}
\def\gsim{\raise0.3ex\hbox{$\;>$\kern-0.75em\raise-1.1ex\hbox{$\sim\;$}}}

\hypersetup{colorlinks=true,linkcolor=blue,citecolor=magenta,filecolor=magenta,
	urlcolor=cyan}
\newcolumntype{P}[1]{>{\centering\arraybackslash}p{#1}}

\hypersetup{bookmarks=true,unicode=true,pdftoolbar=true,pdfmenubar=true,
	pdffitwindow=false,pdfstartview={FitH},pdftitle={SecludedUMSSM},
	pdfauthor={~M.~Frank,Ya\c{s}ar Hi\c{c}y\i lmaz,Stefano Moretti,\"{O}.~\"{O}zdal}, pdfsubject={Subject},
	pdfcreator={Creator},pdfproducer={Producer}, pdfkeywords={Non-minimal 
		supersymmetry}{Beyond the Standard Model Physics}{SUSY},
	pdfnewwindow=true,colorlinks=true,
	linkcolor=blue,citecolor=magenta,filecolor=magenta,urlcolor=cyan}
\newcommand{\be}{\begin{equation}}
\newcommand{\ee}{\end{equation}}
\def\bsp#1\esp{\begin{split}#1\end{split}}
\renewcommand{\figureautorefname}{Figure}

\def\sectionautorefname~#1\null{Sec.~(#1)\null}
\def\subsectionautorefname~#1\null{sub--Sec.~(#1)\null}
\def\figureautorefname~#1\null{Figure~#1\null}
\def\tableautorefname~#1\null{Table~#1\null}
\def\equationautorefname~#1\null{Eq.~#1\null}

\newcommand{\del}{\textsc{Delphes}~3}

\newcommand{\fr}{\textsc{FeynRules}}
\newcommand{\hb}{\textsc{HiggsBounds}}
\newcommand{\hs}{\textsc{HiggsSignals}}

\newcommand{\mg}{\textsc{MG5\_aMC@NLO}}
\newcommand{\ma}{\textsc{MadAnalysis}~5}
\newcommand{\mo}{\textsc{MicrOMEGAs}}
\newcommand{\py}{\textsc{Pythia}~8}
\newcommand{\pysl}{\textsc{PySLHA}}
\newcommand{\sms}{\textsc{SModelS}}
\newcommand{\sa}{\textsc{SARAH}}
\newcommand{\spheno}{\textsc{SPheno}}

\begin{document}

{\title{Leptophobic $Z^\prime$ bosons in the secluded  UMSSM\\}
	
	\author{Mariana Frank$^1$\footnote{Email: mariana.frank@concordia.ca}}
	\author{Ya\c{s}ar Hi\c{c}y\i lmaz$^{2,3}$\footnote{E-mail: y.hicyilmaz@soton.ac.uk}}
	\author{Stefano Moretti$^2$\footnote{Email: s.moretti@soton.ac.uk}}
	\author{\"{O}zer \"{O}zdal$^{1,2}$\footnote{Email: ozer.ozdal@soton.ac.uk}}
	\affiliation{ $^1$Department of Physics,  
		Concordia University, 7141 Sherbrooke St. West ,
		Montreal, Quebec, Canada H4B 1R6}
	\affiliation{$^2$ School of Physics $\&$ Astronomy, University of Southampton, Highfield, Southampton SO17 1BJ,UK}
	\affiliation{$^3$Department of Physics, Bal\i kesir University, TR10145, Bal\i kesir, Turkey}
\date{\today}

\begin{abstract}
 {We perform a comprehensive analysis of the secluded UMSSM model, consistent with present experimental constraints. We find that in this model the additional $Z^\prime$ gauge boson can be leptophobic without resorting to gauge kinetic mixing and, consequently, also $d$-quark-phobic, thus lowering the LHC bounds on its mass. The model can accommodate very light singlinos as DM candidates, consistent with present day cosmological and collider constraints. Light charginos and neutralinos are responsible for muon anomalous magnetic predictions within 1$\sigma$ of the measured experimental value. Finally, we look at the possibility that a lighter $Z^\prime$, expected to decay mainly into chargino pairs and followed by the decay into lepton pairs, could be observed at 27 TeV.}
\end{abstract} 

\maketitle

\section{Introduction}
\label{sec:intro}

With the discovery of the Higgs boson, {the last piece of the Standard Model (SM) construction was fit into place}.  Furthermore, almost all SM predictions have been confirmed by  experimental results, even precision tests involving higher order perturbative Electroweak (EW) and Quantum Chromodynamics (QCD) effects. However, as it stands, the SM cannot be the final theory and the quest for physics Beyond the SM (BSM) is very much alive.  {Among} the many proposed BSM scenarios, Supersymmetry (SUSY) appears to be {one of} the most popular ones, {since it provides elegant solutions to the SM drawbacks, such as the stabilization of the EW scale under radiative corrections, an explanation for the baryon asymmetry of the Universe and for the presence of Dark Matter (DM) in it.} However, {the minimal version of SUSY, the  Minimal Supersymmetric SM (MSSM),} provides no explanation for the $\mu$ problem \cite{Cvetic:1996mf,Suematsu:1994qm,Lee:2007fw,Demir:1998dm}. The $\mu$ parameter, the so-called higgsino mass term,  is expected to be at the SUSY-breaking scale but, for successful EW symmetry breaking, its value should be at the scale of the latter.  Adding a $U(1)^\prime$ gauge symmetry to the MSSM, one solves this  problem by replacing the $\mu$ parameter of the MSSM with an effective one, generated dynamically by the Vacuum Expectation Value (VEV) of the singlet Higgs field responsible for breaking $U(1)^\prime$.  {Furthermore}, the additional $U(1)^\prime$ symmetry is able to generate neutrino masses by allowing right-handed neutrinos into the superpotential and can account for either Majorana- \cite{Demir:2006jj} or Dirac-type neutrinos \cite{Demir:2007dt}.

Normally, it is expected that both EW and $U(1)^\prime$ symmetry breaking are achieved through soft-breaking parameters, which would imply that the mass of the gauge boson associated with $U(1)^\prime$,  a $Z^\prime$, would be of the same order as the EW scale \cite{Langacker:2008yv,Erler:2009jh,Anoka:2004vf}. This conflicts with experimental measurements at the Large Hadron Collider (LHC) \cite{Aad:2019fac}, though, which impose a lower bound on the $Z^\prime$ mass, from the Drell-Yan (DY) channel, i.e., di-lepton hadro-production, of ${\cal O}(4)$ TeV or more. The most natural solution to this inconsistency is that the VEV of the singlet Higgs field is large compared to the EW scale, ${\cal O}(1-10)$ TeV, pushing the SUSY scale very high and rendering it mostly unobservable at the present   LHC. Alternatively, it was observed that fine-tuning the kinetic mixing between the two $U(1)$ groups could yield $Z^\prime$ bosons which do not decay directly into lepton pairs \cite{Araz:2017wbp}. Corresponding $Z^\prime$ gauge boson masses are then limited by its di-jet decays, whose bounds are much weaker in comparisons to DY ones \cite{Sirunyan:2019vgj}. Various aspects of the additional gauge boson and its phenomenological implications have been also studied within non-SUSY and SUSY frameworks \cite{Coleppa:2018fau,Rizzo:1998ut,Babu:1996vt,Chiang:2014yva,Celis:2015ara,Allanach:2019mfl,Alvarado:2019gyh,Mantilla:2016lui,Tang:2017gkz,Kamenik:2017tnu,Alves:2013tqa}. 

An alternative is represented by a $U(1)^\prime$ model where the SUSY-breaking scale and $Z^\prime$ mass are disjoint: the former  is close to the EW scale while a large value for the latter can be generated by the VEVs of additional  Higgs fields ($S_1,\, S_2,\, S_3$, so-called secluded singlets) which are charged under the $U(1)^\prime$ group but couple weakly to the SM fields \cite{Erler:2002pr}. This BSM scenario is known as the {\it secluded} $U(1)^\prime$ model, a realization of the generic class of $U(1)'$-extended MSSMs (UMSSMs). It allows for both explicit and spontaneous CP symmetry breaking and is able to account for baryogenesis \cite{Chiang:2009fs}. Differences between this UMSSM scenario and the MSSM would likely reveal themselves in the nature of DM, as in the extended scenario several additional singlinos as well as sneutrinos could be viable candidates for it \cite{Demir:2010is}. 

In a nutshell, the {secluded} $U(1)^\prime$ model extends the MSSM by an additional Abelian group, to $SU(3)_c \otimes SU(2)_L \otimes U(1)_Y \otimes U(1)^\prime$, and by four Higgs singlets (three in addition to the one needed to break $U(1)^\prime$, to ensure a $Z^\prime - Z$ mass hierarchy \cite{Erler:2002pr}). Exotics with Yukawa couplings to a singlet Higgs field must be introduced to ensure the theory  is anomaly free. However,  despite the presence of these couplings, one can assume their masses to be at the Grand Unification Theory 
(GUT) scale and thus neglect them in TeV scale phenomenology\footnote{Furthermore, their charges are such that they do not mix with ordinary matter.}. (Note, however, that they have been studied extensively in \cite{Kang:2007ib}.) Previous studies of this secluded $U(1)'$ model exist, but since they are older \cite{Demir:2010is,Frank:2012ne}, none of them are consistent with  present experimental data on the discovered Higgs boson mass and signal strengths or with $Z'$ gauge boson mass bounds.   In this work, we revisit this BSM scenario  in detail, with particular interest in addressing the unresolved problems of UMSSMs, by providing light $Z^\prime$ masses yet compatible with current  bounds, an acceptable $(g-2)_\mu$ value and DM relic density plus the viable existence of light SUSY particles, altogether providing one with new distinguishing signals of this BSM realization in  LHC  experiments.

In showing all this, we shall prove first that, in such a $U(1)'$ secluded model,  leptophobia can be achieved easily and without gauge kinetic mixing between the $Z$ and $Z'$, so that a light $Z^\prime$ gauge boson can survive all experimental constraints  in presence of finite width effects. Furthermore, we shall show that this BSM scenario can predict  corrections to $(g-2)_\mu$  within $1 \sigma$ of the experimentally observed value. Finally, we will also find that, in our UMSSM realization,  the Lightest SUSY Particle (LSP), for a large region of its parameter space, is a singlino
consistent with all DM constraints accompanied by very light charginos and neutralinos, with masses of ${\cal O}(100)$ GeV, in turn consistent with collider limits,  into which a $Z'$ can then decay yielding  sizable signals at the LHC. 

Our work is organized as follows. In the next section, Sec. \ref{sec:model}, we provide a description of the  {secluded} $U(1)^\prime$ model, with particular emphasis on the gauge and neutralino sectors, i.e.,  where differences with respect to the MSSM will manifest themselves. We  describe the implementation of this BSM scenario, including the free parameters and the constraints imposed on these, in Sec. \ref{sec:scan}.  Then, we  explain the implications emerging from a wide scan of its parameter space for $Z'$ physics at colliders,  in Sec. \ref{sec:analysis}, and onto the DM candidate  in relic density and  direct detection experiments,  
 in Sec. \ref{sec:darkmatter}. Furthermore, in presence  of all such  constraints on the mass and coupling  spectrum of the model, we analyze the consequences for the muon anomalous magnetic moment in Sec. \ref{sec:muong2}. We  further study the possibility of observing a light $Z^\prime$ boson via chargino/neutralino  decays at the High-Luminosity LHC (HL-LHC) and High-Energy LHC (HE-LHC)
 in Sec. \ref{sec:ZprimeSignal}. Finally, in Sec. \ref{sec:conclusion}, we summarize our findings and draw our conclusions.

\section{The secluded $U(1)^\prime$ Model}
\label{sec:model}

In this section, we review the {secluded} $U(1)^\prime$, known also as the secluded UMSSM. 
In addition to the MSSM superfields, the model has three right-handed neutrino superfields $\hat{N}_i^c$ and four scalar singlets $\hat{S}$, $\hat{S}_1$, $\hat{S}_2$ and $\hat{S}_3$. {An anomaly-free model with an additional $U(1)^\prime$ gauge group can be obtained by embedding it into an $E_6$ GUT.  Breaking $E_6$ yields a combination of two additional $U(1)^\prime$ gauge groups, denoted by $U(1)_\chi$ and $U(1)_\psi$, whose charges mix with angle $\theta_{E_6}$,
\begin{equation}
Q^\prime=Q_\chi \cos \theta_{E_6}+Q_\psi \sin \theta_{E_6} \, ,
\end{equation}
where the orthogonal combination of $U(1)_\chi$ and $U(1)_\psi$ is assumed very heavy and decoupled.  

Three $\bf 27$ representations of $E_6$ are needed to provide three families of  SM fermions, one pair of Higgs doublets,  extra SM singlets plus exotics. In the usual connection established between the breaking of $E6$ and the SM, a pair of ${\bf 27}+{\bf 27^\star}$ (sometimes referred to as ${\bf 27}_L+{\bf 27}_L^\star$) are introduced, in addition to the three ${\bf 27}$ representations, to insure gauge unification without  anomalies \cite{Langacker:1998tc}. These fundamental representations are connected, through the breaking $E6 \to SO(10) \to SU(5)$, to states in the SM.  The breaking of the fundamental representation ${\bf 27}$ of $E6$ yields ${\bf 16+10+1}$ representations of $SO(10)$, with further decay into $SU(5)$ multiplets proceeding as  ${\bf 16} \to 10 (u,d,u^c,e^+) +5^\star (d^c, \nu, e^-)+1 ({\bar N})$, ${\bf 10} \to 5 (D, H_u) +5^\star (D^c, H_d)$ and ${\bf 1}\to 1 (S_L)$, where we have indicated in brackets the remaining particle states. iIn addition to the SM  particles there are  two exotic $SU(2)_L$ singlet quarks of charge $\pm1/3$ and the singlets $S_L$ and $\bar N$, in the conventional $E6$ notation. In our model, $S, \, S_1, \, S_2, \, S_3$ correspond to, respectively,  $S_L, \, S_L^\star, \, S_L^\star$ and $\bar N^\star$ from two partial pairs of $\bf 27$+$\bf 27^\star$. The  two $\bf 27$+$\bf 27^\star$ representations include the extra $S_L$ and $\bar N$ to cancel the $U(1)^\prime$ anomalies \cite{Langacker:1998tc, Kang:2004bz}.

The complete description of $E_6$ SUSY GUTs, including composition of the fundamental {\bf 27} representation has appeared in \cite{Hewett:1988xc,Langacker:1980js}. The secluded model corresponds to $\theta_{E_6}=\arctan \frac{\sqrt{15}}{9}\sim 0.13 \pi$  and a prescribed set of $U(1)^\prime$ charges. The model was shown to, in addition to generating the $\mu$ term dynamically, be anomaly-free \cite{Langacker:2008yv}, solve the $Z-Z^\prime$ mass hierarchy \cite{Erler:2002pr} and facilitate EW baryogenesis \cite{Kang:2004pp,Barger:2004dy}.} In our study, we modify the model by re-assigning the $U(1)^\prime$ charges to allow the model to be leptophobic. As such, we cannot rely on previous restrictions on the model, and shall perform a complete analysis of its parameter space.

The superpotential in this model is described by
\begin{eqnarray}
W&=&Y_{u}^{ij} \hat{Q_{i}} \hat{H_{u}} \hat{u^{c}_{j}} -Y_{d}^{ij} \hat{Q_{i}} \hat{H_{d}} \hat{d^{c}_{j}} -Y_{e}^{ij} \hat{L_{i}} \hat{H_{d}} \hat{e^{c}_{j}} \nonumber \\ &+& Y_{\nu}^{ij} \hat{L_{i}} \hat{H_{u}} \hat{N^{c}_{i}} 
+ \lambda \hat{H_u} \hat{H_d} \hat{S} + \frac{\kappa}{3} \hat{S_1} \hat{S_2} \hat{S_3} + \sum_{n=1}^{n_\varphi} h_{\varphi}^i S \varphi_i \overline{\varphi}_j  + \sum_{n=1}^{n_\Upsilon} h^i_{\Upsilon} S \Upsilon_i \overline{\Upsilon}_j,	  
\label{superpotential}
\end{eqnarray}
where  the  first  line  of  Eq. \ref{superpotential}  contains  the  usual  terms  of  the  MSSM  while the  second line includes the additional interactions of right-handed neutrinos $\hat{N}_i^c$ (assumed to be  Dirac fields here) and $\hat{H}_u$, as well as the singlet superfields $\hat{S}$, $\hat{S}_1$, $\hat{S}_2$ and $\hat{S}_3$, and where $ \Upsilon_i$ and $\varphi_i$ are $n_{\varphi}$,  $n_{\Upsilon}$, respectively,   generations of exotic fermions, vector-like with respect to the MSSM, but chiral under $U(1)^\prime$ symmetry. The $U(1)$ and $U(1)^\prime$ charges associated with these exotics, as well as the number of families are a direct consequence of the anomaly cancellation conditions,  are listed in the Appendix (Sec. \ref{sec:appendix}). The $\varphi_i$ are color-singlet states, while the $ \Upsilon_i$  are color triplet states. Their charges depend on the choices of $U(1)^\prime$ charges of the rest of the particles,  and their mass is restricted by searches for exotic charged particles at the LHC. Although no specific searches for exactly this charge exist, fermions with exotic charges are expected to have masses larger than 1 TeV \cite{Kazana:2016goy}{\footnote{These appear naturally, and have been discussed in the context of $E_6$ gauge groups \cite{Rosner:1999ub}.}.
 The effective $\mu$ term is generated dynamically as  $\mu=  \lambda \langle S \rangle$. The scalar potential includes the $F$-term,  given  by
\begin{eqnarray}
V_F&=&\lambda^2 ( | H_u |^2  | H_d |^2 + |S|^2 |H_u|^2 + |S|^2 |H_d|^2) + 
\kappa^2 (|S_1|^2|S_2|^2 + |S_2|^2|S_3|^2 + |S_3|^2|S_1|^2)\, ,
\end{eqnarray}
while the $D$-term scalar potential is 
\begin{eqnarray}
V_D &=& \frac{g_1^2 + g_2^2}{8}  ( | H_d |^2 - | H_u |^2 )^2  +
\frac{1}{2} g^{\prime \,2} \left( Q_S |S|^2 + Q_{H_u} |H_u|^2 + Q_{H_d} |H_d|^2 + \sum_{\mathclap{n=1}}^{3} Q_{S_i} |S_i|^2 \right)^2,
\end{eqnarray}
where $g_1$, $g_2$ and $g^\prime$ are the coupling constants for the $U(1)_Y$, $SU(2)_L$ and $U(1)^\prime$ gauge groups while  $Q_{\phi}$ is the $U(1)^\prime$ charge of the field $\phi$. Finally, the potential includes the  SUSY-breaking soft terms,
\begin{eqnarray}
V_{\rm soft} &=& m^2_{H_u} | H_u |^2 + m^2_{H_d} | H_d |^2 + m^2_{S} | S |^2  + \sum_{\mathclap{n=1}}^{3} m_{S_i}^2 |S_i|^2  -
(A_\lambda \lambda S H_u H_d + A_\kappa \kappa S_1 S_2 S_3 + h.c) \nonumber \\ &+&
(m_{SS_1}^2 S S_1 + m_{SS_2}^2 S S_2 + m_{S_1 S_2}^2 S_1^\dagger S_2 + h.c.).
\end{eqnarray}
In Table \ref{tab:superfields} we give the complete list of the fields in the model, together with their spin, number of generations and charge assignments under the extended gauge group.
\begin{table}
	\begin{center}	
		\begin{tabular}{|c|c|c|c|c|c|}
			\hline \hline
			SF & Spin 0 & Spin \(\frac{1}{2}\) & Generations & $U(1)_Y\otimes\, SU(2)_L \otimes\, SU(3)_C\otimes\, U(1)^\prime$  \\
			\hline
			\(\hat{q}\) & \(\tilde{q}\) & \(q\) & 3 & \((\frac{1}{6},{\bf 2},{\bf 3}, Q_q) \) \\
			\(\hat{l}\) & \(\tilde{l}\) & \(l\) & 3 & \((-\frac{1}{2},{\bf 2},{\bf 1}, Q_\ell) \) \\
			\(\hat{H}_d\) & \(H_d\) & \(\tilde{H}_d\) & 1 & \((-\frac{1}{2},{\bf 2},{\bf 1}, Q_{H_d}) \) \\
			\(\hat{H}_u\) & \(H_u\) & \(\tilde{H}_u\) & 1 & \((\frac{1}{2},{\bf 2},{\bf 1}, Q_{H_u}) \) \\
			\(\hat{d}\) & \(\tilde{d}_R^*\) & \(d_R^*\) & 3 & \((\frac{1}{3},{\bf 1},{\bf \overline{3}}, Q_d \) \\
			\(\hat{u}\) & \(\tilde{u}_R^*\) & \(u_R^*\) & 3 & \((-\frac{2}{3},{\bf 1},{\bf \overline{3}}, Q_u) \) \\
			\(\hat{e}\) & \(\tilde{e}_R^*\) & \(e_R^*\) & 3 & \((1,{\bf 1},{\bf 1}, Q_e) \) \\
			\(\hat{v}_R\) & \(\tilde{\nu}_R^*\) & \(\nu_R^*\) & 3 & \((0,{\bf 1},{\bf 1}, Q_v) \) \\
			\(\hat{S}\) & \(S\) & \(\tilde{S}\) & 1 & \((0,{\bf 1},{\bf 1}, Q_s) \) \\
			\(\hat{S}_1\) & \(S_1\) & \(\tilde{S}_1\) & 1 & \((0,{\bf 1},{\bf 1}, Q_{s_1}) \) \\
			\(\hat{S}_2\) & \(S_2\) & \(\tilde{S}_2\) & 1 & \((0,{\bf 1},{\bf 1},Q_{s_2}) \) \\
			\(\hat{S}_3\) & \(S_3\) & \(\tilde{S}_3\) & 1 & \((0,{\bf 1},{\bf 1}, Q_{s_3}) \) \\
\hline \hline
		\end{tabular}
		\caption{ \label{tab:superfields} Superfield configuration in the secluded UMSSM.}
	\end{center}
\end{table}
The secluded $U(1)^\prime$ charge assignments and anomaly cancellation conditions allow for some freedom in the choice of the $U(1)^\prime$ charges,  absent in other $U(1)^\prime$ models.  In general, the $U(1)^\prime$ change assignments can be chosen as follows:
\begin{align}
Q_Q = \alpha, \qquad
Q_{H_u} = \beta, \qquad
Q_S = \gamma, \qquad
Q_\ell = - 3\alpha +  \frac{\gamma}{3}, \qquad
Q_{H_d} = -\beta - \gamma, \nonumber \\
Q_u = -\alpha - \beta, \qquad
Q_d = -Q_Q - Q_{H_d} = -\alpha + \beta + \gamma, \qquad
Q_e = -Q_\ell - Q_{H_d} =  3\alpha + \beta +\frac{2\gamma}{3}, \nonumber\\
Q_N = -Q_\ell - Q_{H_u} = 3\alpha - \beta -\frac{\gamma}{3}, \qquad
Q_{S_1} = Q_{S_3} = \delta,  \qquad
Q_{S_2} = -2Q_{S_1} = -2Q_{S_3} = -2\delta. \qquad
\label{U1Charges}
\end{align}
Here,  $Q_{H_d}$ = 0  dictates $\gamma = -\beta$. 
From the conditions above we can choose, for simplicity,  $Q_e = Q_\ell$.
The leptophobic condition $Q_\ell = Q_e = 0$ requires $\alpha = -\frac{\beta}{9}$, so that   the leptophobia condition can be achieved without resorting to  kinetic mixing between the two $U(1)$ groups\footnote{This is unlike models where 
	the $U(1)^\prime$ charges are derived from the mixing of, e.g., $\theta_{E_6}$ angles \cite{Frank:2020pui}.}. 
Thus, Eq. \ref{U1Charges} can be rewritten in terms of $\alpha$ and $\delta$ only as:
\begin{align}
Q_Q = \alpha, \qquad
Q_{H_u} = -9\alpha, \qquad
Q_S = 9\alpha, \qquad
Q_\ell = 0, \qquad
Q_{H_d} = 0, \qquad
Q_u = 8\alpha, \qquad
Q_d = -\alpha,  \nonumber \\
Q_e = 0, \qquad
Q_N = 9\alpha,  \qquad
Q_{S_1} = Q_{S_3} = \delta, \qquad
Q_{S_2} = -2Q_{S_1} = -2Q_{S_3} = -2\delta.
\label{U1ChargesIII}
\end{align}
After the spontaneous breaking of the extended gauge symmetry group down to electromagnetism (EM), the $W^\pm, Z$ and $Z^\prime$ bosons acquire  masses while the photon remains massless. At tree level, the squared masses of the $Z$ and $Z^\prime$ bosons are given by
\begin{eqnarray}
M_Z^2 &=& \frac{g_1^2 + g_2^2}{2}  \left( \langle H_u^0 \rangle^2 + \langle H_d^0\rangle^2 \right),  \nonumber \\
M_{Z^\prime}^2 &=&g^{\prime\, ^2} \left( Q_S \langle S \rangle ^2 + Q_{H_u} \langle H_u^0 \rangle^2 + Q_{H_d} \langle H_d^0\rangle^2 + \sum_{\mathclap{n=1}}^{3} Q_{S_i} \langle S_i \rangle^2 \right),
\label{eq:HeavyZmasses}
\end{eqnarray}
where $\displaystyle H^0_d \equiv \frac{v_d}{\sqrt{2}}$ and $ \displaystyle H^0_u \equiv \frac{v_u}{\sqrt{2}}$ stand for the neutral components of the down-type and up-type Higgs fields $H_d$ and $H_u$.

While the chargino sector is unaltered, {the neutralino sector of the secluded $U(1)^{\prime}$ model includes}  five additional fermion fields: the $U(1)^{\prime}$ gauge fermion
$\widetilde{Z}^{\prime}$ and four singlinos $\widetilde{S}$,
$\widetilde{S_1}$, $\widetilde{S_2}$, $\widetilde{S_3}$, in total,
nine neutralino states $\widetilde{\chi}_i^0$ ($i=1,\dots,9$)
\cite{Erler:2002pr}:
\begin{eqnarray}
\label{neutralino-def1} \widetilde{\chi}_i^0 = \sum_{a} {\cal N}^0_{i a}
\widetilde{G}_a\,,
\end{eqnarray}
where the mixing matrix ${\cal N}^0_{i a}$ connects the gauge-basis neutral
fermion states to the physical-basis
neutralinos $\widetilde{\chi}_i^0$. The neutralino masses
$M_{\widetilde{\chi}_i^0}$  are
obtained through the diagonalization
\noindent
${\cal N}^0 {\cal{M}} {\cal N}^{0\ T}
= \mbox{Diag}$ $\Big\{M_{\widetilde{\chi}_1^0},$ $\dots,$
$M_{\widetilde{\chi}_9^0}\Big\}$. The $9 \times 9$ neutral fermion mass matrix is
\begin{widetext}
	\begin{eqnarray}\label{mneut}
	{\cal M}=\left(
	\begin{array}{ccccccccc}
	M_{\tilde Z}&0&-M_{\tilde Z\tilde H_d}&M_{\tilde Z \tilde H_u}&0&M_{\tilde Z
		\tilde Z'}&0&0&0 \\[1.ex]
	0&M_{\tilde W}&M_{\tilde W \tilde H_d}&-M_{\tilde W \tilde
		H_u}&0&0&0&0&0\\[1.ex]
	-M_{\tilde Z \tilde H_d}&M_{\tilde W \tilde
		H_d}&0&-\mu&-\mu_{H_u}&\mu'_{H_d}&0&0&0\\[1.ex]
	M_{\tilde Z \tilde H_u}&-M_{\tilde W \tilde
		H_u}&-\mu&0&-\mu_{H_d}&\mu'_{H_u}&0&0&0\\[1.ex]
	0&0&-\mu_{H_u}&-\mu_{H_d}&0&\mu'_S&0&0&0\\[1.ex]
	M_{\tilde Z \tilde Z'}&0&\mu'_{H_d}&\mu'_{H_u}&\mu'_S&M_{\tilde
		Z'}&\mu'_{S_1}&\mu'_{S_2}&\mu'_{S_3}\\[1.ex]
	0&0&0&0&0&\mu'_{S_1}&0&-\frac{\kappa v_{3}}{3\sqrt{2}}&-\frac{\kappa v_{2}}{3\sqrt{2}}\\[1.ex]
	0&0&0&0&0&\mu'_{S_2}&-\frac{\kappa v_{3}}{3\sqrt{2}}&0&-\frac{\kappa v_{1}}{3\sqrt{2}}\\[1.ex]
	0&0&0&0&0&\mu'_{S_3}&-\frac{\kappa v_{2}}{3\sqrt{2}}&-\frac{\kappa v_{1}}{3\sqrt{2}}&0\\[1.ex]
	\end{array}
	\right)\, ,\\
	\end{eqnarray}
\end{widetext}
where the lightest eigenvalue is the DM candidate.
In the neutralino mass matrix, the mass mixing terms are defined in terms of  $\displaystyle \tan \beta =\frac{v_d}{v_u}\, , \langle S \rangle = \frac{v_S}{\sqrt{2}}$ and $ \displaystyle  \langle S_i \rangle = \frac{v_{i}}{\sqrt{2}}~(i=1, 2, 3),$
as
\begin{eqnarray}
M_{\widetilde{Z}\, \widetilde{H}_d} &=& M_Z \sin\theta_W
\cos\beta\,,\quad
M_{\widetilde{Z}\, \widetilde{H}_u} = M_Z \sin\theta_W \sin\beta\,,\nonumber\\
M_{\widetilde{W}\, \widetilde{H}_d} &=& M_Z \cos\theta_W
\cos\beta\,, \quad
M_{\widetilde{W}\, \widetilde{H}_u} = M_Z \cos\theta_W \sin\beta\, ,
\end{eqnarray}
where $\mu_i$,  $\mu^\prime_j$ stand for  the effective  couplings in each sector, given in terms of  $h_s$ or $g^\prime$, the coupling constant of $U(1)^{\prime}$, as
\begin{eqnarray}
\mu_{H_d} &=& h_s \frac{v_d}{\sqrt{2}}\,,\qquad \mu_{H_u} = h_s
\frac{v_u}{\sqrt{2}}\,,\quad \mu^{\prime}_{H_d}=g^{\prime}
Q_{H_d} v_d,\nonumber\\
\mu^{\prime}_{H_u} &=& g^{\prime} Q_{H_u} v_u\,,\quad
\mu^{\prime}_{S} = g^{\prime} Q_{S} v_S\,,\quad
\mu^{\prime}_{S_i} = g^{\prime} Q_{S_i} v_{i}\,.
\end{eqnarray}
{In our further analysis, we impose gauge coupling unification by setting $g_1 = g_2 = g^\prime \approx g_3$ at  the GUT scale. }


\section{Computational Setup}
\label{sec:scan}

Following the development of the model as in Sec. \ref{sec:model}, to enable our analysis and impose constraints coming from  experimental data, we implement the model within a computational framework. We have then made use of \sa~ (version 4.13.0) \cite{Staub:2008uz,Staub:2010jh,Staub:2015kfa} to generate CalcHep \cite{Belyaev:2012qa} model files and a UFO \cite{Degrande:2011ua} version of the model \cite{Christensen:2009jx}, so that we could employ \mo~ (version 5.0.9) \cite{Belanger:2018ccd} for the computation of the predictions relevant for our dark matter study and \spheno~ (version 4.0.4) \cite{Porod:2003um,Porod:2011nf} package for spectrum analysis. Note that  \sa~ (version 4.13.0) includes all RGE corrections to model  parameters to second order, and these are intrinsically dependent on our choice of $U(1)^\prime$ charges.  In this package, the weak scale values of the gauge and Yukawa couplings present in secluded UMSSM are evolved to the unification scale $M_{GUT}$ via the RGEs. After $M_{GUT}$ is determined by the requirement of the gauge coupling unification (by setting $g_1 = g_2 = g^{\prime} \approx g_3$) through their RGE evolutions, all the soft supersymmetry breaking (SSB) parameters along with the gauge and Yukawa couplings are evolved back to the weak scale with the boundary conditions given at $M_{GUT}$.


In  order  to  apply  the  LHC  constraints  on  the  properties  of $Z^\prime$ bosons,  we  calculate the $Z^\prime$ production cross section at next-to-leading order (NLO) accuracy in QCD \cite{Fuks:2007gk,Fuks:2017vtl}. This relies on the joint use of FeynRules version 2.3.36 \cite{Alloul:2013bka} and the included NLOCT package \cite{Degrande:2014vpa}, as well as FeynArts \cite{Hahn:2000kx}, for the automatic generation of a UFO library \cite{Degrande:2011ua} containing both tree-level and counter term vertices necessary at NLO. This UFO model is then used by MG5aMC@NLO (version 2.7.3) \cite{Alwall:2014hca}  for the numerical evaluation of the hard-scattering matrix elements, which are convoluted with the NLO set of NNPDF 3.1 parton distribution functions (PDF) \cite{Ball:2017nwa}. Using the decay table provided by the SPheno package and assuming the narrow-width approximation, we compare our predictions with the ATLAS and CMS limits on $Z^\prime$ bosons in the dilepton \cite{Aad:2019fac} and dijet \cite{Sirunyan:2018xlo,Sirunyan:2019vgj} modes in order to estimate the impact of supersymmetric decay channels in the secluded UMSSM.

We make use of \hb~ \cite{Bechtle:2013wla} to constrain the possibility of BSM Higgs bosons detection at colliders and \hs~ \cite{Bechtle:2013xfa} to test the signal strengths of the SM-like Higgs state. During the numerical analysis performed in this work, we have used the \pysl~ 3.2.4 package \cite{Buckley:2013jua} to read the input values  for  the  model  parameters  that  we  encode  under  the  SLHA  format \cite{Skands:2003cj}, and to  integrate the various employed programmes into a single framework.




Using our interfacing and following the Metropolis-Hastings technique, we performed a random scan over the parameter space,  illustrated in Table \ref{tab:scan_lim}, where we restrict ourselves only to universal boundary conditions. Here $m_0$ denotes the Spontaneous Symmetry Breaking (SSB) mass term for all the scalars while $M_{1/2}$ stands for the SSB mass terms for the gauginos including the one associated with the U(1)$^\prime$ gauge group. As before,  $\tan \beta$ is the ratio of VEVs of the MSSM Higgs doublets,  $A_0$ is the SSB trilinear scalar interacting term{\footnote{Note that, while we scan $A_0/m_0$ between $[-3,3]$, most of our solutions lie near $A_0 \approx 0$.}}} $\lambda$ is the coupling associated with the interaction of the $\hat{H}_u$, $\hat{H}_d$ and $\hat{S}$ fields while $\kappa$ is the coupling of the interaction of the $\hat{S}_1$, $\hat{S}_2$ and $\hat{S}_3$ fields. Trilinear couplings for $\lambda$ and $\kappa$ are defined as $ A_\lambda \lambda$ and  $A_\kappa \kappa$, respectively, at the SUSY scale. Here, $Y^{ij}_\nu$ is the Yukawa coupling of the term $\hat{L}_i \hat{H}_u \hat{N}_i^c$ and we vary only the diagonal elements in the range of $1 \times 10^{-8}$ -- $1\times 10^{-7}$ while  setting the off-diagonal elements to zero.

The desired distribution here is to designed to generate a collection of secluded UMSSM solutions consistent with all constraints along with the relic density constraint and muon $g-2$ within 2$\sigma$. 
%

\begin{table}
	\setlength\tabcolsep{20pt}
	\renewcommand{\arraystretch}{2.0}
	\small
	\centering
	\begin{tabular}{|c|c||c|c|}
		\hline
		Parameter      & Scanned range& Parameter      & Scanned range \\
		\hline
		$m_0$        & $[0., 3.]$ TeV     & $v_S$   & $[0.97, 15.8]$ TeV  \\
		$M_{1/2}$   & $[0., 3.]$ TeV     & $v_1$   & $[1.6, 15.]$ TeV  \\
		$\tan\beta$   & $[1., 55.]$        & $v_2$  & $[0.8, 11.2$] TeV  \\
		$A_0/m_0$   & $[-3., 3.]$    & $v_3$  & $[1.6., 15.]$ TeV \\		
		$\lambda$   & $[3. \times 10^{-2}, 0.6]$ & $\kappa$   & $[0.3, 2.65]$ \\		
		$A_\lambda$   & $[1.8, 7.5]$ TeV & $A_\kappa$   & $[-8.3, -0.2]$ TeV    \\
		$Y_\nu^{ij}, (i = j)$   & $[1\times 10^{-8}, 1\times 10^{-7}]$  & $Y_\nu^{ij}, (i \neq j)$   & 0. \\																						
		\hline
	\end{tabular}
	$\vspace{0.5cm}$
	\caption{\label{tab:scan_lim} Scanning range of parameter space of the secluded $U(1)^\prime$ model.}
\end{table}

We followed \cite{Cacciapaglia:2006pk} where a simple method for analyzing the impact of precision EW data above and below the $Z$ peak on flavor-conserving heavy new physics is implemented. There,  the  corrections to all leptonic data can be converted into oblique corrections to the vector boson propagators and condensed into  seven parameters.  Numerical fits for the new physics parameters are included and  the method is applied to generic  $Z^\prime$ gauge bosons highlighting parameter  combinations  most strongly constrained. The authors report the 99\% Confidence Level
(CL) iso-contours of bounds on $M_{Z^\prime}/g^\prime$ for a set of $Z^\prime$'s. Their constraints  depend only on the leptonic and Higgs $U(1)^\prime$ charges, $Q_{H_u}$, $Q_{H_d}$, $Q_\ell$, $Q_e$, and the assumption that their arbitrary overall normalization is fixed, $Q^2_{H}+Q^2_\ell+Q^2_e= 2$. Given that we fix $Q_\ell = Q_e =Q_{H_d} = 0$,  the  $Z^\prime$ gauge boson in our model cannot be considered as one of the given set of $Z^\prime$'s, so that the bounds on  $M_{Z^\prime}/g^\prime$ given  
by \cite{Cacciapaglia:2006pk} are not applicable in a straightforward way. Therefore, we  require  a  2$\sigma$ (i.e. 95$\%$  CL)  agreement  with  EW   precision  observables, parametrized through the  oblique parameters $S, T, U$ \cite{Altarelli:1990zd, Peskin:1990zt, Peskin:1991sw, Maksymyk:1993zm}. The constraints from the latter are included by evaluating
\begin{equation}
\chi^2_{\rm STU} = X^T C^{-1} X\, ,
\end{equation}
with $X^T= (S-\hat{S}, T-\hat{T}, U-\hat{U})$. The observed parameters deviations are given by \cite{Baak:2014ora}
\begin{equation}
\hat{S} = 0.05, \hspace{1cm}  \hat{T} = 0.09, \hspace{1cm}  \hat{U} = 0.01,
\end{equation}	
where the unhatted quantities denote the model predictions. The covariance matrix is \cite{Baak:2014ora}
\begin{equation}
\textbf{C}_{ij}=
\begin{bmatrix}
0.0121 & 0.0129 & −0.0071 \\
0.0129 & 0.0169 & −0.0119 \\
−0.0071 & −0.0119 & 0.0121 \\
\end{bmatrix}.
\end{equation}
We then require $\chi^2_{\rm STU} \leq $ 8.025,  corresponding to a maximal $2\sigma$ deviation given the 3 degrees of freedom.

We also verified that the vertex corrections due to loops with supersymmetric particles are small. For the  parameter space which survives all constraints, BR$(Z \to b {\bar b}) \in (0.1508-0.1510) $, which is consistent with the experimental requirement BR$(Z \to b {\bar b})= (15.12 \pm 0.005)$\%  \cite{Tanabashi:2018oca}.

\section{Gauge boson mass constraints}
\label{sec:analysis}
After imposing the constraints from the previous section, we turn our attention to gauge bosons. From the SSB of the $SU(2)_L\otimes U(1)_{Y}\otimes U(1)^\prime$ symmetry, the gauge bosons $Z$ and $Z^\prime$ mix to form physical mass eigenstates. The $Z-Z^\prime$ mixing mass matrix is 
\begin{eqnarray}
\mathbf{M_{Z}^2} &=&
\left(
\begin{array}{cc}
M_{ZZ}^2	&	M_{ZZ^\prime}^2	\\
M_{ZZ^\prime}^2	&	M_{Z^\prime Z^\prime}^2	
\end{array}
\right).	
\end{eqnarray}
As the mixing between the $Z$ and $Z^\prime$ bosons is very small, to a good approximation, these  are good physical states, with masses given in Eq. \ref{eq:HeavyZmasses}. 
Following the methodology described in the previous section, we scan the parameter space imposing constraints on SUSY particles, rare $B$-meson decays and oblique parameters so that the SM $Z $ gauge boson properties are consistent with experimental data, as indicated in Table \ref{tab:constraints}. 
\begin{table}{
		\setlength\tabcolsep{7pt}
		\renewcommand{\arraystretch}{1.6}
		\begin{tabular}{l|c|c||l|c|c}
		\small
		\centering			
			Observable & Constraints & Ref. & Observable & Constraints & Ref.\\
			\hline
			$m_{h_1} $ & $ [122,128] $ GeV                     & \cite{Chatrchyan:2012xdj} &
			$m_{\widetilde{t}_1} $                                    & $ \geqslant 730 $ GeV & \cite{Tanabashi:2018oca}\\
			$m_{\widetilde{g}} $                                       & $ > 1.75 $ TeV & \cite{Tanabashi:2018oca} &
			$ m_{\widetilde{\chi}_1^\pm} $                     & $ \geqslant 103.5 $ GeV & \cite{Tanabashi:2018oca} \\
			$m_{\widetilde{\tau}_1} $                               & $ \geqslant 105 $ GeV & \cite{Tanabashi:2018oca} & 
			$m_{\widetilde{b}_1} $                                    & $ \geqslant 222 $ GeV & \cite{Tanabashi:2018oca}\\
			$m_{\widetilde{q}} $                                       & $ \geqslant 1400 $ GeV & \cite{Tanabashi:2018oca} &
			$m_{\widetilde{\mu}_1} $                               & $ > 94 $ GeV & \cite{Tanabashi:2018oca} \\
			$m_{\widetilde{e}_1} $                                    & $ > 107 $ GeV & \cite{Tanabashi:2018oca} &
			$ \lvert \alpha_{Z Z^\prime} \rvert $                            & $\mathcal{O}(10^{-3})$ & \cite{Erler:2009jh} \\
			$\chi^2_{\rm STU}$                                        & $\leq 8.025 $ & - &
			BR$(B^0_s \to \mu^+\mu^-) $ & $[1.1,6.4] \times10^{-9}$  &
			\cite{Aaij:2012nna} \\ 
			$\displaystyle  \frac{{\rm BR}(B \to \tau\nu_\tau)}
			{{\rm BR}_{SM}(B \to \tau\nu_\tau)} $ & $  [0.15,2.41] $ &
			\cite{Asner:2010qj} &
			BR$(B^0 \to X_s \gamma) $ & $  [2.99,3.87]\times10^{-4} $ &
			\cite{Amhis:2012bh}\\
		\end{tabular}
		\caption{\label{tab:constraints}  Current experimental and theoretical bounds used to determine consistent solutions in our scans.}}
\end{table}
In the following, we analyze the properties of the gauge sector for all scenarios accepted in our scanning procedure. In Figure \ref{fig:oblique}, we depict the relations between the parameters $M_{Z^\prime}$, $g^\prime_{\rm SUSY}$, $Q_Q$,  the ratio of $M_{Z^\prime} /g^\prime_{\rm SUSY}$ and $\chi^2_{\rm STU}$. Here, $g^\prime_{\rm SUSY}$ is the coupling constant for the $U(1)^\prime$ group at the SUSY-breaking scale. The color bar of the upper panels shows the $\chi^2_{\rm STU}$ values for  solutions with $\chi^2_{\rm STU} \leq $ 8.025 while the color bar of the left bottom panel represents the gauge coupling $g^\prime_{\rm SUSY}$. According to the top left panel of Figure \ref{fig:oblique}, the ratio  $M_{Z^\prime}/g^\prime_{\rm SUSY}$ can be as low as 2.2 TeV when the charge $Q_Q$ is small (i.e., $[1.-3.]\times 10^{-2}$) while the bound on $M_{Z^\prime} /g^\prime_{\rm SUSY}$ tends to increase up to 8 TeV for larger  $Q_Q$ values (i.e., $ 1\times 10^{-1}$). Further, the top right and bottom left panel of Figure \ref{fig:oblique} shows that light $Z^\prime$ solutions consistent with the constraints given in Table  \ref{tab:constraints} can be found to lie around 1.5 TeV.  For heavier $Z^\prime$ masses, the range for the ratio  $M_{Z^\prime}/g^\prime_{\rm SUSY}$ opens up to a larger interval. As seen from the bottom panels of the figure, the lowest bound on the ratio  $M_{Z^\prime}/g^\prime_{\rm SUSY}$ can be fulfilled at 2117 GeV when $M_{Z^\prime} =$ 1388 GeV, the corresponding gauge coupling being $g^\prime_{\rm SUSY} \simeq$ 0.66, $Q_Q = 1.11 \times 10^{-2}$ and $\chi^2_{\rm STU} = $ 2.64. The lowest bound on $M_{Z^\prime}/g^\prime_{\rm SUSY}$ increases drastically, up to 15.7 TeV,  when $g^\prime_{\rm SUSY}$ has its minimum value 0.25,  $M_{Z^\prime} = $ 3940 GeV and $\chi^2_{\rm STU} = $ 6.01.

The modules created by {\tt SARAH} for {\tt SPheno} calculate the full one-loop and partially two-loop-corrected mass spectrum. While the experimental value for the Higgs mass is very precise, {\tt SARAH/SPheno} maintains the uncertainty estimate around 2-3 GeV for sparticle masses \cite{Staub:2015kfa}. It was shown that the sparticle spectrum can shift the Higgs boson mass by 1-2 GeV \cite{Bahl:2019hmm,Gogoladze:2011db,Gogoladze:2011aa,Ajaib:2013zha,Un:2016hji}. 
	
For each solution with Higgs boson mass between 122-128 GeV, we make use of {\tt HiggsBounds}, which takes the Higgs sector predictions for each solution as input and then uses the values of production cross sections and decays from Higgs searches at LEP, the Tevatron and the LHC to determine if each parameter point has been excluded, at 95\% C.L. We accept all solutions with ratio ($k_0$) less than 1 where $k_0$ is defined as $k_0 = O_{\rm model}/O_{\rm obs}$, for $O$ a relevant observable,  for the process with highest statistical sensitivity.

Moreover, we also make use of {\tt HiggsSignals} which is the complement to {\tt HiggsBounds} and checks how good a solution reproduces the Higgs mass and rate measurements. It performs a statistical test of the Higgs sector predictions for the secluded UMSSM using measurements of Higgs boson signal rates and masses from the Tevatron and the LHC. To do this, we have applied peak-centered $\chi^2$-squared method along with a box-shaped pdf with Gaussian tails for the SM-like Higgs mass uncertainty. Then, we assume only solutions with total $\chi^2$ value less than 90, which is obtained by the peak-centered $\chi^2$ method for the SM-like Higgs boson.

\begin{figure}	
	$\begin{array}{cc}
	\includegraphics[width=.48\columnwidth]{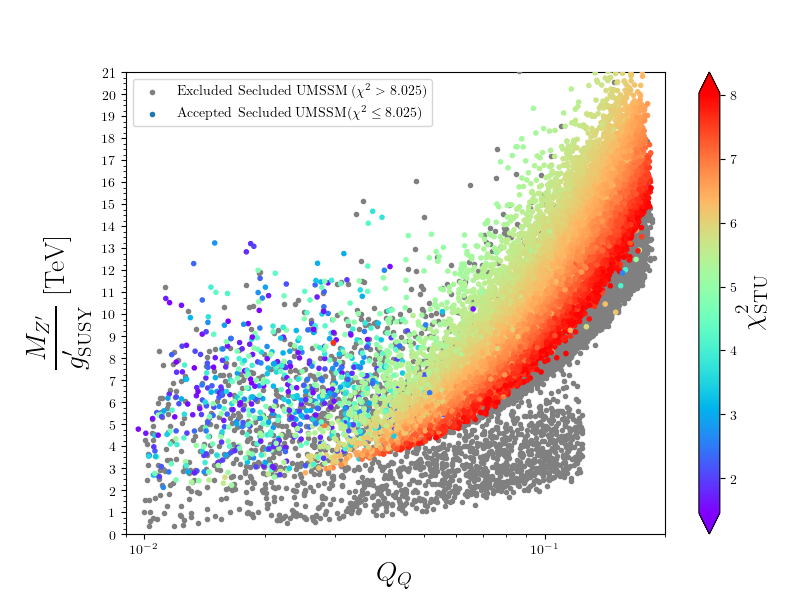}		
	\includegraphics[width=.48\columnwidth]{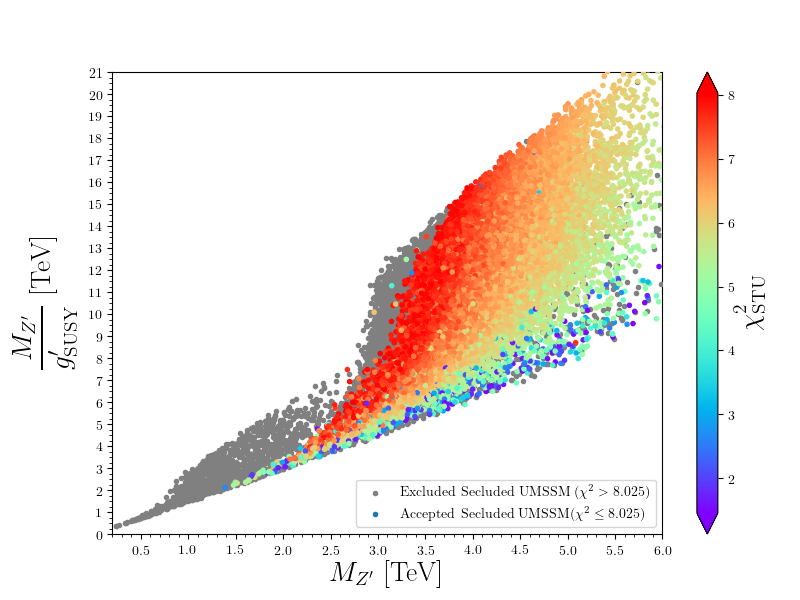}	\\
	\includegraphics[width=.48\columnwidth]{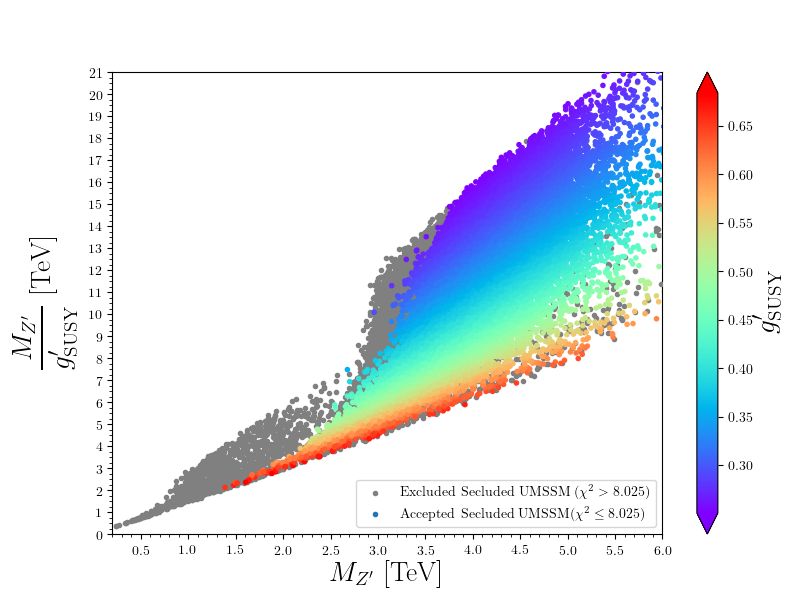}
	\includegraphics[width=.48\columnwidth]{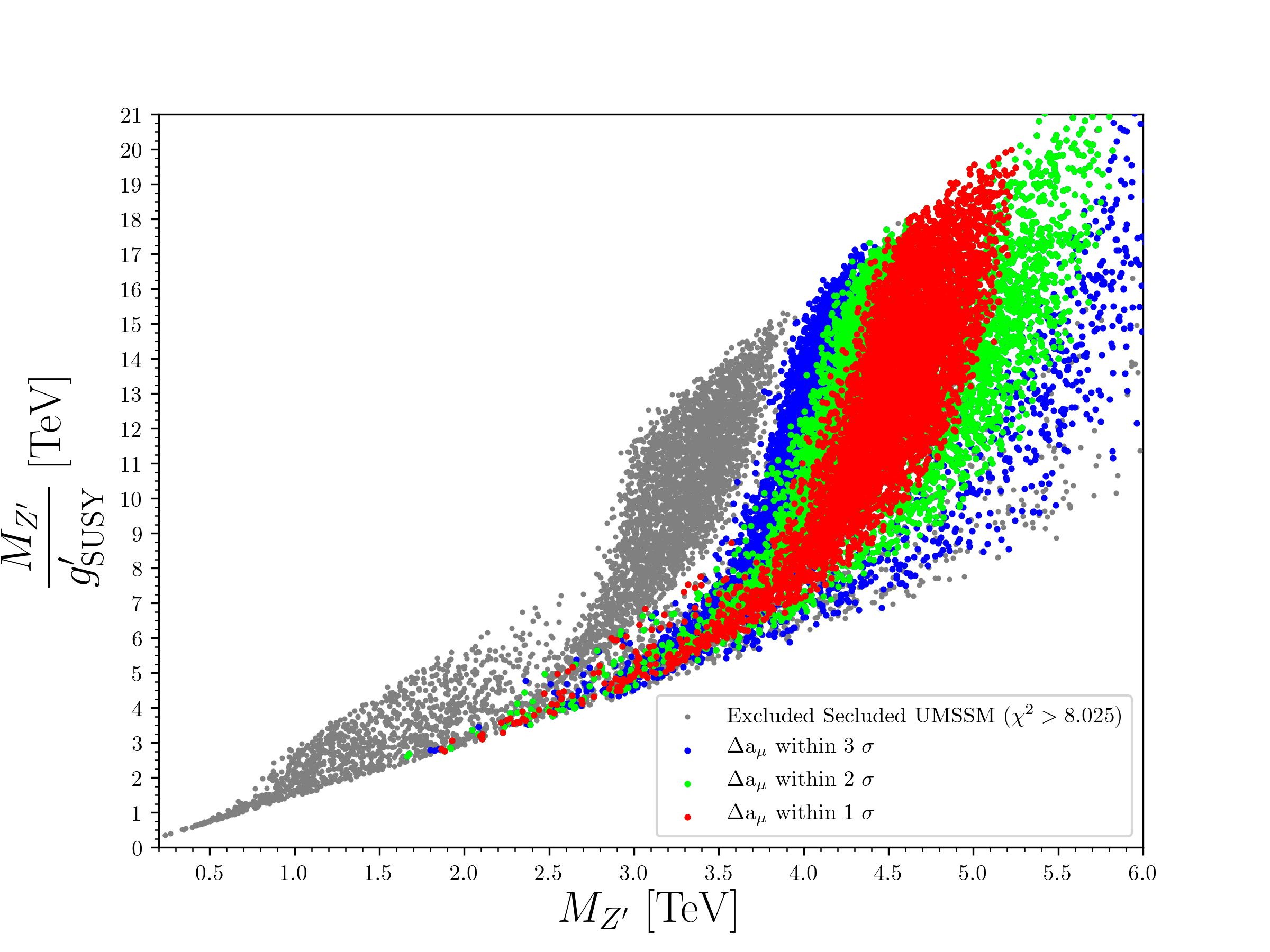}	
	\end{array}$
	\caption{The effect of oblique parameters and $(g-2)_\mu$ experimental bounds on the ratio  $M_{Z^\prime} /g^\prime$.}
	\label{fig:oblique}
\end{figure}

In Figure \ref{fig:ZpLimits} top left panel, we present the comparison of $\sigma(pp \to Z^\prime) \times$ BR$(Z^\prime \to \ell \ell)$ vs $M_{Z^\prime}$ consistent with the ATLAS data of \cite{Aad:2019fac}, scanning through the whole parameter space and displaying the values of BR($Z^\prime \to \ell \ell)$ in  different  color  codes. The experimental constraints are the same as in Figure \ref{fig:oblique} except that we  relax the $\chi^2_{\rm STU}$ value,   since we want to plot the branching ratios (BR) also for light $Z^\prime$ solutions which are  excluded by the $\chi^2_{\rm STU} $ bound. Since we fix $Q_\ell = Q_e = 0$, the $Z^\prime$ state  does not couple to $\ell \ell$. However, the small mass mixing  $Z-Z^\prime$ still allows the $Z^\prime$ to decay into $\ell \ell$ states, but only with BRs of $0.01\%$ for $M_{Z^\prime} \simeq $ 600 GeV while the BR decreases drastically for heavier $Z^\prime$ masses. The ATLAS observed limit on the fiducial cross section times BR ranges from 3.6 (13.1) fb at 250 GeV to about 0.014 (0.018) fb at 6 TeV for a zero (10\%) relative width signal in the combined di-lepton channel \cite{Aad:2019fac}. Therefore, our results imply a lower limit of $\sim$ 700 GeV at the 95\% CL on $M_{Z^\prime}$ for the $Z^\prime$ boson in the combined di-lepton channel. In the top right panel of Figure \ref{fig:ZpLimits} we compare the CMS high-mass di-jet yield from Ref. \cite{Sirunyan:2019vgj} with our predictions for $\sigma(pp \to Z^\prime) \times$  BR$(Z^\prime \to q \bar{q})$, obtained after scanning the secluded UMSSM parameters as described in Table \ref{tab:scan_lim} and imposing the constraints of Table \ref{tab:constraints}. For the sake of consistency with the experimental analysis, the $\sigma \times {\rm BR}$ rate  is  multiplied  by  an  acceptance  factor  $A=0.5$  and the  fraction  of $Z^\prime \to t\bar{t}$ events is not included in the calculation. 

These results are similar to those found in  $Z^\prime$ models which employ gauge kinetic mixing to achieve leptophobia. However, there are some differences. One is that, while in these other scenarios the di-jet BR of the $Z'$ cannot be lowered below 36\%,  in the secluded UMSSM  it can be lowered to 5\%. Another important aspect is that the model is also $d$-quark-phobic (the BR of $Z^\prime$ to $d$-type quarks is only about 1.4\%). This is a direct consequence of different $U(1)^\prime$ charge assignments, {in particular of the fact that imposing leptophobia results in $Q_d=\alpha=\frac{Q_u}{8}$,  Eq. \ref{U1ChargesIII}}.  Leptophobia and $d$-quark-phobia have thus further lowered the bound on the $Z^\prime$ mass by lowering its production cross section. Also, we benefit from new experimental acceptance ($A=0.5$ with the new data at ${\cal L}=137$ fb$^{-1}$ \cite{Sirunyan:2019vgj}, compared to $A=0.6$  at ${\cal L}= 27$ fb$^{-1}$ and $36$ fb$^{-1}$ \cite{Sirunyan:2018xlo}). From the top right panel of Figure \ref{fig:ZpLimits}, one learns that the computed $\sigma \times {\rm BR}$ is always below the CMS exclusion limits \cite{Sirunyan:2019vgj,Sirunyan:2018xlo} in the range 1.5 TeV $ < M_{Z^\prime} <$ 6 TeV at the 95\% CL, with the exception of a  tiny region around $M_{Z^\prime} \simeq 2.3$ TeV. One can, therefore, conclude that much  lighter $Z^\prime$ bosons consistent with the constraints given in Table \ref{tab:constraints} could  be  allowed  by  data  when leptophobic secluded UMSSM realizations, such as the one introduced in section \ref{sec:model}, are considered. In the middle left panel, we check the ratio $\Gamma(Z^\prime)/M_{Z^\prime}$ to assure that the Narrow Width Approximation (NWA) can be used consistently while in the middle right panel we investigate the  variation of the $Z^\prime$  mass limit with the $Q_Q$ charge,  $Q_Q=\alpha$, the  free parameter for the matter fields in the secluded $U(1)^\prime$ group. As seen from the color bar in the middle left panel, the $Z^\prime$  is quite narrow for the solutions found while the color bar of the middle right panel indicates that also the $\alpha$ parameter  should be quite small (less than $\alpha$ $<$ 2 $\times$ 10$^{-1}$). Moreover, one can see the correlation between $\alpha$ and $\Gamma(Z^\prime)/M_{Z^\prime}$. When $\alpha$ is increased, the $\Gamma(Z^\prime)/M_{Z^\prime}$ ratio also increases and approaches the CMS observed limits. As seen from the bottom left panel of Figure \ref{fig:ZpLimits},  $M_{Z^\prime}/g^\prime$ ratios below $\sim 3$ TeV require a decay width smaller than 1\% and a $Q_Q$ value smaller than $\sim 2\times 10^{-2}$. Finally, the bottom right panel of Figure \ref{fig:ZpLimits} shows the relation between various $Z^\prime$ masses and the $U(1)^\prime$ charges for the $S_1$, $S_2$ and $S_3$ secluded singlets, where we set $Q_{S_1} = Q_{S_3} = - Q_{S_2}/2 = \delta$ for simplicity. Solutions with lighter $Z^\prime$ masses necessitate smaller $\delta$ values while $\delta$ values increase for heavier $Z^\prime$ masses.  This relation can be understood via Eq. \ref{eq:HeavyZmasses}. 

\begin{figure}	
	\centering
	\includegraphics[width=.48\columnwidth]{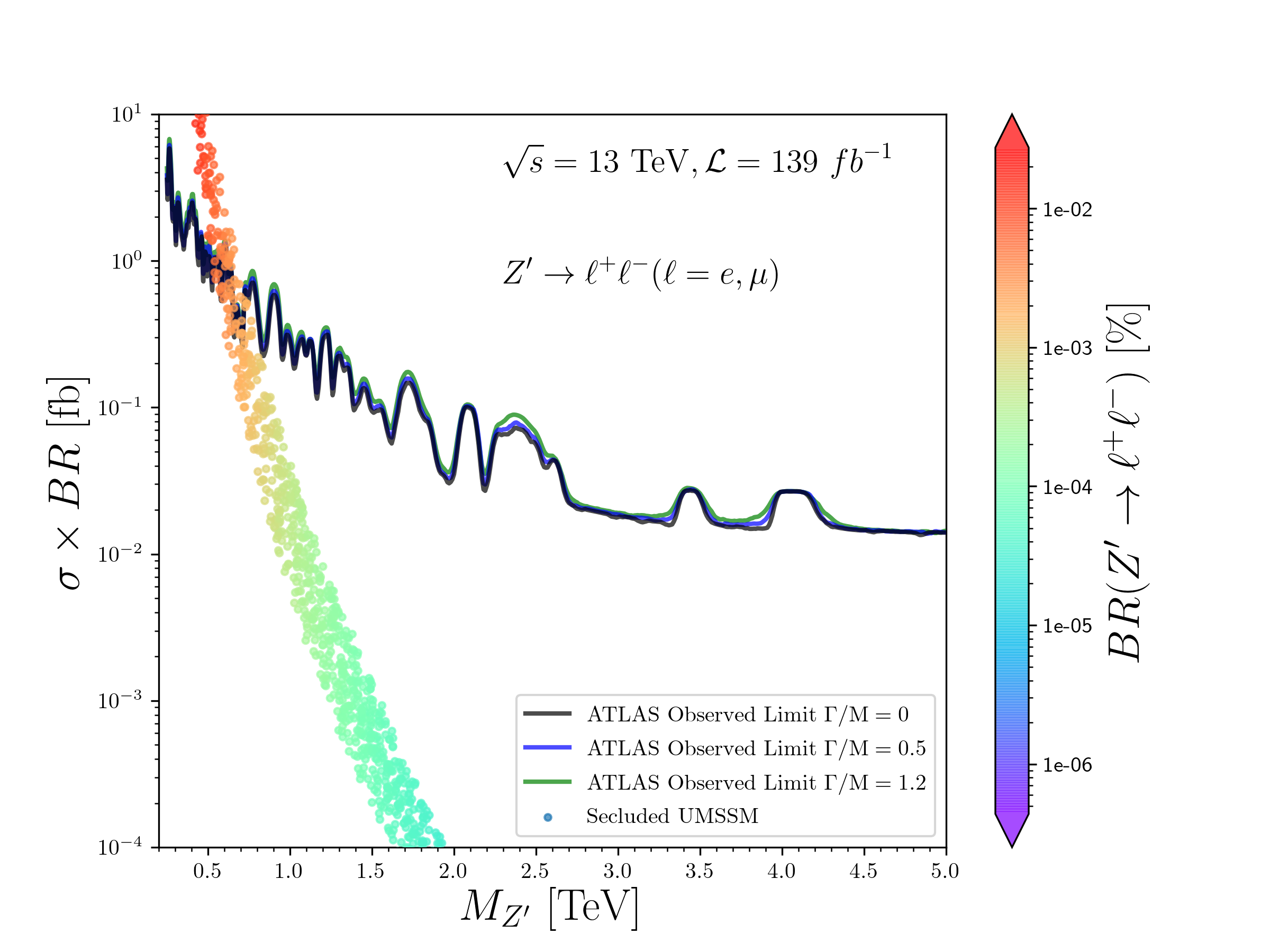} 
	\includegraphics[width=.48\columnwidth]{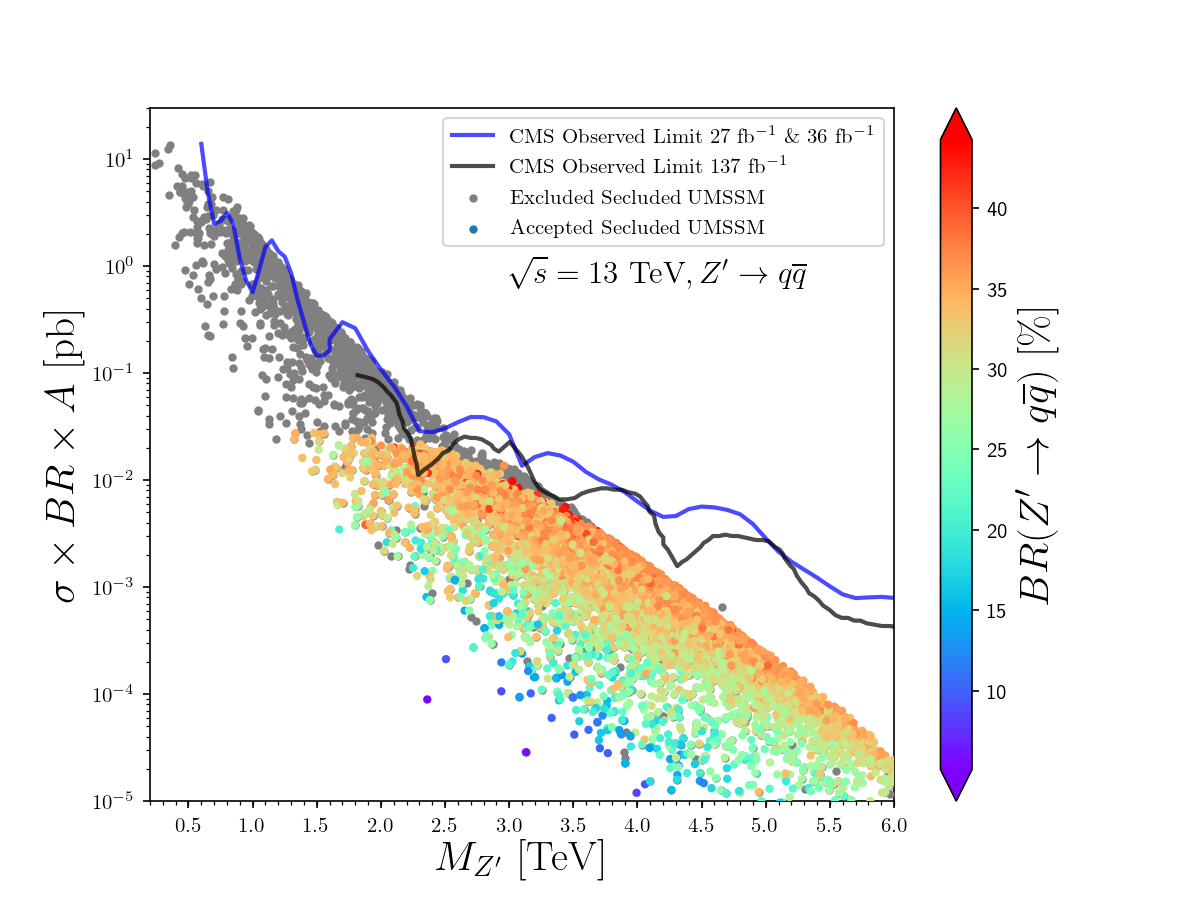} \\
	\includegraphics[width=.48\columnwidth]{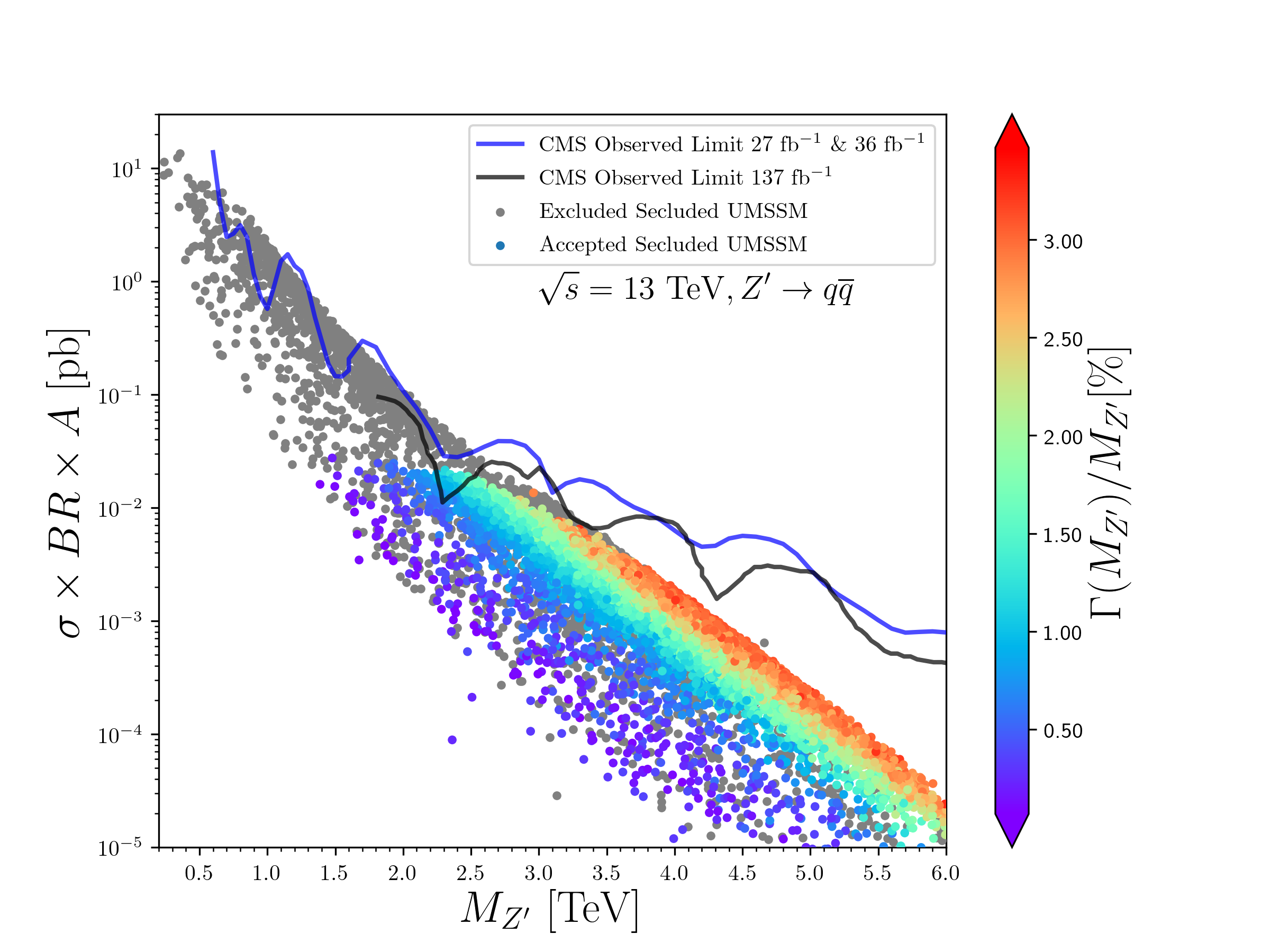} 
	\includegraphics[width=.48\columnwidth]{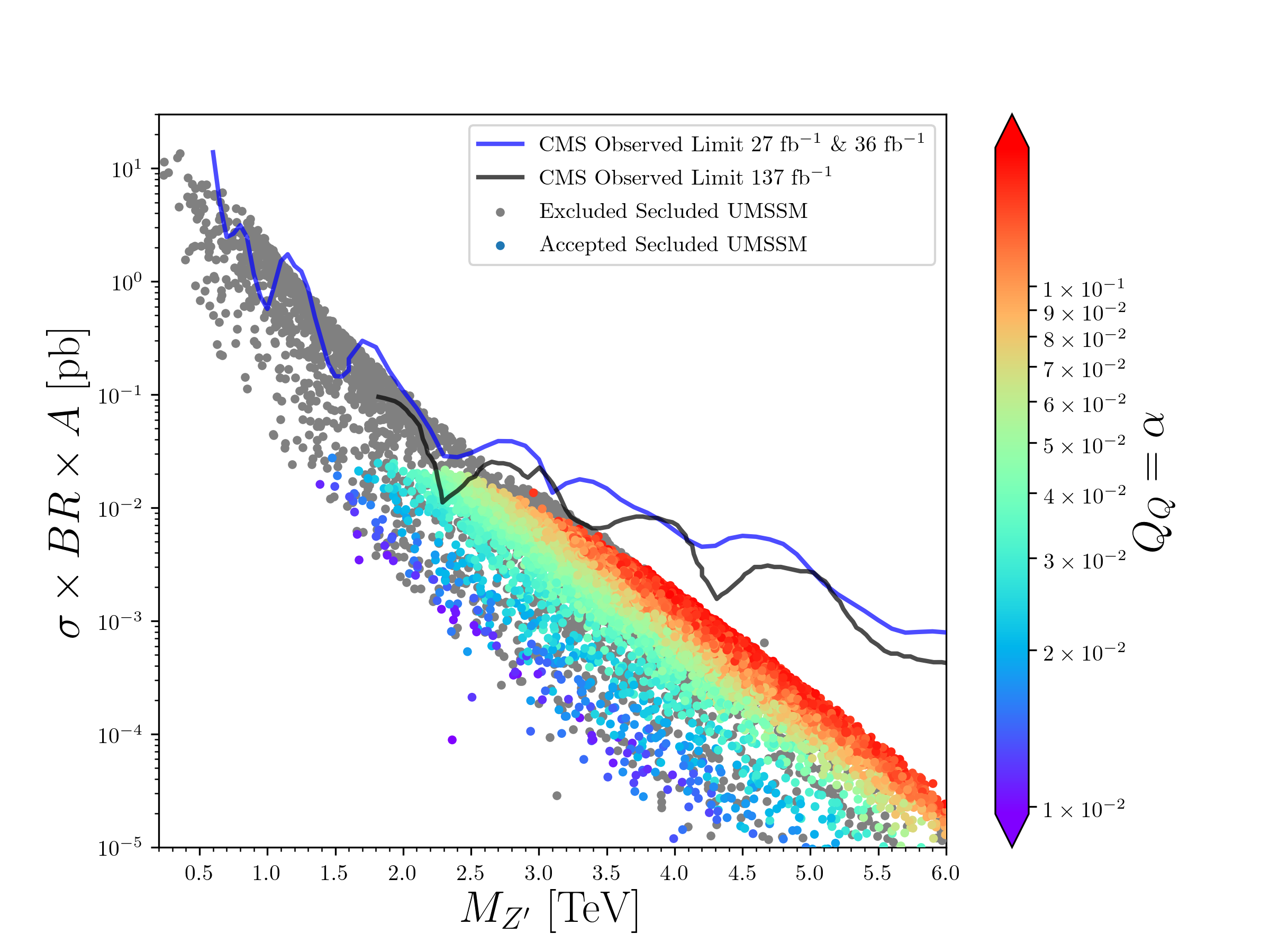} \\	
	\includegraphics[width=.48\columnwidth]{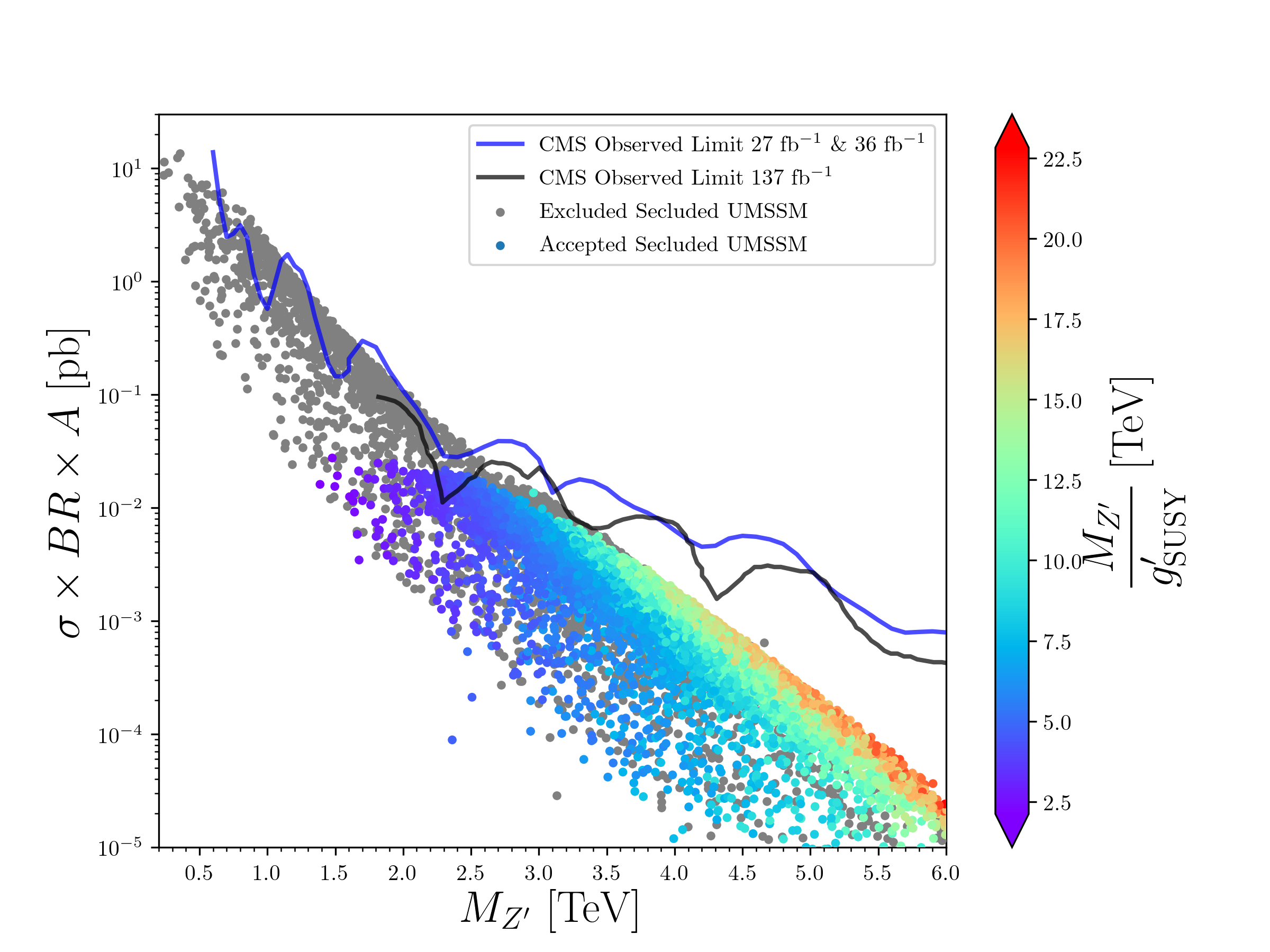} 
	\includegraphics[width=.48\columnwidth]{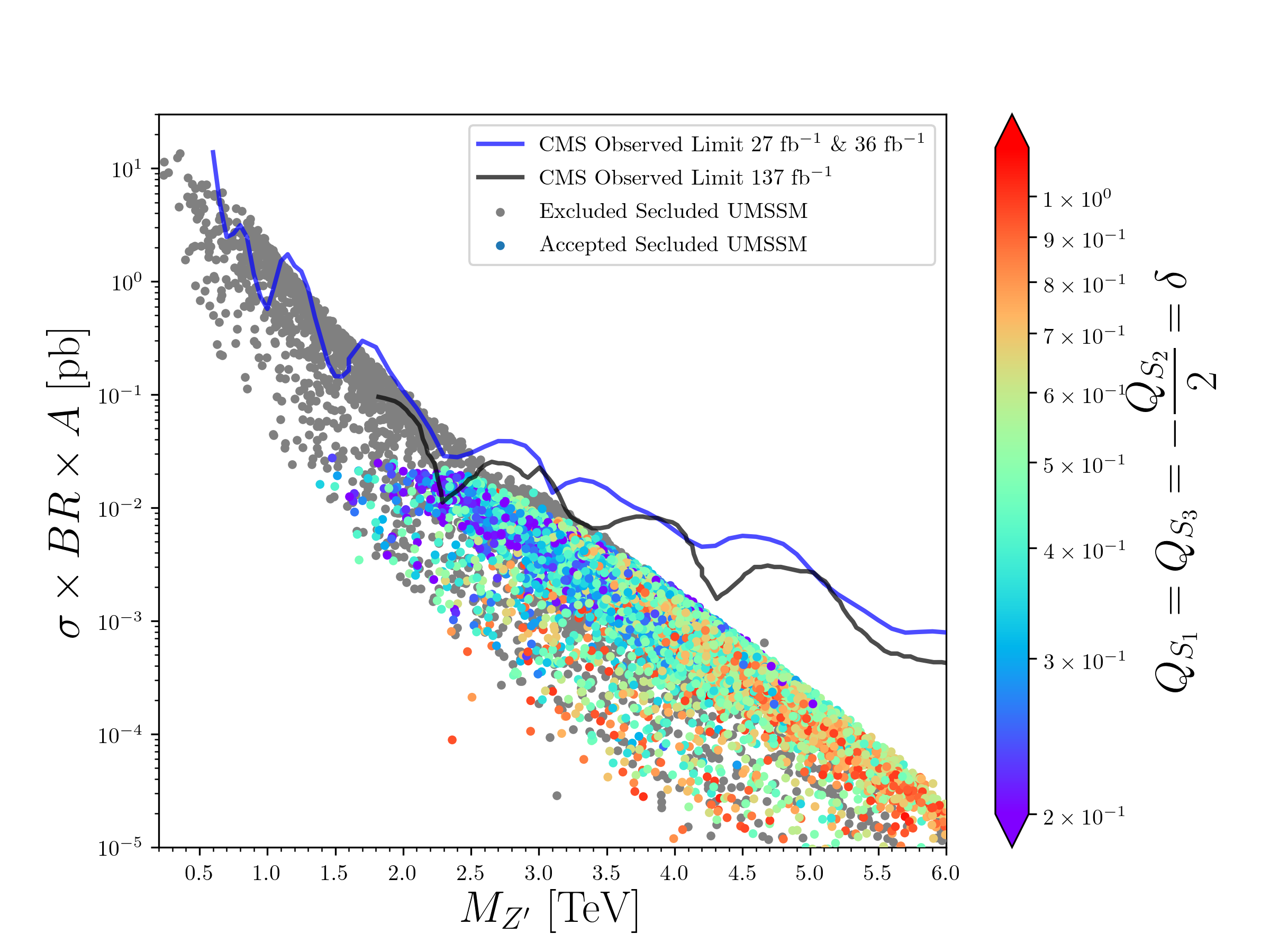}		
	\caption{Leptophobic $Z^\prime$ mass limits $(Q_\ell = Q_e = 0)$. We investigate the $Z^\prime$ production cross section multiplied by the di-lepton and di-jet BR (and by the acceptance $A=0.5$ for the latter), respectively. We compare theoretical predictions of the secluded UMSSM to the bounds obtained by the ATLAS \cite{Aad:2019fac} and CMS  \cite{Sirunyan:2018xlo,Sirunyan:2019vgj} collaborations.}
	\label{fig:ZpLimits}
\end{figure}

\section{Neutralino Dark Matter}
\label{sec:darkmatter}
In this section, we analyze the model parameters which survive cosmological bounds from the DM experiments. We investigate the constraints on the model arising from requiring the lightest neutralino  to be a  viable  DM  candidate,  with  properties  compatible  with  current  cosmological  data. 

{In the MSSM, the neutral higgsinos (nearly mass degenerate with the higgsino-like charginos)
could then play the role of the LSP. But the relic density upper limit
favors a neutralino with a large higgsino or wino component as the LSP.  A pure higgsino LSP cannot saturate the relic density constraint unless its mass is $\sim 1 $ TeV \cite{ArkaniHamed:2006mb}.  Consequently, one needs the admixture  of light binos and higgsinos  to form a light DM candidate, as the minimal ingredient of a natural MSSM \cite{Cao:2015efs}. For  DM lighter than about 100 GeV, in the MSSM, the chargino mass limit from the LEP experiments requires the DM to be bino-dominated. Then the weak interaction of the DM, together with a significant mass splitting of the DM from the other sparticles,  typically lead to
the overproduction of  DM in the early universe \cite{Cao:2015efs}. As a result, only a small corner of the MSSM parameter space survives. A DM candidate lighter than about 30 GeV has  been excluded in the MSSM \cite{Calibbi:2013poa}.
In the NMSSM, instead, it would be possible for the singlino to be quite light but, there, correct relic density  is obtained in the case when  a small singlino mass results only from mixing with the neutral higgsinos \cite{Ellwanger:2016sur}.

Previous studies of $U(1)^\prime$'s discussed light neutralino DM  \cite{Belanger:2015cra,Barger:2004bz}, before imposing limits from Higgs data and/or $Z^\prime$ mass and BR  constraints and outside a leptophobic scenario.  We revisit the light neutralino sector in our leptophobic scenario, while including all relevant constraints.}
First,  we demand that the predicted relic density agrees within 20\% (to conservatively allow for uncertainties on  the  predictions)  with  the  recent  Planck  results, $\Omega_{DM} h^2 = $  0.12 \cite{Ade:2013zuv,Aghanim:2018eyx}. We  calculate,  for  all points returned by our scanning procedure in Table \ref{tab:scan_lim} that are in addition compatible with current experimental bounds given in Table \ref{tab:constraints},  the  associated  DM  relic  density. We  present  our  results  in  Figure \ref{fig:relicdensity}.  

In all the subfigures, the relic density is plotted as a function of the mass of the lightest neutralino, denoted by $M_{\tilde{\chi}^0_1}$. As seen from the panels, solutions consistent with the relic density constraint emerge for almost all values of $M_{\tilde{\chi}^0_1}$ depending on the $\tilde{\chi}^0_1$ composition, which is given in the following basis: $ (\tilde{B^\prime}, \tilde{B}, \tilde{W}, \tilde{H}_u, \tilde{H}_d, \tilde{S}, \tilde{S}_1, \tilde{S}_2, \tilde{S}_3) $. The color bar of the top left panel of Figure \ref{fig:relicdensity} shows the $\tilde{S}$ content, as we are particularly interested in singlinos  as non-MSSM LSP candidates. One can learn from this panel that the relic  density  observed  by  the  Planck  collaboration  can be accommodated by $\tilde{S}$-like  $\tilde{\chi}^0_1$'s lying roughly in the [25, 300] GeV window, region largely {disfavored for MSSM neutralinos where universal boundary conditions are applied at the GUT scale \cite{Cao:2015efs,Frank:2017ohg,Araz:2017qcs}}. Once the lightest neutralino spectrum becomes heavier, the contribution of the combination of $\tilde{S}_1$, $\tilde{S}_2$ and $\tilde{S}_3$ singlets   increases, so as to become dominant for $M_{\tilde{\chi}^0_1}$ heavier than 400 GeV, as seen from the upper right panel of Figure \ref{fig:relicdensity}. In the middle left panel, we focus on the combined contribution of all singlinos, that is, $\tilde{S}$,  $\tilde{S}_1$, $\tilde{S}_2$ and $\tilde{S}_3$. As seen from the panel, singlino-like LSP solutions largely dominate the  parameter space. The middle right panel shows the higgsino-like neutralino content. As observed from the panel, the relic density is at the scale of $10^{-3}$ for higgsino-like neutralino with $M_{\tilde{\chi}^0_1} \sim 100$ GeV, but it increases dramatically for heavier higgsino-like neutralino masses. As in the MSSM, the relic density observed by the Planck collaboration can be accommodated by higgsino-like solutions at roughly $\sim 1$ TeV \cite{ArkaniHamed:2006mb}. Since TeV scale neutralino solutions are naturally less appealing from a collider point of view and we want to pay particular attention to singlino LSP scenarios, we did not increase the scanned neutralino mass range beyond 1 TeV. Although potentially viable scenarios could be obtained for  even  heavier  neutralinos (in particularly, for winos), for the purpose of this work, we ignore  this  regime  throughout. The bottom left panel of Figure \ref{fig:relicdensity} represents the bino composition of the lightest neutralino. Note that only solutions with bino contribution larger than 20\% are represented in the panel. Although there are some bino dominated $\tilde{\chi}^0_1$ solutions in our spectrum, their corresponding relic density mostly tends to lie in the [10, 100] range. An important fact is that the lightest bino-like solutions can be obtained  near 300 GeV. Bino contributions start to decrease, yielding  lower values of the relic density, and giving a maximum 50\% contribution, when the relic density constraint is satisfied and $M_{\tilde{\chi}^0_1} \sim 400$ GeV. The other $\sim$ 50\% contributions to mostly bino-like solutions consistent with the relic density constraint mainly come from higgsinos and winos, both of which contribute  more significantly for heavier $\tilde{\chi}^0_1$ masses, up to roughly 850 GeV, where we can classify the DM  as mixed neutralino states. We summarize the various lightest neutralino DM compositions  in the bottom right panel of Figure \ref{fig:relicdensity}. As seen from this panel, bino dominated neutralino solutions cannot be  good candidates for DM since they  do not satisfy the relic density constraints. Viable mixed (mostly bino and higgsino) neutralino DM solutions can be found with a mass lying in the 400--800 GeV range. When the spectrum is heavier, i.e., with a lightest neutralino  $M_{\tilde{\chi}^0_1} \in $ [0.8--1.0] TeV, the relic density as  observed  by  the  Planck  collaboration can be accommodated  by higgsino or singlino dominated solutions. It should be noted that $\tilde{B^\prime}$ contributions are no more than 5\% in the whole parameter space. Given that we mostly focus on small $Q_Q$ values,  this leads to small couplings with the gaugino $\tilde{B^\prime}$ associated with the $U(1)^\prime$ gauge group, so relatively small $\tilde{B^\prime}$ contributions are expected.
\begin{figure}	
	\centering
	\includegraphics[width=.48\columnwidth]{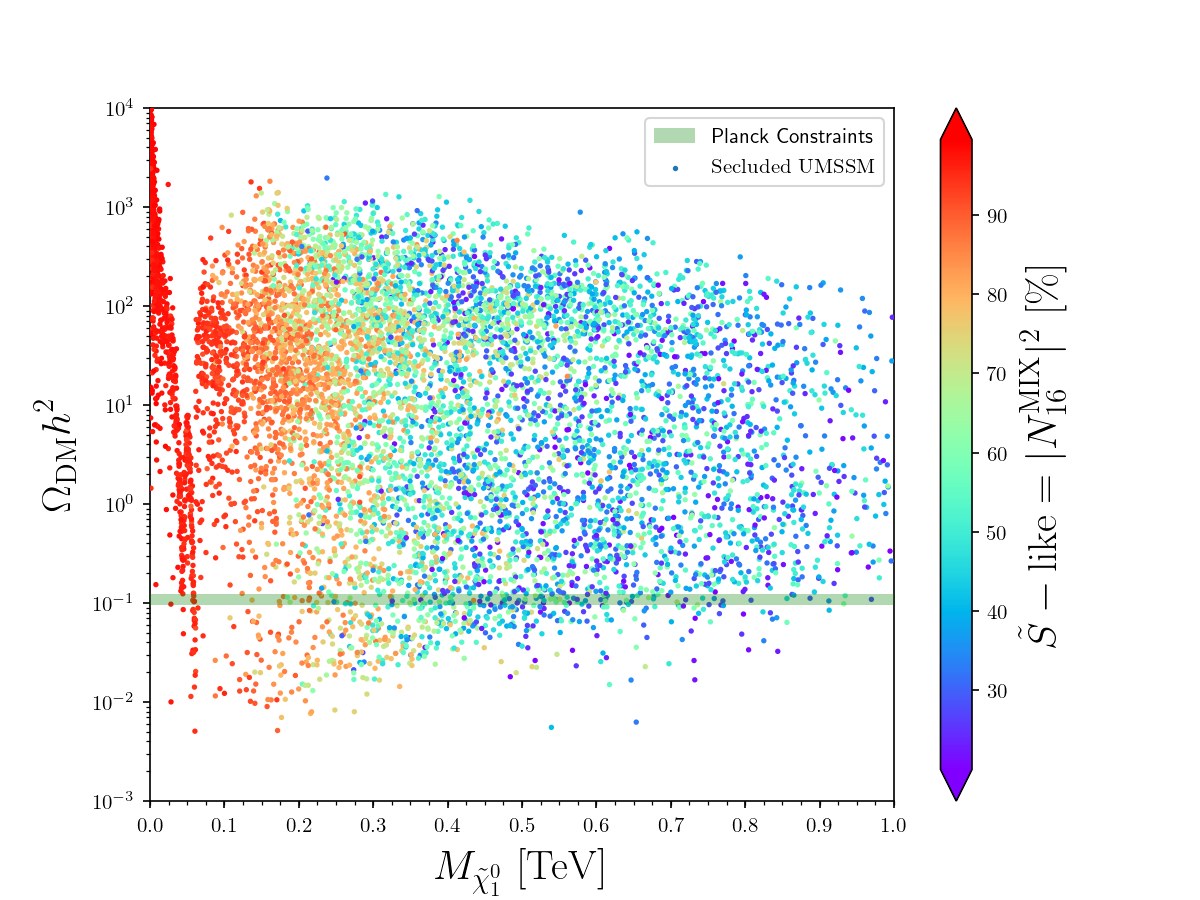} 
	\includegraphics[width=.48\columnwidth]{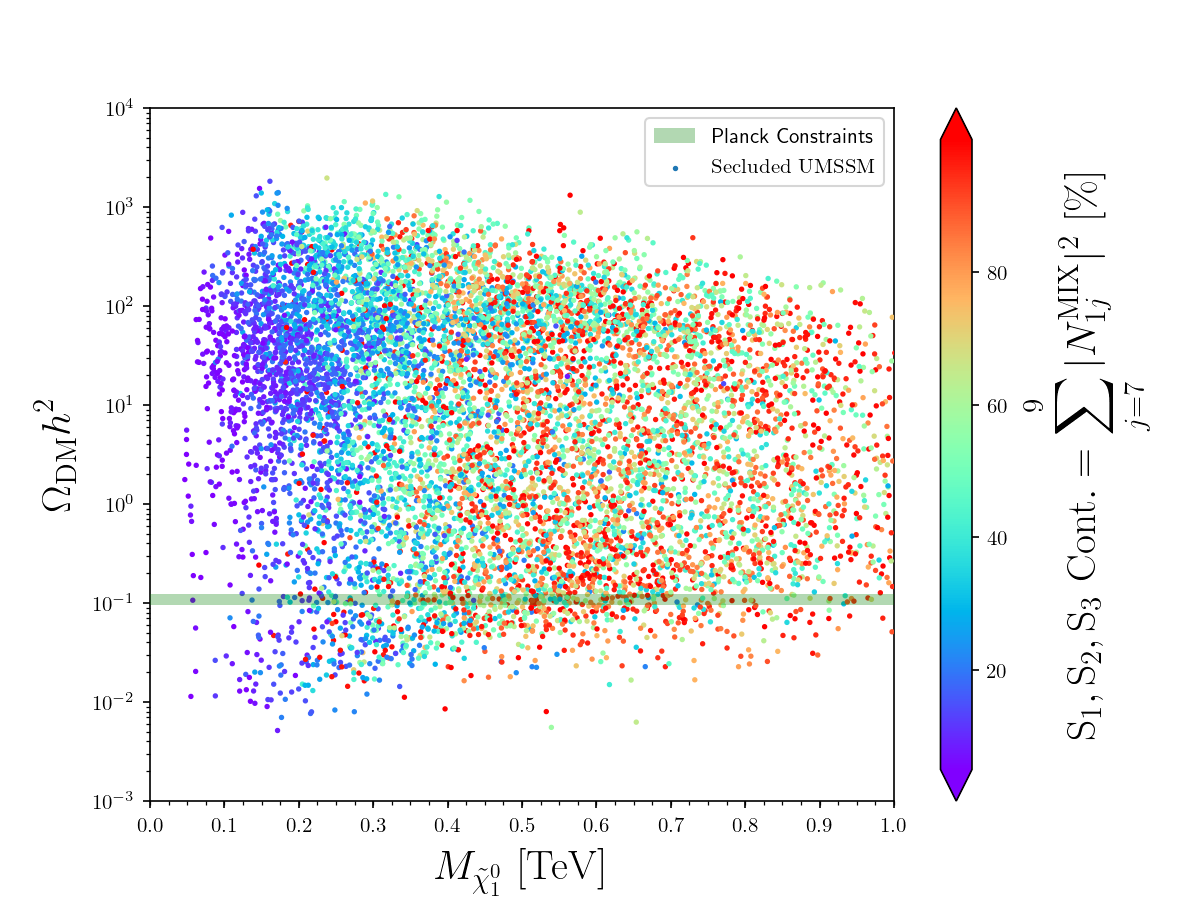}\\
	\includegraphics[width=.48\columnwidth]{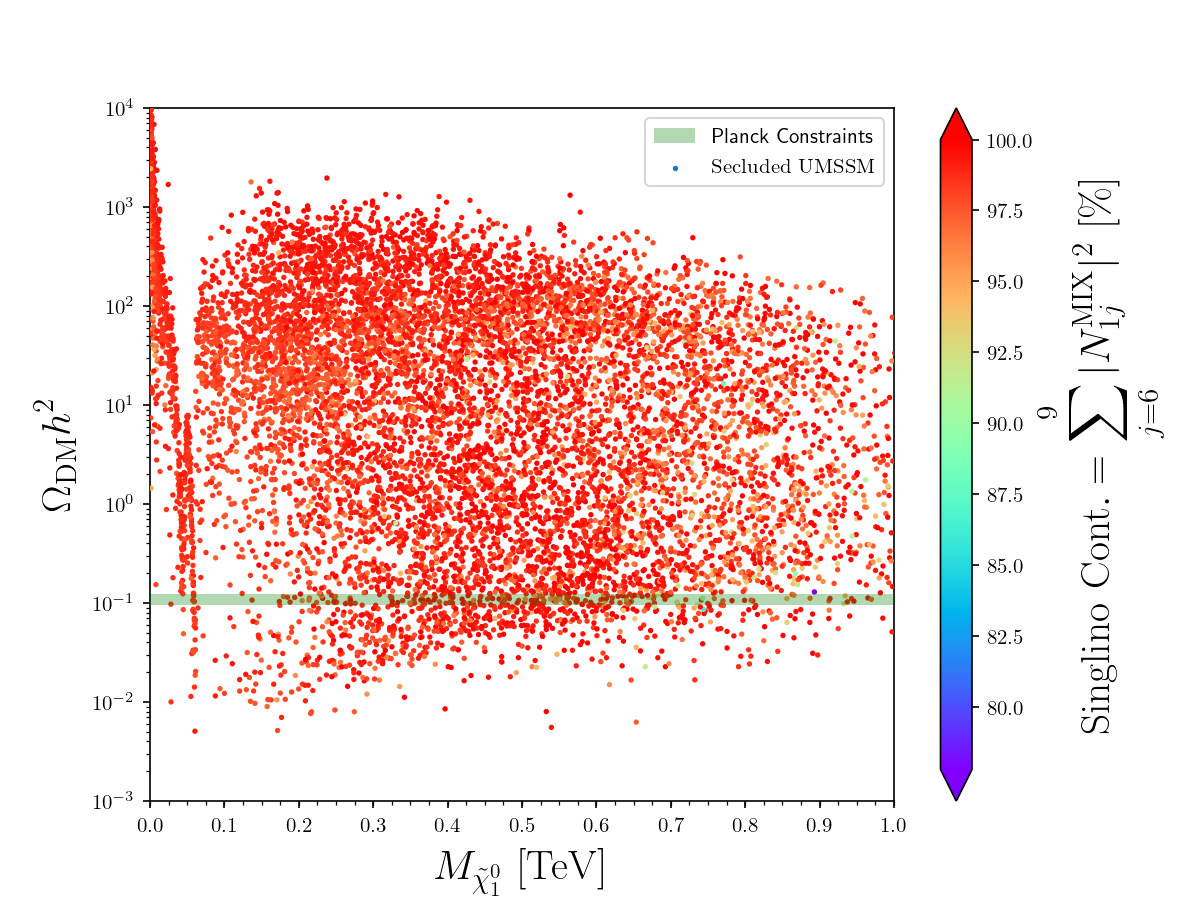}
	\includegraphics[width=.48\columnwidth]{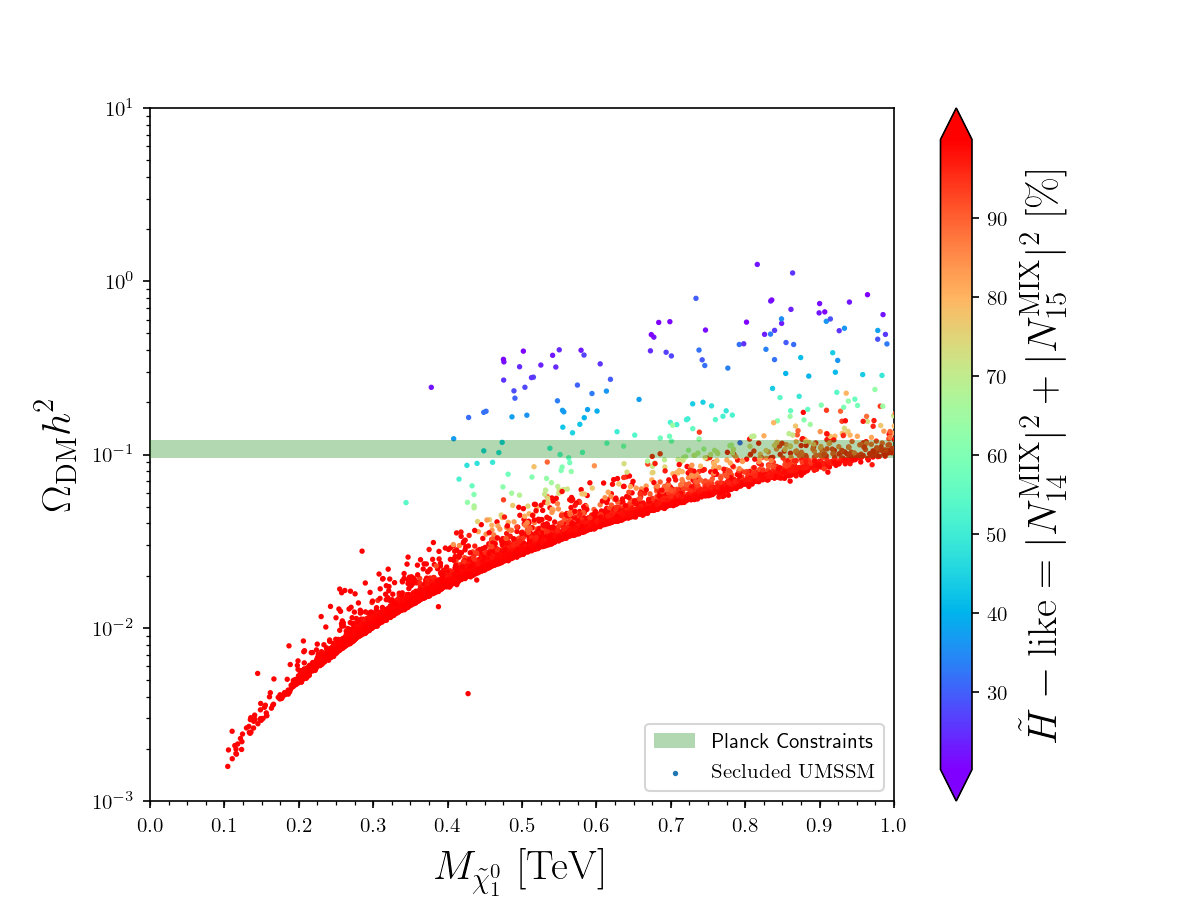} \\
	\includegraphics[width=.48\columnwidth]{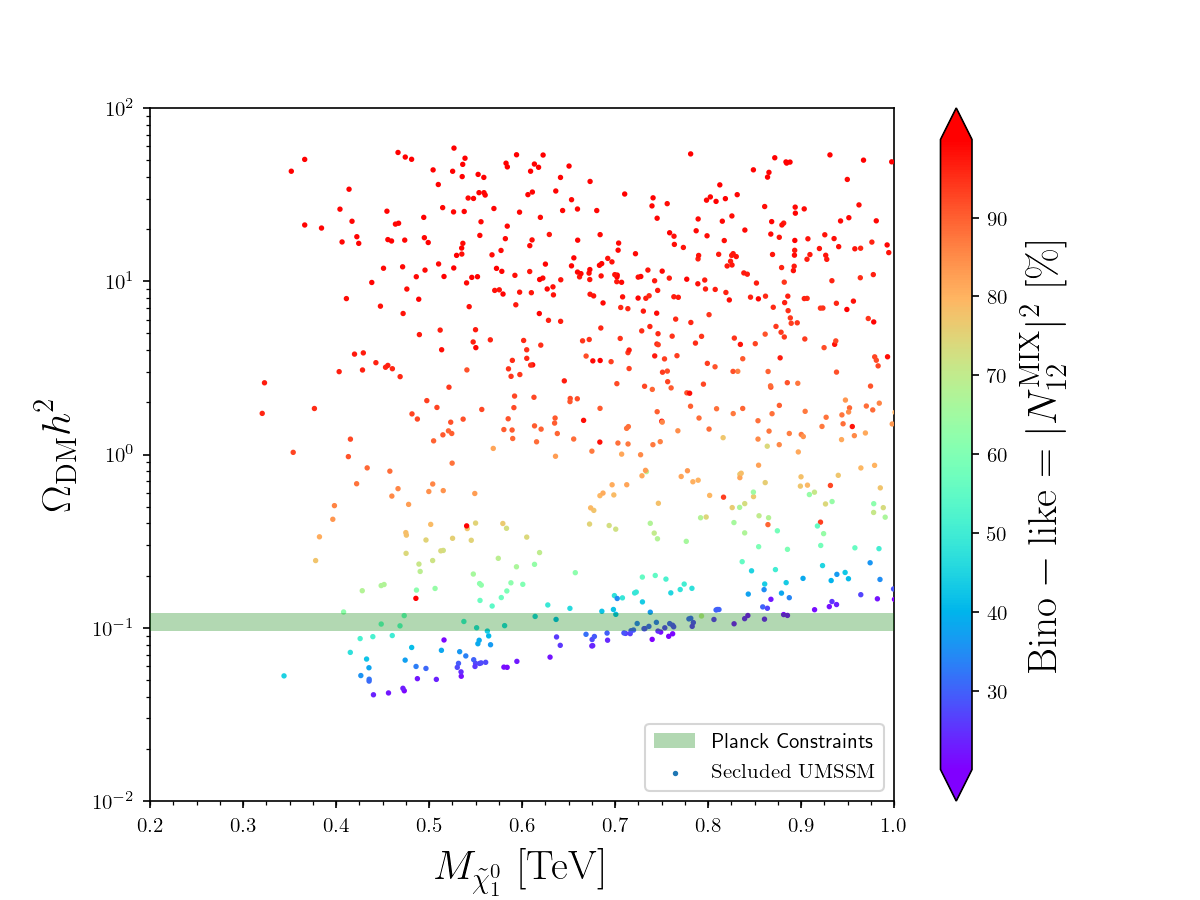} 
	\includegraphics[width=.48\columnwidth]{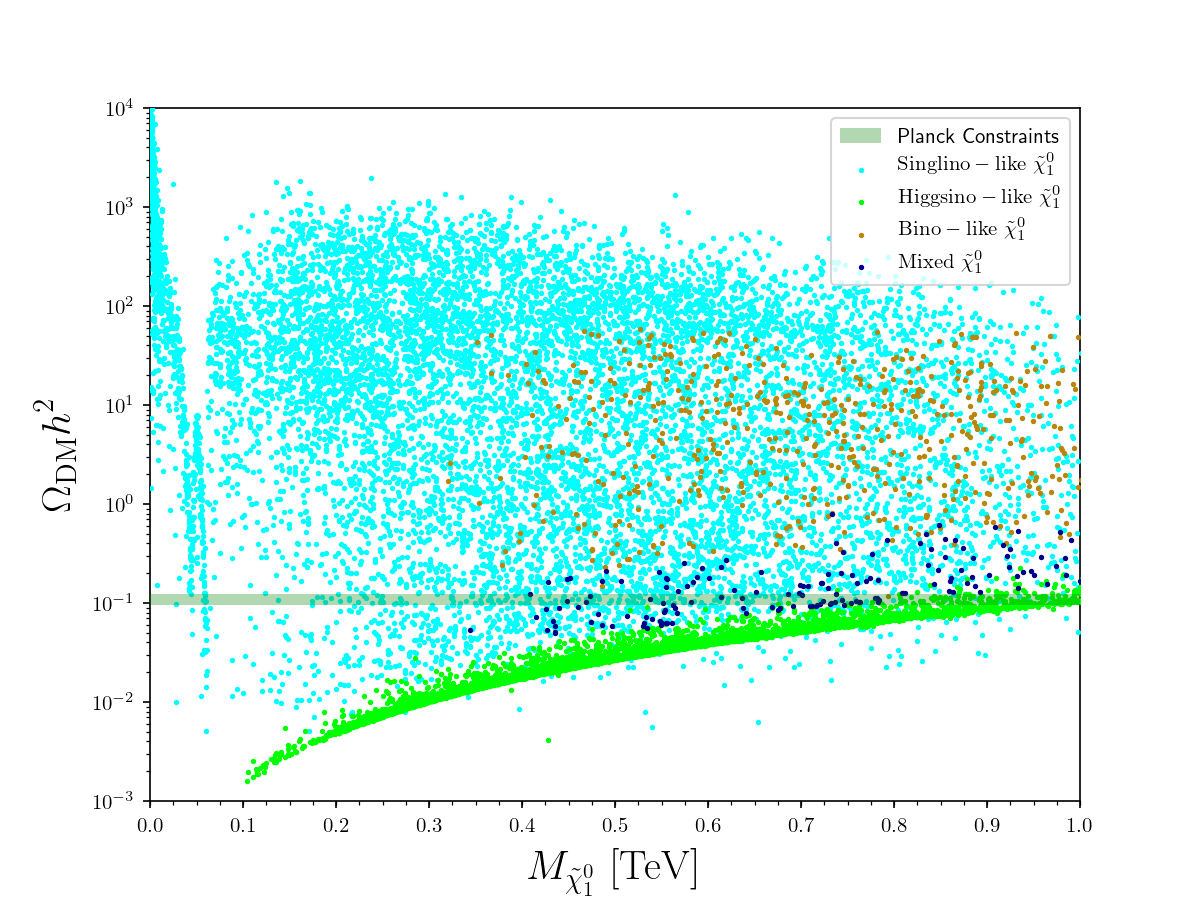} \\
	\caption{Relic  density  predictions  for   secluded UMSSM scenarios satisfying  all the constraints  imposed during our  scan  and  compatible with   $Z^\prime$ bounds from the LHC,  indicating the   dependence  on  the  mass  of  the  lightest neutralino. In each panel of the figure, we analyze the composition of the LSP for different parameter regions.  In the upper left panel, we represent by a color code the $\tilde{S}$-like contribution,  whilst in the upper right panel, we show the combined contribution of $\tilde{S}_1$, $\tilde{S}_2$ and $\tilde{S}_3$. In the middle left panel, we show the total contribution from the singlinos while, in the middle right panel, we present the composition of MSSM-like higgsinos. The bottom left panel shows the contributions of the mostly-bino solution while, in the bottom right panel, we indicate the parameter space populated by all the solutions. The horizontal green band in all panels indicates the measured value of the relic density,  consistent at $2\sigma$ with  the Planck experiment  \cite{Ade:2013zuv,Aghanim:2018eyx}.}
	\label{fig:relicdensity}
\end{figure}

Finally, we depict, in Figure \ref{fig:sIcrosssections}, the constraints coming from  direct detection experiments. The top panels show the spin-independent cross section for the nucleon as a function   of  the  mass  of  the  lightest  neutralino. Note that the results for spin-independent cross sections for proton and neutron are almost the same. Therefore, we denoted it as $\sigma_{\rm SI}^{\rm nucleon}$ and normalised it to the present-day relic density. The top left plane shows how the spin-independent cross section for the nucleon depends on the composition of the lightest neutralino for solutions which survive all the constraints given in Table \ref{tab:constraints}. Blue solutions in the top right panel refer to all solutions represented in the top left plane whilst all the other colors are subsets of blue and  represent solutions consistent with the relic density constraint in addition to the ones in Table \ref{tab:constraints}. The black line indicates the limits from the  Xenon 1T \cite{Aprile:2018dbl} with the region above the curve being excluded. In addition, the blue and red lines show the prospects for XENON nT and DARWIN \cite{Aalbers:2016jon} collaborations, respectively. As seen from the top left plane, almost all singlino solutions survive the results of the Xenon 1T experiment \cite{Aprile:2018dbl} while some portion of higgsino and bino dominated solutions are excluded. Another important feature is that all mixed neutralino solutions are strictly excluded by Xenon 1T. Once we compare our solutions consistent with the relic density bound to the result of Xenon 1T, a large fraction of higgsino dominated solutions consistent with the former are  excluded by the latter as seen from the top right figure. In contrast, singlino DM solutions consistent with the relic density bound are always below the excluded region by Xenon 1T and   can be probed by the next generation of DM experiments such as Xenon nT and Darwin.

Whilst we have demonstrated that the singlino-type lightest neutralino could be a viable DM candidate from the point of view of the relic density and direct detection bounds, at  the  same  time  it is  important to verify that DM indirect detection bounds are  also satisfied. In the bottom panels of Figure \ref{fig:sIcrosssections},  we  present  the  value  of  the  total DM  annihilation  cross  section  at  zero  velocity  as  a  function  of  the  lightest LSP neutralino mass for  all  scanned scenarios satisfying the $Z^\prime$ boson  limits from the LHC. Configurations for which the relic density is found in agreement with Planck data are shown along with their  higgsino, singlino and mixed compositions in the bottom right panel, whilst any other setup returned by the scan is shown in light sky-blue and tagged as ``Main Constraints'', referring to those given in Table \ref{tab:constraints}. In our predictions, we rescaled also the DM annihilation cross section to its present-day density.  We compare our predictions to the latest bounds derived from the Fermi-LAT data \cite{Ackermann:2015zua, Ahnen:2016qkx}. We depict, as a yellow area, the parameter space region that is found out to be excluded. The bottom panel of Figure \ref{fig:sIcrosssections} indicates that, unlike relic density and direct detection bounds, which impose strong constraints on the model parameters, indirect detection experiments are easily satisfied for a large portion of the parameter space. Most singlino DM scenarios naturally feature an annihilation cross section that is at least 3 or 4 orders of magnitude too small to leave any potentially visible signal in Fermi-LAT data. Therefore, singlino DM solutions are  unaffected  by  current  indirect  detection  limits  and will potentially stay so for some time by virtue of their correspondingly small annihilation cross sections.  In contrast, the annihilation cross sections of higgsino and mixed neutralino solutions are about 10$^{-26}$ cm$^3$ s$^{-1}$, hence, they are more likely to be probed by Fermi-LAT when the precision of the annihilation cross section measurement will be improved. 

\begin{figure}	
	\centering
	\includegraphics[width=.48\columnwidth]{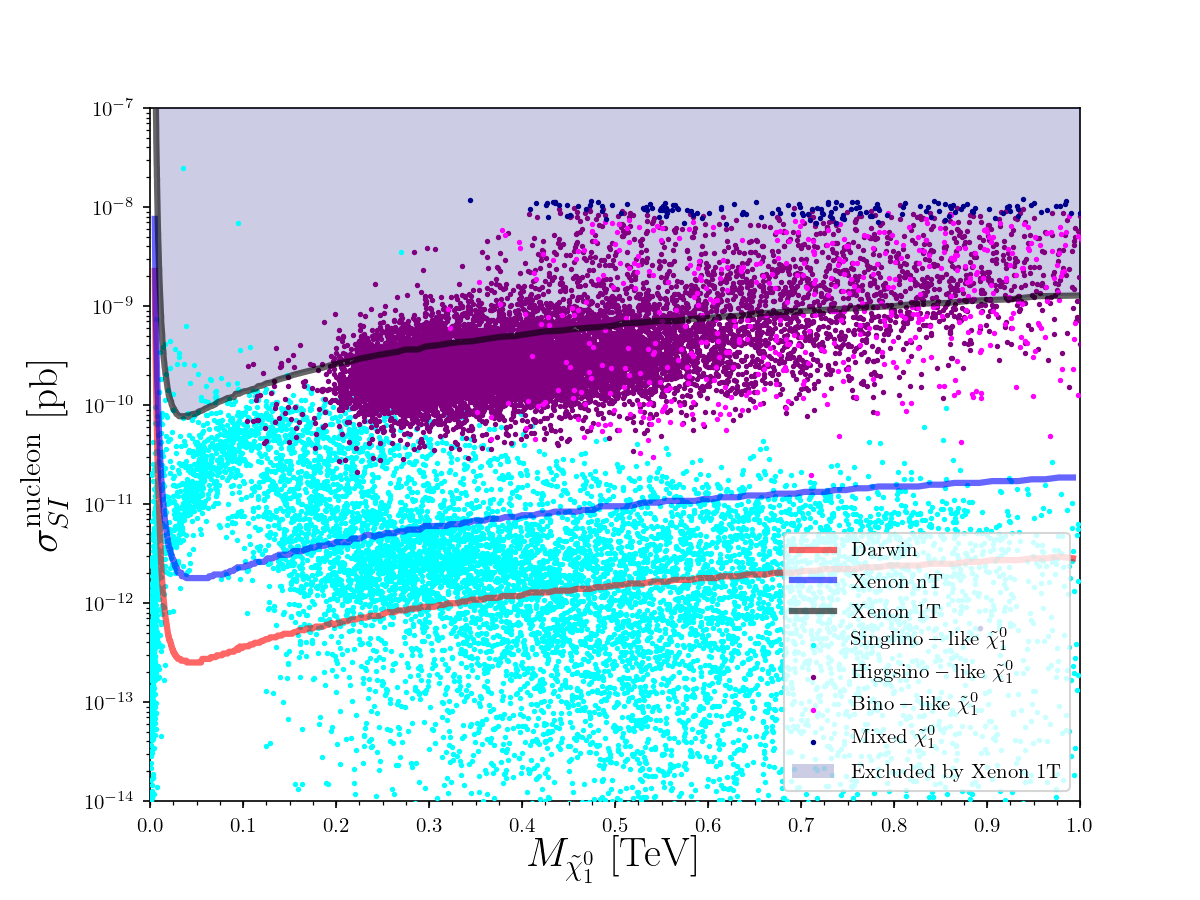} 
	\includegraphics[width=.48\columnwidth]{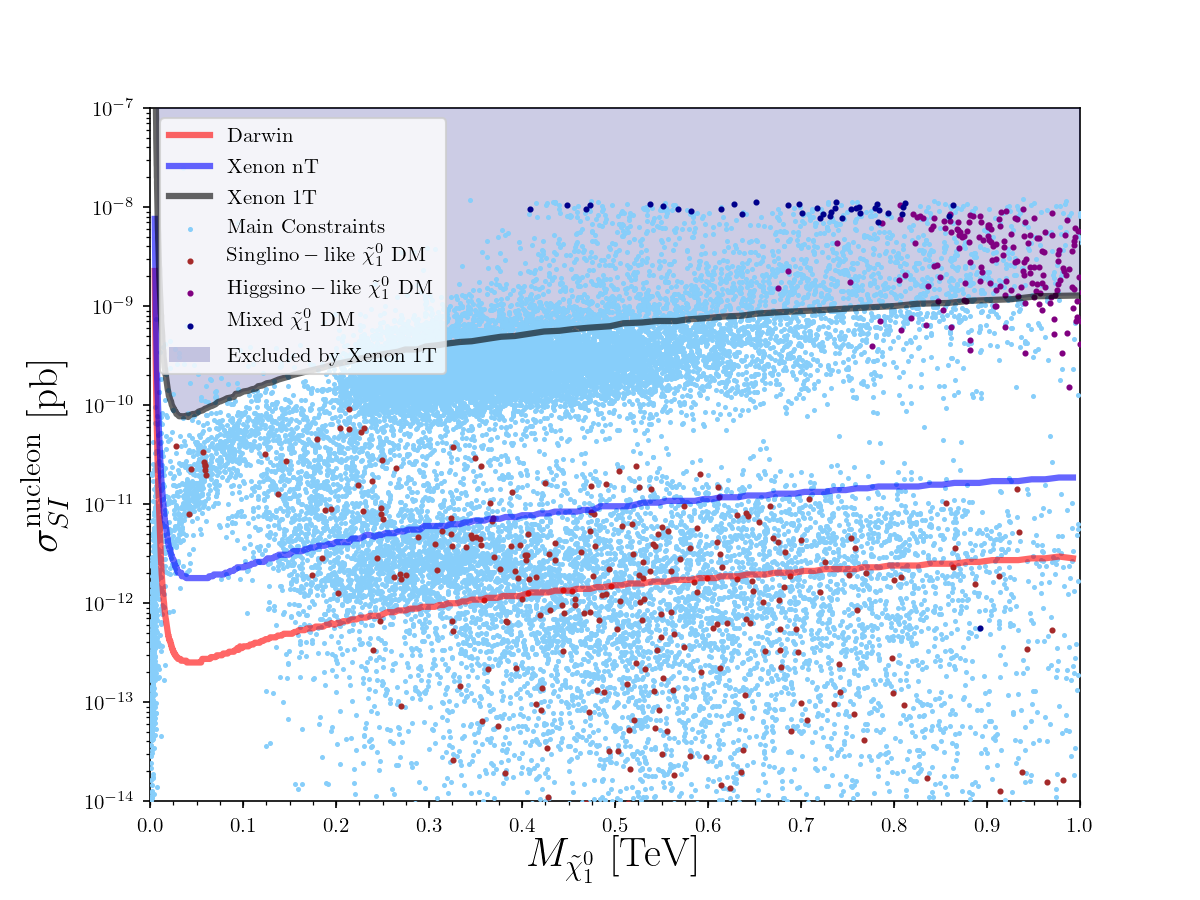} \\
	\includegraphics[width=.48\columnwidth]{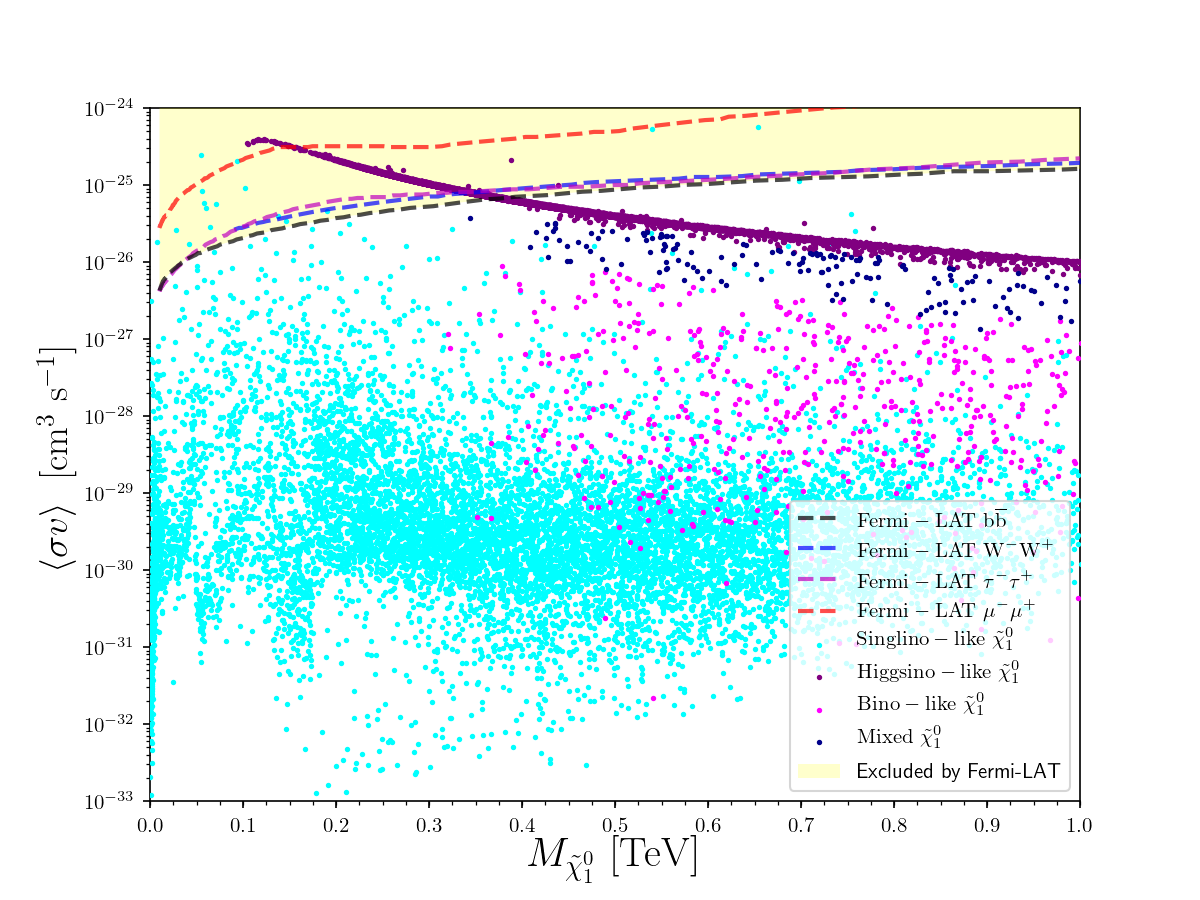}	
	\includegraphics[width=.48\columnwidth]{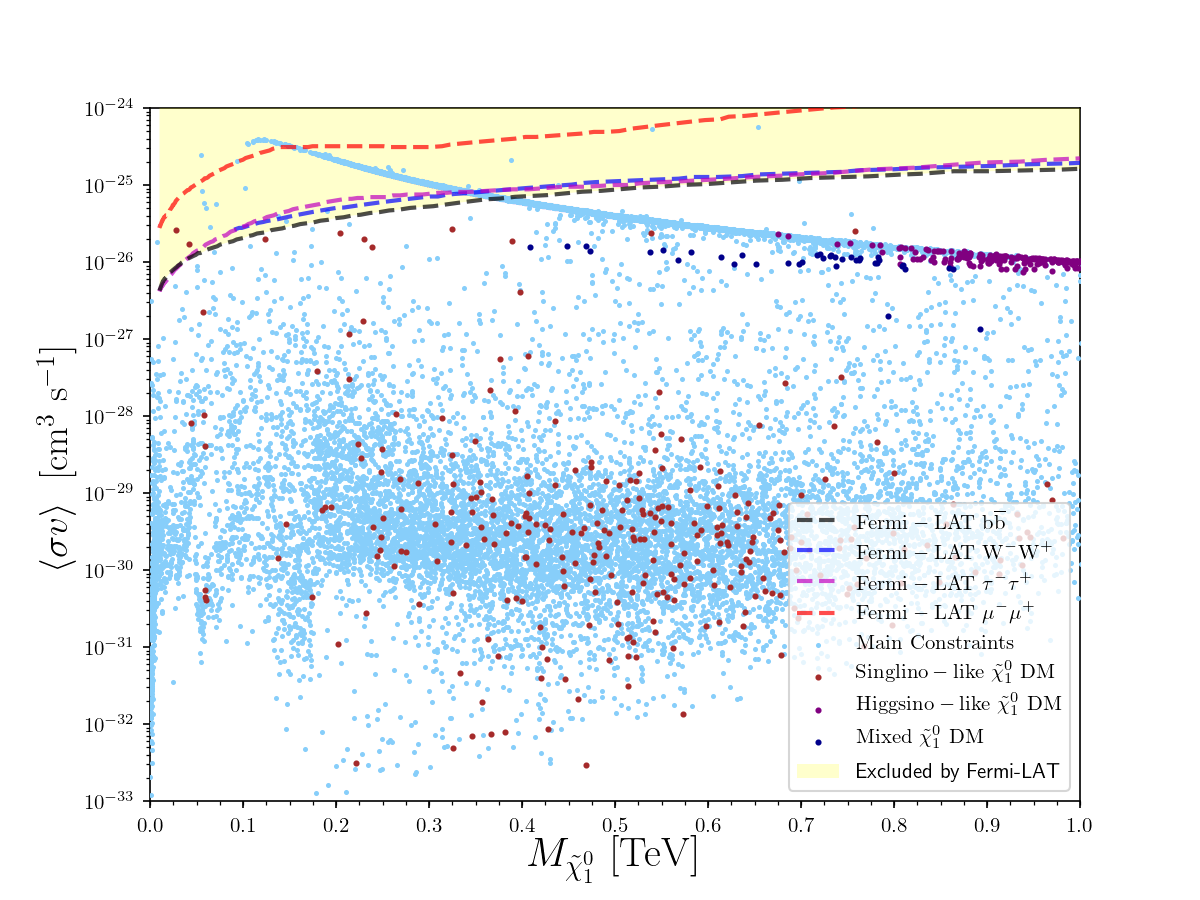}
	\caption{DM direct and indirect detection constraints on the parameter space on the secluded UMSSM model. The top panels show the constraints from the spin-independent cross section for the nucleon while the bottom panels show the corresponding annihilation cross sections.}
	\label{fig:sIcrosssections}
\end{figure}


In addition to the neutralino, in the secluded UMSSM, the sneutrino can be the LSP and thus a candidate for DM. Unfortunately, for most of the sneutrino LSP solutions  the relic density is overabundant compared to the requirements of the Planck Collaboration \cite{Ade:2013zuv,Aghanim:2018eyx}. In addition, the spin independent cross sections of all sneutrino LSP solutions are populated in the region excluded by XENON1T \cite{Aprile:2018dbl}. This is because, when the sneutrino is the LSP, it includes  more $\tilde{\nu}_L$ components than  $\tilde{\nu}_R$. Inevitably, then, the LSP sneutrino interacts more with $SU(2)_L$ doublets, and the spin-independent (SI) dark matter (DM)-nucleon cross section increases into the region excluded by XENON1T. Therefore, the LSP sneutrino is not a promising candidate in the secluded UMSSM.

\section{Muon anomalous magnetic moment}
\label{sec:muong2}
The measurement of the muon anomalous magnetic moment exhibits an intriguing discrepancy between the value found from the E821 experiment at BNL \cite{Bennett:2006fi} and the value predicted by the SM. Adding uncertainties, the deviations amount to 3.5 $\sigma$ \cite{Tanabashi:2018oca,Parker_2018} while recent theory predictions for $a_\mu$ find values as large as 4.1$\sigma$,
$$\Delta a_\mu \equiv a_\mu^{\rm exp}-a_\mu^{\rm SM}= 268(63)(43) \cdot 10^{-11}.$$
Several models  have been constructed and dedicated entirely to explain this discrepancy. Conversely, whether the discrepancy is real or not\footnote{Leading order hadronic vacuum polarization contributions represent the main limitation of theoretical calculations of non-perturbative  low-energy QCD behavior.}, it has been used as a test of how well BSM scenarios perform.

In the secluded UMSSM, loop diagrams with additional neutralinos and sleptons as well as with (right) sneutrinos and charginos provide additional contributions to the  $(g-2)_\mu$ observable. The parameter space is  restricted by limits on slepton masses from LHC. While these are not as restrictive as gluino or squark mass limits, bounds on selectron and smuon masses are 550 GeV and 560 GeV, respectively \cite{Sirunyan:2019ctn,Sirunyan:2018nwe},  whilst staus are allowed to be as light as 390 GeV \cite{Sirunyan:2019mlu,ATLAS:2019ucg}.

We present the results of our analysis in  Figure \ref{fig:muong2}, where we show  solutions consistent with the muon anomalous magnetic moment within $1\sigma$ of the experimental value. Here, we indicate the model solutions over the following planes:
($M_{\tilde\chi^\pm_1}, M_{\tilde\chi^0_1}$) (top left); ($M_{\tilde\chi^\pm_1}, M_{\tilde\chi^0_2}$) (top right); ($M_{\tilde\chi^\pm_1}, M_{\tilde\chi^0_3}$) (bottom left) and ($M_{\tilde\nu_1}, M_{\tilde\tau_1}$) (bottom right). When the lightest neutralino is singlino, the second and the third lightest ones are higgsino-like, rather light and almost degenerate in mass. The main contribution to the muon anomalous magnetic moment comes from these two heavier states as well as (albeit more marginally) from the lightest  (right) sneutrino and  (through slepton-mixing) stau states, in the appropriate diagrammatic combinations.  As seen from the figure, a large portion of the solution satisfies the $\Delta a_\mu$ bound within 1$\sigma$.  The grey region below the black curve represents the parameter region ruled out by ATLAS searches \cite{Aad:2019vvf,Aad:2019vnb}, close to which most solutions are found.

 At the same time, the anomalous magnetic moment of the electron was also measured  precisely to be \cite{Aoyama:2014sxa}

\noindent
$ a_e^{\rm exp}=1.15965218076(28) \times 10^{-3}$, while calculations within the SM, considering QED contributions up to ten loops,  obtain 
$ a_e^{\rm SM}=1.159652181643(25)(23)(16)(763) \times 10^{-3}$ \cite{Volkov:2018jhy,Aoyama:2017uqe}, yielding a difference close to 2.4$\sigma$ between experiment and theory for $\Delta a_e$, and of the opposite sign than the corresponding one for  the muon:
$$\Delta a_e \equiv a_e^{\rm exp}- a_e^{\rm SM}=-(8.8\pm 3.6) \times 10^{-13}.$$

The discrepancy was studied recently in the literature \cite{Volkov:2018jhy,Volkov:2017xaq,Aoyama:2019ryr}, and specifically in Two-Higgs Doublet Models \cite{Chun:2019oix} and $3-3-1$ models \cite{DeConto:2016ith}.
Unfortunately, in the context of our model, we cannot explain both discrepancies. The experimental observation is 
$$\frac{\Delta a_e}{\Delta a_\mu} \sim (-14) \frac{m_e^2}{m_\mu^2}\, ,$$
while is known that if the BSM scenario chosen is flavor-blind, as is in our case 
$$\frac{\Delta a_e}{\Delta a_\mu} \sim  \frac{m_e^2}{m_\mu^2}\, .$$
The latter is  consistent with our results.  A way out of this impediment would be to consider non-universal soft masses for smuons or selectrons. Moreover, the contributions to the electron and muon magnetic moments would have to be dominated by different diagrams with different signs. The latter would be possible if $M_1M_2 <0$, where $M_1, M_2$ are $U(1)_Y$ and $SU(2)_L$ gaugino masses, respectively, as chargino-sneutrino loops contribution is proportional to sign$(\mu M_2)$ while the neutralino-slepton contribution is proportional to sign$(\mu M_1)$ \cite{Badziak:2019gaf}. Our model has neither of these features. Thus  for the parameter regions consistent with $\Delta a_\mu$, even at 3$\sigma$, values for $\Delta a_e$ have the wrong sign and magnitude to satisfy the discrepancy between theory and experiment.

\begin{figure}	
	\centering
	\includegraphics[width=.40\columnwidth]{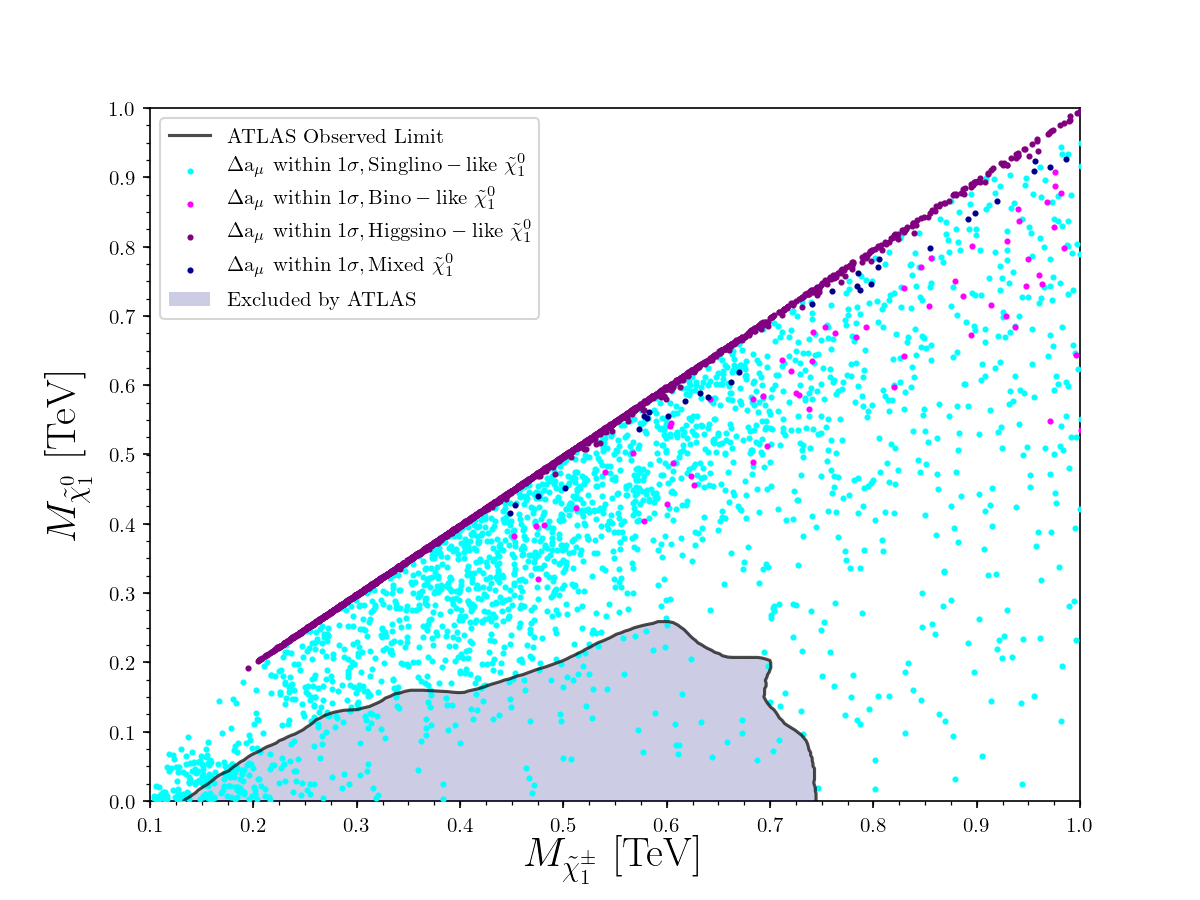}
	\includegraphics[width=.40\columnwidth]{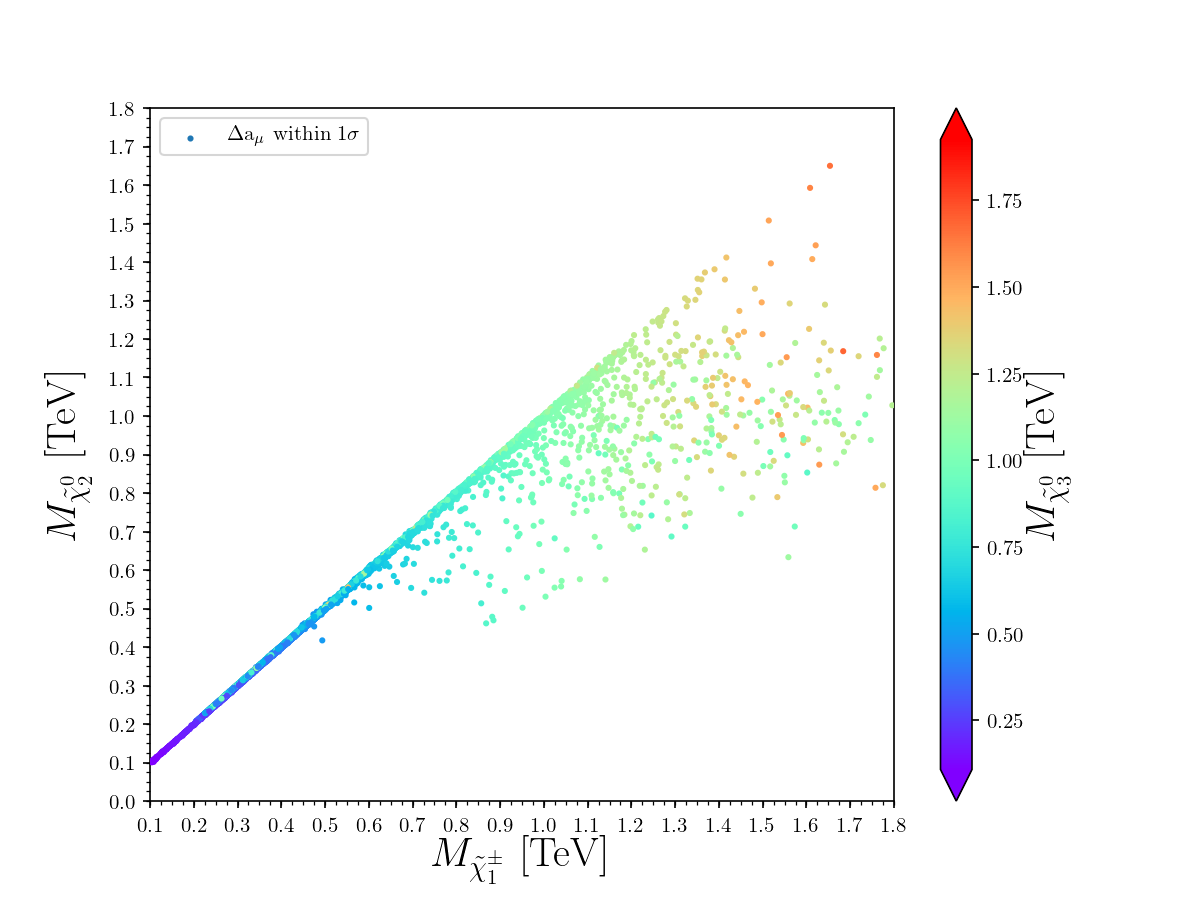} \\
	\includegraphics[width=.40\columnwidth]{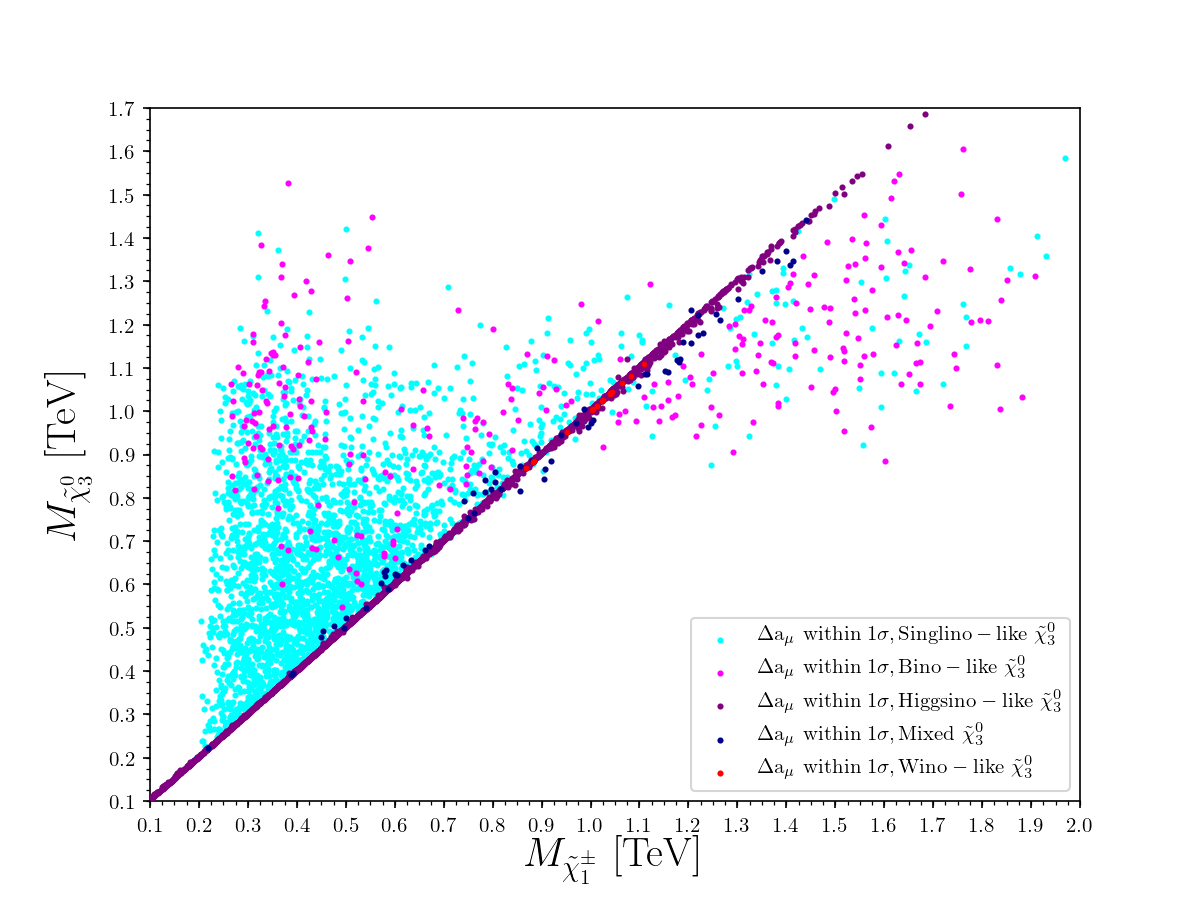} 
	\includegraphics[width=.40\columnwidth]{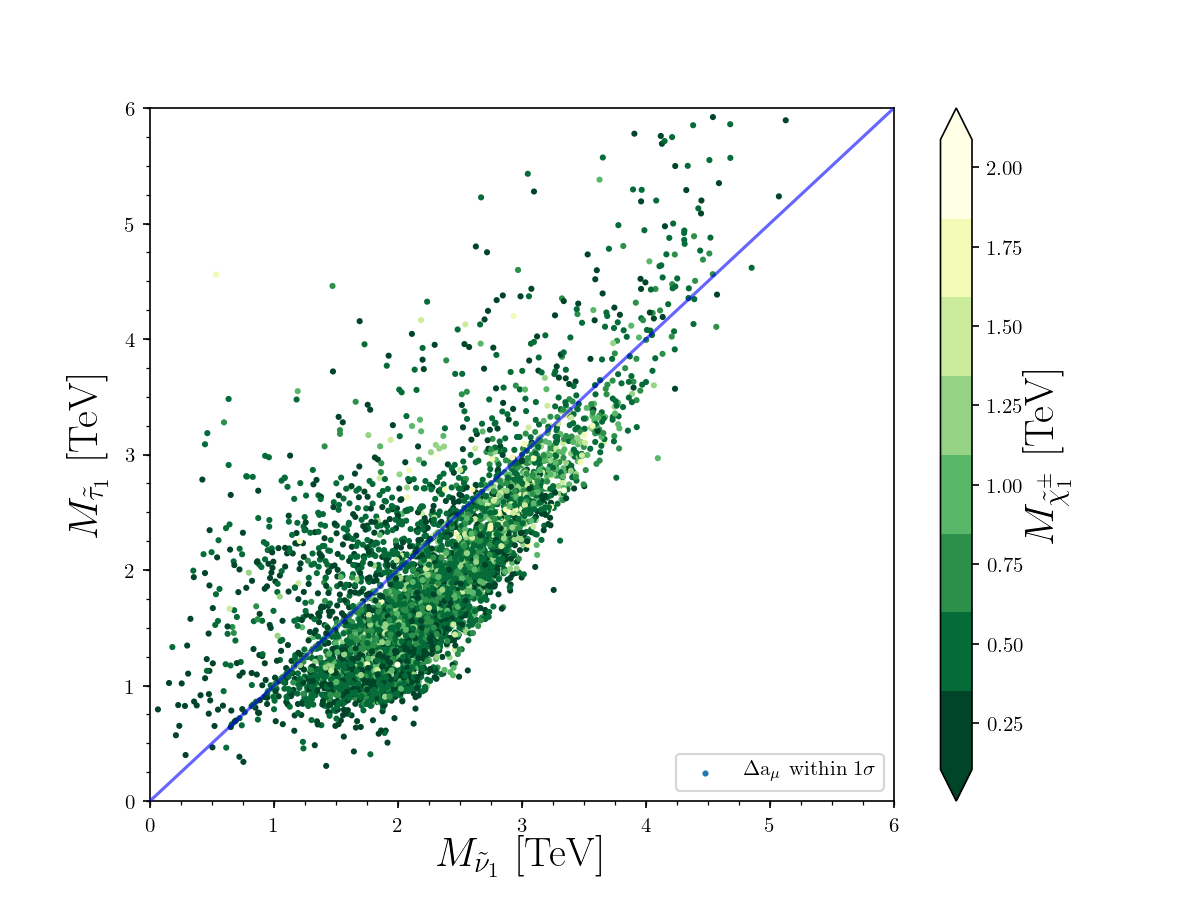}
	\caption{Parameter regions of chargino, neutralino, (right) sneutrino and stau masses consistent with $\Delta a_\mu$ within 1$\sigma$.  We show the following mass  mappings: (top left) lightest chargino versus lightest neutralino; (top right) lightest chargino versus   second lightest neutralino;  (bottom left) lightest chargino versus third lightest neutralino; (bottom right) lightest (right) sneutrino versus  lightest stau. The grey region is ruled out by ATLAS searches for chargino-neutralino states \cite{Aad:2019vvf,Aad:2019vnb}. The model  solutions to the $(g-2)_\mu$ discrepancy are dominated by the neutralino (higgsino-like)-slepton and chargino-sneutrino loop contributions, where, in particular,  the contributing neutralinos and charginos are  light yet consistent with all experimental constraints.}
	\label{fig:muong2}
\end{figure}

\section{$Z^\prime$ signal at colliders}
\label{sec:ZprimeSignal}

In this section, we investigate the observability of a secluded UMSSM scenario with light $Z^\prime$ masses at LHC.  To choose correct benchmarks, we first compare the range of chargino and neutralino masses with restrictions from the ATLAS searches for chargino/neutralino states \cite{Aad:2019vvf,Aad:2019vnb}. We make use of $\sms$ (version 1.2.2) \cite{Ambrogi:2017neo,Ambrogi:2018ujg,Dutta:2018ioj,Khosa:2020zar} in order to calculate the upper limit on the chargino-neutralino cross sections based on ATLAS-SUSY-2019-08 \cite{Aad:2019vvf} and ATLAS-SUSY-2018-32 \cite{Aad:2019vnb} implemented and validated with the $\sms$ authors. Figure \ref{fig:SModelSResults} showcases our results in terms of the lightest chargino and neutralino masses, as functions of the ratio between our calculated cross sections versus the upper limit on the chargino-neutralino cross sections. We exclude all solutions with signal strength value exceeding 1. This plot is complementary to the one shown in Figure \ref{fig:muong2} top left panel, with the grey region in that plot corresponding to the area below the curve. While in the former plot we indicate muon $g-2$ values consistent with experiment, here we explore neutralino and chargino masses constrained by bounds given in Table \ref{tab:constraints}, with the aim to choose benchmarks compatible with allowed EW-ino masses. Our plot indicates, however, that the parameter space allowed by this model is less restrictive than the one in the ATLAS analysis. We rule out some points for low chargino-neutralino masses (in red, lower left-hand corner) but allow the purple-blue points in the upper right-hand corner. {The reason why we can obtain light chargino masses, without introducing new charged particles in the model is the following. The $\mu$ parameter, which affects both chargino and neutralino masses, is generated dynamically in the model, and obtained by solving the Renormalization Group Equations (RGEs). This parameter, which affects chargino and neutralino masses, is obtained using the software \spheno \cite{Porod:2003um,Porod:2011nf}. The parameter space for EW-ino masses is consistent  with collider bounds from PDG \cite{Zyla:2020zbs} and the DM constraints from the previous section.}


We shall concentrate our analysis in this parameter region. 

\begin{figure}	
	\centering
	\includegraphics[width=.48\columnwidth]{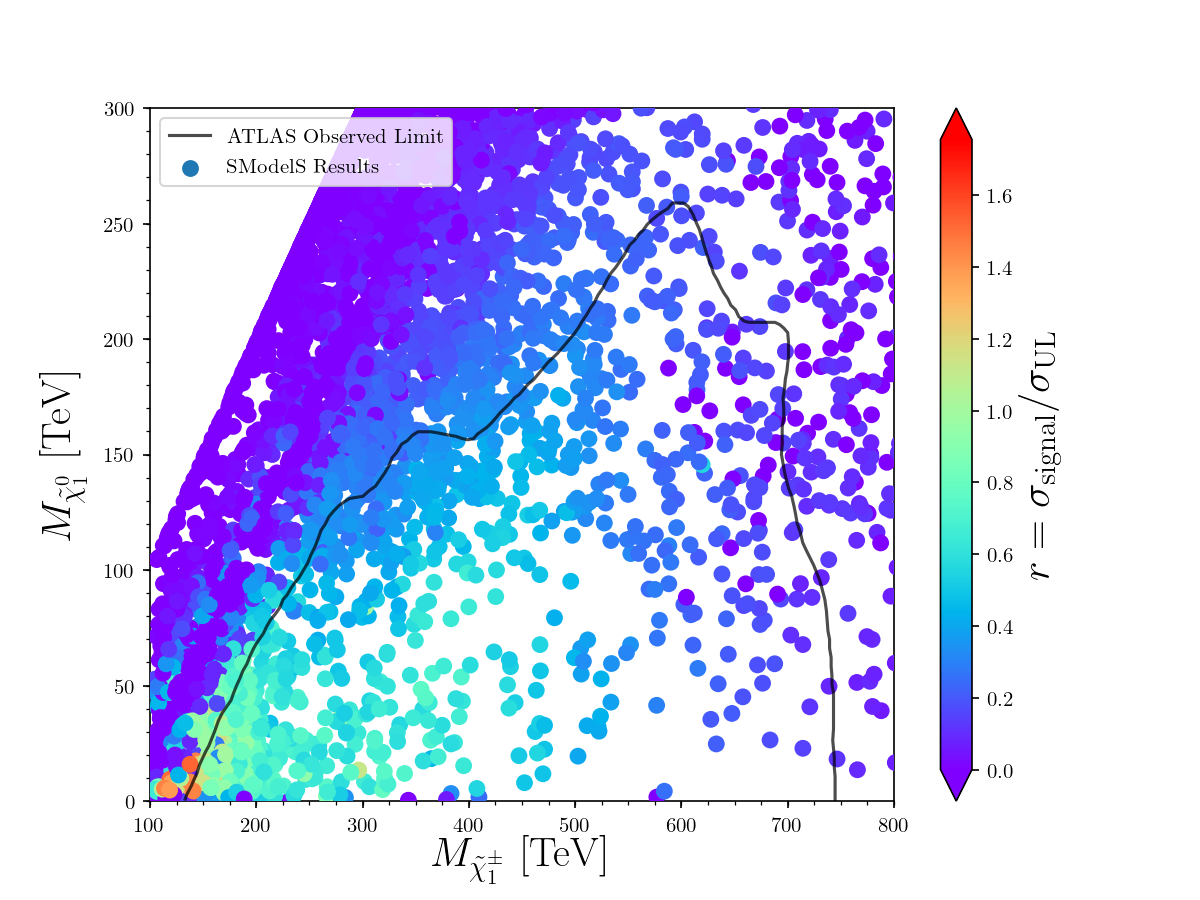}
	\includegraphics[width=.48\columnwidth]{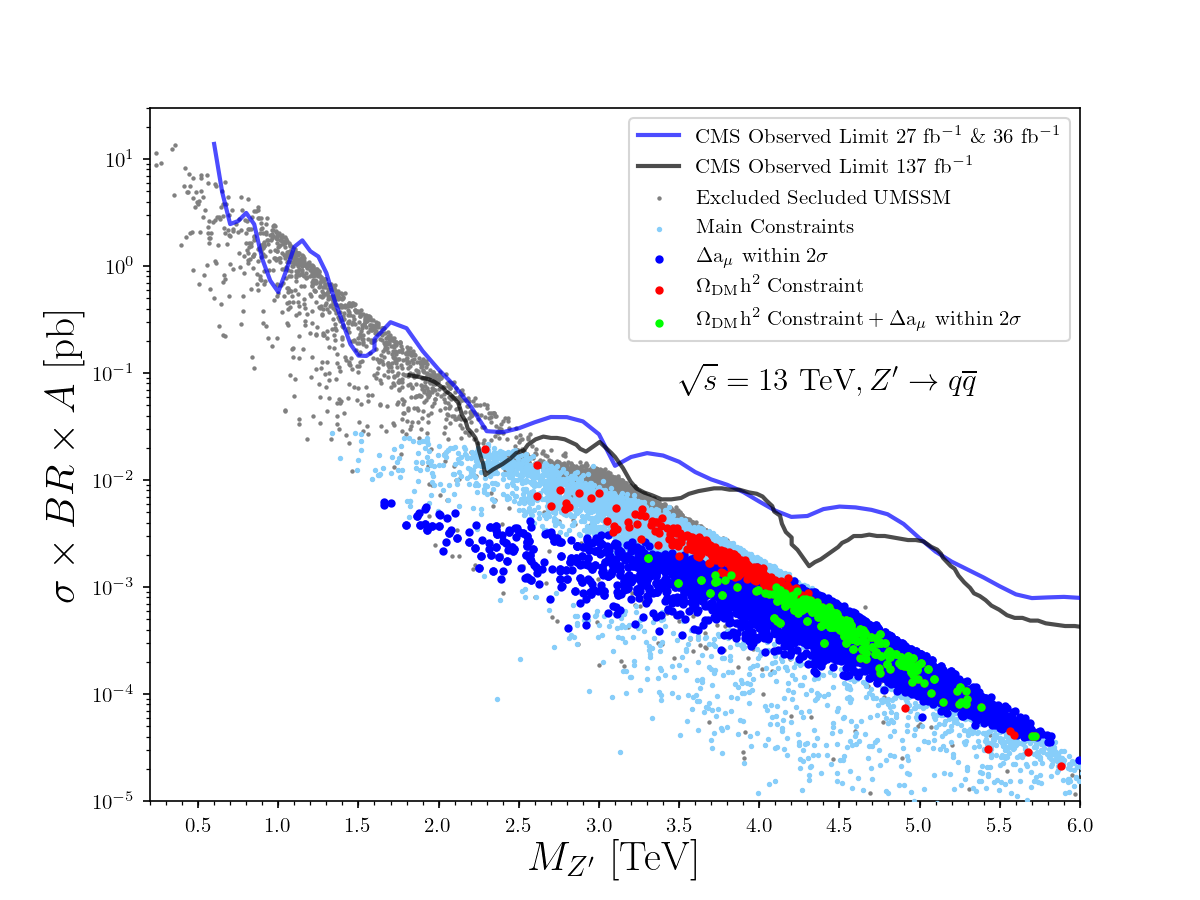}
	\caption{(Left) Neutralino-chargino mass limits in secluded UMSSM. The black curve represents mass limits from ATLAS \cite{Aad:2019vvf,Aad:2019vnb}, while our analysis rules out only points which exceed the upper limits on the chargino-neutralino cross sections, as indicated on the right-side color bar (which gives our predicted cross section measured against the limits from ATLAS). (Right) $Z^\prime$ production cross sections multiplied by the di-jet BRs (and by the acceptance $A=0.5$).} 
	\label{fig:SModelSResults}
\end{figure}


Scanning over the whole range of allowed $Z^\prime$ mass values, we find that  consistency with ATLAS production and di-lepton decay results allows $M_{Z^\prime}$ to be quite light. However, for the parameter space to satisfy both DM and muon anomalous magnetic moment constraints to at least 2$\sigma$, the ${Z^\prime}$ mass must be $M_{Z^\prime}  \gsim {\cal O}(3)$ TeV as seen from the right plane of Figure \ref{fig:SModelSResults}. To highlight the model characteristics, we chose  two benchmarks, {\bf BM I} and {\bf BM II}. The first benchmark is consistent with all constraints, including relic density, and satisfies the bounds on the $g-2$ factor of the muon at $1\sigma$. The second benchmark  satisfies the same constraints, except that we relax requirements on consistency with the anomalous magnetic moment of the muon.  We list the values of the relevant free parameters in the model in Table \ref{tab:benchmark_free} and the corresponding mass values for the fermions and bosons in the model in Table \ref{tab:benchmarks_mass}.


\begin{table}[t]
	\renewcommand{\arraystretch}{1.3}\setlength\tabcolsep{6pt}
	\begin{center}
		\begin{tabular}{c|c c c c c c c }
			\small
			[GeV] &$m_0$& $M_{1/2}$ & $A_0$ & $v_S$  & $v_{S_1} = v_{S_2} = v_{S_3}$ \\
			\hline\hline
			{\bf BM I}   & 942 & 2821 & 662 & 2421 & 5401 \\
			{\bf BM II}  & 1722 & 2568 & -1092 & 2282 & 6935  \\			
		\end{tabular}\\[.2cm]
		\begin{tabular}{c|c c c c c c c c c c}
			\small
			& $\tan \beta$   & $\lambda$  & $A_\lambda$
			& $\kappa$   & $A_\kappa$
			& $\alpha$ & $\delta$ & $Y_v^{ij}$ \\
			\hline\hline
			{\bf BM I}   & 11.9 & 2.04 $\times 10^{-1}$ & 3469  & 1.81  & -4781 & 4.48$\times 10^{-2}$ & 4.44$\times 10^{-1}$ & 1.63$\times 10^{-8}$  \\
			{\bf BM II}  & 20.1  & 9.70 $\times 10^{-2}$  & 3051  & 6.73$\times 10^{-1}$  & -3910 & 4.44$\times 10^{-2}$ & 4.00$\times 10^{-1}$ & 6.71$\times 10^{-8}$\\			
		\end{tabular}\\[.2cm]
		\caption{Set values for the free secluded UMSSM parameters defining our 
			benchmark scenarios {\bf BM I} and {\bf BM II}. Here, $m_0$ is the universal scalar mass and $M_{1/2}$ the gaugino mass.}
		\label{tab:benchmark_free}
	\end{center}
\end{table}
\begin{table}[t]
	\renewcommand{\arraystretch}{1.3}\setlength\tabcolsep{6pt}
	\begin{center}
		\begin{tabular}{c|c c c c c c c c  c c c}
			[GeV] & $M_{Z'}$ & $M_{H_1^0}$   & $M_{H_2^0}$  & $M_{H_3^0}$ & $M_{H_4^0}$ & $M_{H_5^0}$ & $M_{H_6^0}$	& $M_{A_1^0}$   & $M_{A_2^0}$	& $M_{H_1^\pm}$ \\
			\hline\hline
			{\bf BM I}   & 3307  & 126 & 332 & 2559 & 3405 & 3535 & 4148 & 3405 & 5066 & 3407 \\
			{\bf BM II} & 2291 & 123 & 394 & 758    & 2474 & 3138 & 3332 & 3138 & 3580 & 3139  \\								
		\end{tabular}\\[.2cm]
		\begin{tabular}{c|c  c c c c c c c c c c c}
			[GeV]  & $M_{\tilde{\chi}_1^0}$
			& $M_{\tilde{\chi}_2^0}$ & $M_{\tilde{\chi}_3^0}$ & $M_{\tilde{\chi}_4^0}$
			& $M_{\tilde{\chi}_5^0}$  & $M_{\tilde{\chi}_6^0}$  & $M_{\tilde{\chi}_7^0}$ & $M_{\tilde{\chi}_8^0}$ & $M_{\tilde{\chi}_9^0}$ & $M_{\tilde{\chi}_1^\pm}$ & $M_{\tilde{\chi}_2^\pm}$ & $M_{\tilde{g}}$\\
			\hline\hline
			{\bf BM I}   & 45  & 358  & 363 & 1247 & 2295 & 2321 & 3595 & 4106 & 4590 & 359 & 2321 & 5761 \\
			{\bf BM II} & 44  & 160  & 165 & 1100 & 1133 & 2122 & 2201 & 2325 & 3025 & 162 & 2121 & 5316\\									
		\end{tabular}
		\begin{tabular}{c|c c c c c c c c c c c c}
		    [GeV]  & $M_{\tilde{d}_1}$ & $M_{\tilde{d}_2}$ & $M_{\tilde{d}_3}$ & $M_{\tilde{d}_4}$
	    	& $M_{\tilde{d}_5}$  & $M_{\tilde{d}_6}$ & $M_{\tilde{u}_1}$ & $M_{\tilde{u}_2}$ & $M_{\tilde{u}_3}$ & $M_{\tilde{u}_4}$ & $M_{\tilde{u}_5}$ & $M_{\tilde{u}_6}$\\
			\hline\hline
			{\bf BM I}  & 4765 & 4952 & 4989 &  4989 & 5235 & 5235 & 3896 & 4772 & 4918 & 4918  & 5234 & 5234 \\
			{\bf BM II} & 4421 & 4692 & 4817 & 4817 & 5021 & 5021 & 3499  & 4429 & 4731 & 4731  & 5021 & 5021 \\									
	\end{tabular}
		\begin{tabular}{c|c  c c c c c c c c c c c}
			[GeV]  & $M_{\tilde{\ell}_1}$ & $M_{\tilde{\ell}_2}$ & $M_{\tilde{\ell}_3}$ & $M_{\tilde{\ell}_4}$
			& $M_{\tilde{\ell}_5}$  & $M_{\tilde{\ell}_6}$ & $M_{\tilde{\nu}_1}$ & $M_{\tilde{\nu}_2}$ & $M_{\tilde{\nu}_3}$ & $M_{\tilde{\nu}_4}$ & $M_{\tilde{\nu}_5}$ & $M_{\tilde{\nu}_6}$ \\
			\hline\hline
			{\bf BM I}  & 1333 & 1382 & 1383 & 2055 & 2071 & 2071 & 180 & 180 & 180  & 2053 & 2069 & 2069 \\
			{\bf BM II} & 1766 & 1912 & 1913 & 2366 & 2421 & 2422 & 1374 & 1374 & 1374 & 2364 & 2420  & 2420\\									
		\end{tabular}				
		\caption{Particle spectrum of {\bf BM I} and {\bf BM II}: bosons (top), fermions (middle), squarks and sleptons (bottom). All masses are given in GeV.}
		\label{tab:benchmarks_mass}
	\end{center}
\end{table}

While scanning over the parameter space consistent with all constraints, we were unable to find any allowed parameter space for which $M_{Z^\prime}<3.3$ TeV ({\bf BM I}). Relaxing the imposed constraints on the anomalous magnetic moment of the muon  completely (for {\bf BM II}), while requiring agreement with the measured relic density, still poses rigid constraints on the parameter space, but allows a lower $M_{Z^\prime} \sim 2.3$ TeV. The relevant predictions for  {\bf BM I} and {\bf BM II} for the DM and $(g-2)_\mu$ observables discussed in the above sections are shown in Table \ref{tab:BenchmarkRelic}.  We note that slepton masses do not necessarily need to be light to yield significant contributions to muon $g-2$. Indeed, slepton masses are mostly at TeV scale. As seen from Table \ref{tab:benchmark_free}, the lightest slepton mass is 1333 GeV for {\bf BM I} and 1766 GeV for {\bf BM II}. We also included the lightest slepton decays of {\bf BM I} and {\bf BM II}. Therefore, the current slepton searches cannot easily restrict our parameter space. We also show the stau masses in the right bottom panel of Figure \ref{fig:muong2}. As seen from the graph, the lightest stau masses are mostly $M_{\tilde{\tau}_1} >$ 750 GeV. The right-sneutrino contribution is really significant for muon g-2 because the dominant contribution to muon g-2 comes from the diagram with right-sneutrinos and charginos running in the loop. This can be also seen from mass values in {\bf BM I} and {\bf BM II}. The light sneutrino and chargino states in  {\bf BM I}  give significant contribution to muon g-2. However, the same loop effect is suppressed in the scenario {\bf BM II} due to heavy sneutrino masses, and this is the reason that {\bf BM II} does not contribute significantly  to the muon $g-2$ as seen from Table \ref{tab:benchmarks_mass}.

To test the signal coming from production and decay of the  leptophobic $Z^\prime$ boson, we use its decay into supersymmetric particles, here into chargino pairs, followed by the decay into lepton pairs or jets plus missing energy\footnote{The decay into chargino pairs is not the only one yielding the required di-lepton (or jets) + missing $E_T$  signal, but it dominates other intermediate steps by a few orders of magnitude.}. The decay of the lightest chargino yielding lepton or jet final states is into $\tilde{\chi}_1^\pm \to \tilde{\chi}_1^0 W^\pm$  and we choose points for which this BR is almost 1, as shown in Table \ref{tab:lhc}. In the same table, we show predictions for the LHC phenomenology  of our two  benchmark  scenarios,  including the production cross sections at a centre-of-mass energy  $\sqrt{s}=13, 14, 27 $ and $100$ TeV, plus the dominant BRs of the $Z^\prime$. For both scenarios, $Z^\prime$ boson production is small enough relatively to the LHC limits at a centre-of-mass energy of 13 TeV. The cross section is about 0.016 fb for  {\bf BM I} and 0.1889 fb for {\bf BM II} after accounting for the $Z^\prime$ boson decaying into electron and muon pairs through two chargino states. Consequently this makes the $Z^\prime$ signal  difficult to observe, even with more luminosity at a centre-of-mass energy of 13 TeV.
 
The $Z^\prime$ can also decay into right-handed sneutrinos.  Such signature has tiny SM backgrounds and would be advantageous  for $Z^\prime$ model searches. We give, in Table \ref{tab:lhc}, the BR($Z^\prime \to \tilde{\nu}_i^* \tilde{\nu}_i$) for {\bf BM I} and {\bf BM II} where the index i can be 1, 2 and 3. As seen from the the table, $Z^\prime$ boson of {\bf BM I} decays to $\tilde{\nu}_i^* \tilde{\nu}_i$ at a rate of 3.011\%. Then, each sneutrino decays to a neutrino and a neutralino LSP at a rate of 100\%. This signature is not significant since the final state is nothing but only missing energy. In addition, the decay width of the lightest sneutrino is 2.18 $\times 10^{-3}$ GeV, and the corresponding flight time is quite short. Therefore, the sneutrino states of {\bf BM I} are not long lived particles. On the other hand, the BR($Z^\prime \to \tilde{\nu}_i^* \tilde{\nu}_i$) for {\bf BM II} is smaller than $\mathcal{O}(<10^{-4})$.

The $Z'$ production cross section is therefore about 0.33 fb for  {\bf BM I} and 3.82 fb for  {\bf BM II} at 13 TeV, after accounting for the Z$^\prime$ bosons decaying into all SM fermions (quarks + leptons) via two chargino states,  giving rise to a multi-jet plus missing energy signature. The latter is also typically expected from supersymmetric squark/gluino production and decay, so that the results of SUSY searches in the multi-jet plus missing energy mode could be reinterpreted to constrain the secluded UMSSM. We therefore recast these results from  \cite{Aaboud:2016zdn,Aaboud:2017vwy,Sirunyan:2017cwe,ATLAS:2019vcq} with \ma. However, such a rate is far beyond the reach of typical multi-jet plus missing transverse momentum searches at the LHC, as confirmed by reinterpreting and extrapolating the results of the CMS search in  \cite{Sirunyan:2017cwe} and the results of the ATLAS search in  \cite{Aaboud:2016zdn,Aaboud:2017vwy,ATLAS:2019vcq} targeting superpartner production and decay in the jets plus missing transverse momentum mode to integrated luminosity of 3 ab$^{-1}$ with \ma. Consequently, this makes the $Z^\prime $ signal difficult to observe in  di-jet final states, even with more luminosity. We therefore focus on $Z^\prime$ signals that instead involve di-leptons in the final state at a centre-of-mass energy of 14 TeV and 27 TeV. 

The  study of \cite{Araz:2017wbp} provides a prescription for finding leptophobic $Z^\prime$ bosons at the center-of-mass energy $\sqrt{s}=$ 14 TeV and 3 ab$^{-1}$ of luminosity in the di-lepton  channel. The  signal  process  consists  of  the resonant  production  of  a  chargino  pair,  followed  by  the  decay  of  each  chargino  into  a charged lepton and missing energy,

\begin{equation}
pp \to	Z^\prime \to \tilde{\chi}_1^\pm \tilde{\chi}_1^\mp \to \ell^+ \ell^- + \cancel{\it{E}}_{T}.
\end{equation}

We followed the same procedure and carried out a full Monte Carlo (MC) event simulation at the LHC, for a center-of-mass energy $\sqrt{s}=$ 14 TeV and applied the cuts as in  \cite{Araz:2017wbp}. The  production cross  section of $Z^\prime$ boson is 15.8 fb for  {\bf BM I} and 154.4 fb for  {\bf BM II} for a center-of-mass energy $\sqrt{s}=$ 14 TeV as given in Table \ref{tab:lhc}. We have made use of \fr~ to generate a UFO \cite{Degrande:2011ua} version of the model, so that we could employ \mg~ (version 2.7.3) \cite{Alwall:2014hca} for generating the hard-scattering signal event samples necessary for our collider study. These events, obtained by convoluting the hard-scattering matrix elements with the NLO set of NNPDF 3.1 parton densities \cite{Ball:2017nwa}, were subsequently matched with \py~ (version 8.244) \cite{Sjostrand:2014zea} parton showering and hadronisation algorithms, plus we simulated the typical response of an LHC detector by means of the \del~ \cite{deFavereau:2013fsa} programme (version 3.4.2) employing the \textsc{Snowmass} parameterization \cite{Anderson:2013kxz,Avetisyan:2013onh} that  relies on the anti-$k_T$ algorithm \cite{Cacciari:2008gp} with a radius parameter $R$ = 0.6 as implemented into \textsc{FastJet} \cite{Cacciari:2011ma} (version 3.3.3) for event reconstruction. We have employed \ma~ \cite{Conte:2012fm} (version 1.8.23) and normalized our results to an integrated luminosity of 3 ab$^{-1}$ for the collider analysis. 

We select events featuring two well-separated muons and veto the presence of jets, by requiring
\begin{equation}
N^{\ell} = 2, \hspace{1cm} \Delta R(\ell_1, \ell_2) > 2.5, \hspace{1cm}  N^{j} = 0. 
\end{equation}
The transverse momenta of the two leptons and the missing transverse energy are required  to fulfill
\begin{equation}
p_T(\ell_1) > 300 {\rm\ GeV}, \hspace{1cm} p_T(\ell_2) > 200 {\rm\ GeV}, \hspace{1cm}  \cancel{\it{E}}_{T} > 100 {\rm\ GeV}.
\end{equation}
To investigate the observability of the two benchmarks at the HL-LHC, we use of two standard significance parameters, labelled as $s$ and $Z_A$ (the Asimov significance),  defined as:
\begin{eqnarray}
s  &=&\ \frac{S}{\sqrt{B+\sigma_B^2}}\ ,\\
Z_A&=&\ \sqrt{ 2\left(
	(S+B)\ln\left[\frac{(S+B)(S+\sigma^2_B)}{B^2+(S+B)\sigma^2_B}\right] -
	\frac{B^2}{\sigma^2_B}\ln\left[1+\frac{\sigma^2_BS}{B(B+\sigma^2_B)}\right]
	\right)} \ ,
\label{eq:signfigance}
\end{eqnarray}
where $S$ is the number of signal events,  $B$ of  background events and $\sigma_B$ is the standard deviation of background events. 

The corresponding cutflows are shown in Table \ref{tab:Signal}, where we give our original and final number of signal events, and the ones surviving  each cut,  shown in the left-handed column. We assume that we would get the same cut efficiency of the background as in  \cite{Araz:2017wbp}. Therefore, we first estimate the final number of background events (after imposing the cuts in Table  \ref{tab:Signal}) at 27 TeV by using a boost factor calculated from  the dominant background channel, the di-boson production. We expand more on this choice. Background events at 14 TeV were generated by \cite{Anderson:2013kxz,Avetisyan:2013onh} and adapted from that work without regenerating them. We wanted to get an estimation about detectability of our model at the LHC. To do this, we assumed that the cut efficiency for background events would be the same when the same cuts are applied at 27 TeV instead of 14 TeV.  More explicitly, \cite{Araz:2017wbp} clearly shows that the dominant background comes from the  di-boson channel. Therefore, we assume that di-boson production cross sections at 27 TeV  divided by the di-boson production cross section at 14 TeV would give us a boost factor which is found to be 2.19.  Also,  the number of final background events at 27  TeV,  after applying all cuts,  is estimated as number of final background events at 14  TeV multiplied by the boost factor, which is found to be 21.96. One can see that the significance of the benchmarks at 14 TeV and with integrated luminosity 3 ab$^{-1}$ is very small, making it unlikely to be observed, even at the HL-LHC. Therefore, we extend the analysis of our benchmark scenarios at 27 TeV, and in Table  \ref{tab:Signal}, we give our original and final number of signal events in parentheses. {The significance plots, as  functions of luminosity, in Figure \ref{fig:Significance27} are obtained by using the number of final background events, which is estimated as described above.} While {\bf BM I} remains below the 3$\sigma$ minimum significance required for a positive identification, the {\bf BM II} significance rises above 3$\sigma$ at $\sqrt{s}= 27$ TeV and integrated luminosity 3 ab$^{-1}$, making this benchmark promising at the HE-LHC. That this indeed so is seen in Figure \ref{fig:Significance27}, where we plot significance curves for $s$ and $Z_A$ at $\sqrt{s}=27$ TeV, for both {\bf BM I} and {\bf BM II}, as a function of the total integrated luminosity $\cal{L}$. While {\bf BM I} would be observable at high integrated luminosity 3 ab$^{-1}$ at $3\sigma$ under only the most optimistic scenario, in which we assume small systematic errors ($\Delta_{\rm syst}=5\%$), {\bf BM II} shows promise for observability even for larger systematic errors, $\Delta_{\rm syst}=20\%$. Of course, we stress that, while {\bf BM II} is promising, it was obtained by relaxing the condition that the model satisfies $(g-2)_\mu$ to (1-2)$\sigma$.

For more information about the signal, we simulate the SM background events leading to final states with two charged leptons and missing energy:  $t\bar{t}$,  single top events, as well as single vector bosons $V$+jets, and di-bosons $VV$, with $V$ being a $W$ boson or a $Z$ boson decaying leptonically at 27 TeV. We include the NLO effects of the signal through a $K$ factor. The whole QCD $K$-factor comes from the initial state and depends on the  $Z^\prime$-boson mass and the set of PDFs used. Previous work provides an NLO implementation of the $Z^\prime$ in the $U(1)^\prime_\chi$ model \cite{Fuks:2007gk}.  The gauge boson mass is assumed to be 1 TeV, and the width is calculated to be  $\Gamma_{Z^\prime}=12.04$ GeV, justifying a narrow width approximation. The NLO $K$-factor for $pp \to \gamma, Z, Z^\prime \to l^+ l^-$ obtained at 1 TeV for $\sqrt{s}=14$ TeV is 1.26.   
 
We  calculate the $Z^\prime$ production cross section at next-to-leading order (NLO) accuracy in QCD for $\sqrt{s}=27$ TeV. We verified that the $K$-factor can be inclusively calculated depending on the PDF choice and $Z^\prime$ mass, which enters the $Q^2 \approx M_{Z^\prime}^2$ dependence of the PDFs. For $\bold{BM\ I}$ and $\bold{BM\ II}$, the $K$-factor is found to be 1.17 and 1.15, respectively. Therefore,  the NLO corrections are  small, and they are included in Table \ref{tab:lhc}.

We include plots of the transverse momentum of the leading muon $p_T(\ell_1)$, the next-to-leading muon $p_T(\ell_2)$ and of the missing di-lepton transverse energy spectrum (after applying all cuts of Table \ref{tab:Signal}) for the benchmarks in Figure \ref{fig:Significance27} and compare to the SM backgrounds. The effects of single boson and single top are rendered negligible by the cuts imposed in Table \ref{tab:Signal}. The more promising scenario {\bf BM II} is seen to rise consistently above the SM backgrounds, confirming the promise indicated in the significance plots.

\begin{table}
	\renewcommand{\arraystretch}{1.3}\setlength\tabcolsep{6pt}
	\begin{center}
		\small		
		\begin{tabular}{c|c c c c | c}
			& $\Omega_{\rm DM} h^2$
			& $\sigma_{\rm SI}^{\rm proton}$ [pb]
			& $\sigma_{\rm SI}^{\rm neutron}$ [pb]
			& $\langle\sigma v\rangle$ [cm$^3$s$^{-1}$]
			& $\Delta a_{\mu} \times 10^{10} $\\
			\hline\hline
			{\bf BM I}  & 0.131 & 1.84$\times 10^{-13}$ & 1.89$\times 10^{-13}$
			& 5.58 $\times 10^{-29}$ & 36.4 (within 1$\sigma$) \\
			{\bf BM II}  & 0.124  & 2.21$\times 10^{-11}$ & 2.26$\times 10^{-11}$ & 8.17$\times 10^{-29}$ & 173.4 (outside 3$\sigma$) \\									
		\end{tabular}
		\caption{Predictions  for the {\bf BM I} and {\bf BM II}
			scenarios, of the observables discussed in our dark matter analysis.}
		\label{tab:BenchmarkRelic}
	\end{center}
\end{table}
\begin{table}
	\renewcommand{\arraystretch}{1.3}\setlength\tabcolsep{6pt}
	\begin{center}
		\small
		\begin{tabular}{c|cccc|c c c }
			&$\sigma (pp \to Z')$ [fb]	 &&& & BR($Z'\to \tilde{\chi}_1^\pm \tilde{\chi}_1^\mp$) & BR($Z'\to j j$)  & BR($\tilde{\chi}_1^\pm \to \tilde{\chi}_1^0 W^\pm$)  \\
			\hline\hline
			& 13 TeV & 14 TeV & 27 TeV & 100 TeV &&&\\ \hline
			{\bf BM I}  & 11.13 & 15.8 & 156.6 & 1942 & 0.059 & 0.309 & 0.99 \\
			{\bf BM II} & 119.7 & 154.4 & 856.2 & 7375 & 0.066 & 0.340 & 1.0  \\							
		\end{tabular}\\[.2cm]	
		\centering
		\begin{tabular}{c|c c c c c}
		& BR($Z'\to \tilde{\nu}_1^* \tilde{\nu}_1$)
		& BR($Z'\to \tilde{\nu}_2^* \tilde{\nu}_2$)
		& BR($Z'\to \tilde{\nu}_3^* \tilde{\nu}_3$)
		& BR($ \tilde{\nu}_{1,2,3} \to \nu_{2,1,3} \tilde{\chi}_1^0$) 
		& BR($ \tilde{\ell}_{1} \to \nu_{1} \tilde{\chi}_1^\pm$) \\
		\hline\hline
		{\bf BM I}  & 3.011$\times 10^{-2}$  & 3.011$\times 10^{-2}$ &  3.011$\times 10^{-2}$ & 1.0 (Each) & 0.426 \\
		{\bf BM II} & -  & - & - & 0.99 (Each) & 0.237\\									
	\end{tabular}
		\caption{$Z^\prime$ production cross section at $\sqrt{s}=13, 14, 27 $ and $100$ TeV and branching ratios  for the {\bf BM I} and {\bf BM II} scenarios,  relevant for the associated LHC phenomenology. NLO QCD corrections to the production cross sections $\sigma(pp \to Z^\prime)$ are included.}
		\label{tab:lhc}
	\end{center}
\end{table}
\begin{table}[]
	\renewcommand{\arraystretch}{1.3}\setlength\tabcolsep{6pt}
	\begin{center}
		\small
		\begin{tabular}{c|c||c|c|}
			Step &  Requirements &  $\bold{BM\ I}$ & $\bold{BM\ II}$  \\ \hline
			0      &  Initial                 & 71 (92)                 &  726 (3854)                   \\ \hline
			1       & $N^{\ell} = 2 $    &45	 (61)                & 386  (2310)                    \\ \hline
			2       &  Electron Veto     &13  (18)                & 115  (712)                        \\ \hline
			3       &  $|\eta^{\ell}| < 1.5$   &13 (18)        & 112  (685)                         \\ \hline
			4   	&  $I_{\rm rel}^{\mu} < 0.15$   &13 (18)  & 107 (663)                    \\ \hline
			5       &  $\Delta R(\ell_1, \ell_2) > 2.5$   &11 (18)	& 107 (662)             \\ \hline	
			6       &  $N^{j} = 0$                  &11 (18)   & 60 (330)                            \\ \hline		
			7       &  $p_T(\ell_1) > 300$ GeV   &6	(18)              & 17 (107)             \\ \hline	
			8       &  $p_T(\ell_2) > 200$ GeV   &2	(17)              & 6 (36)                  \\ \hline	
			9       &  $\cancel{\it{E}}_{T} > 100$ GeV   & 2 (15)	& 4 (25)                 \\ \hline \hline												
			s	    &	 $(\Delta_{\rm syst}=20\%)$         &0.53 (2.33)  & 1.09 (3.89)  \\ \hline 
			$Z_A$  & $(\Delta_{\rm syst}=20\%)$	        &0.51 (2.03)  & 0.99 (3.16)  \\
		\end{tabular}
		\caption{Events surviving after each cut (as given in the left column) and significance of  {\bf BM I} and {\bf BM II} at 14 (27) TeV and integrated luminosity 3 ab$^{-1}$.}
		\label{tab:Signal}
	\end{center}
\end{table}

\begin{figure}
	\centering
	\includegraphics[width=.42\columnwidth]{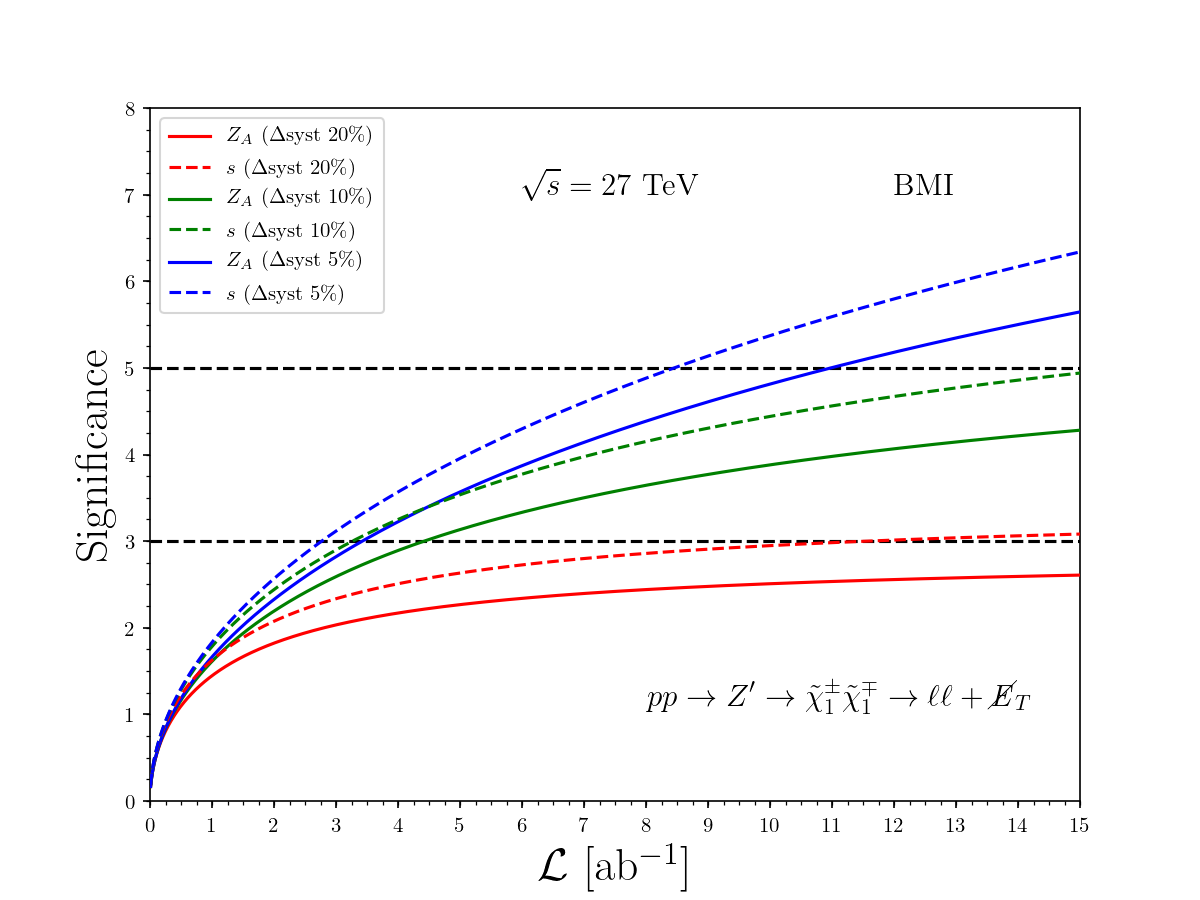}	
	\includegraphics[width=.42\columnwidth]{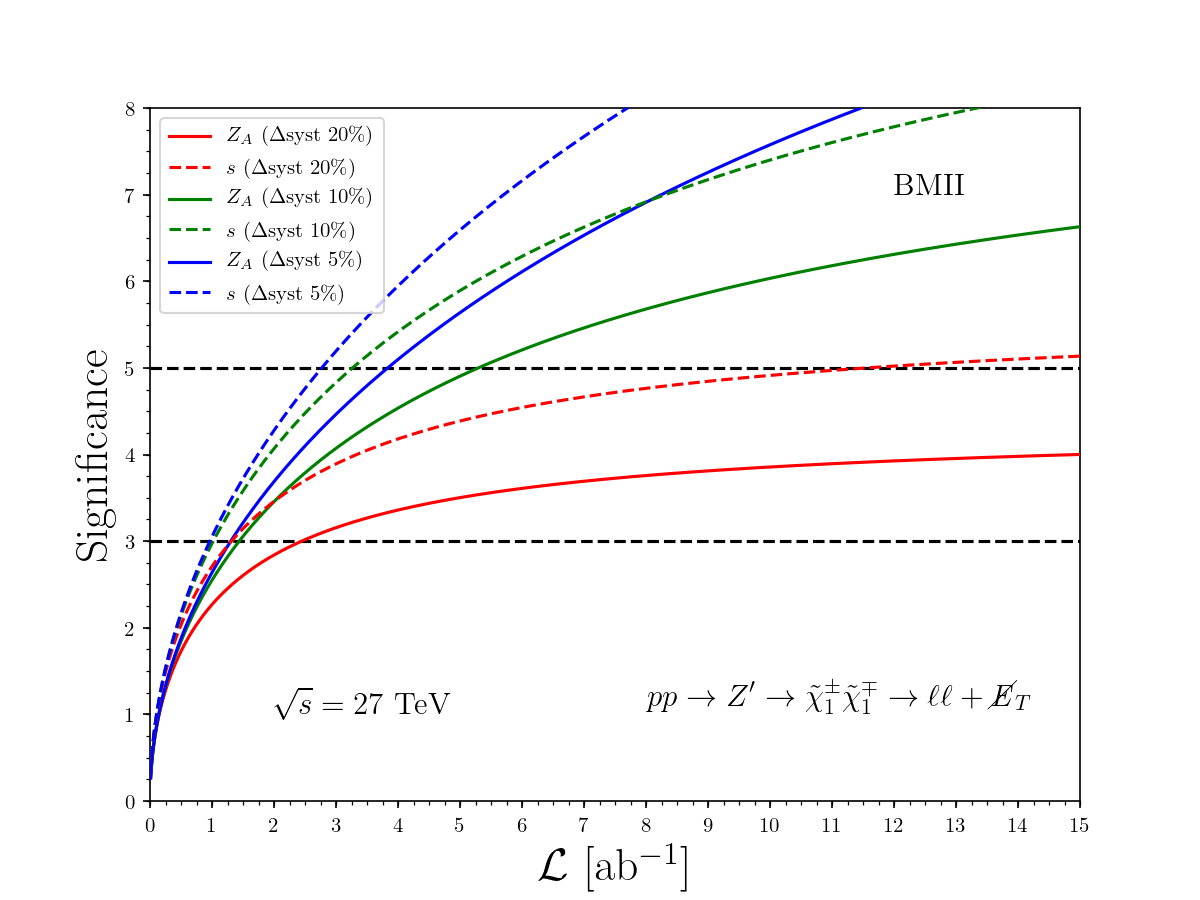} \\
	\includegraphics[width=.31\columnwidth]{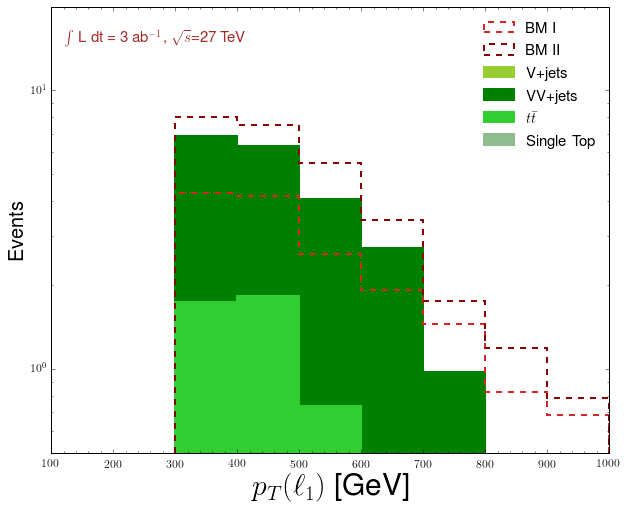} 
	\includegraphics[width=.31\columnwidth]{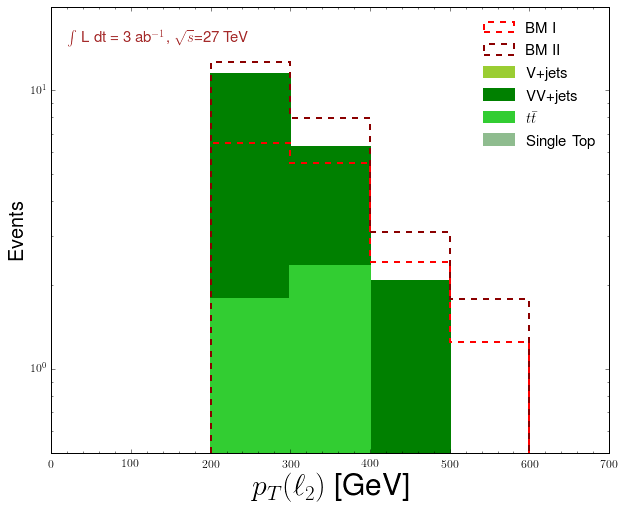} 
	\includegraphics[width=.31\columnwidth]{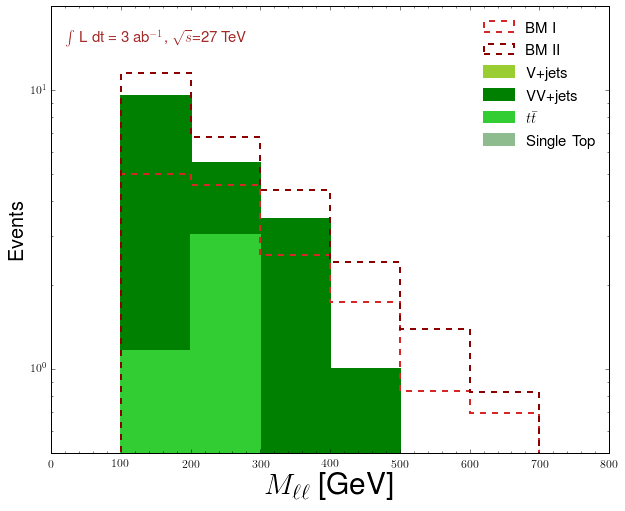}\\
	\caption{ Significance of  benchmarks {\bf BM I} (top left panel) and {\bf BM II} (top right panel) at $\sqrt{s}=27$ TeV, as a function of  the luminosity ${\cal L}$. In each panel  we  plot the usual significance $s$ and the Asimov significance $Z_A$. Different curves are obtained assuming different systematic errors, as indicated in the upper left-hand panel. {Bottom panels: Transverse momentum distribution of the leading muon $\ell_1$ (left) and  next-to-leading muon $\ell_2$ (middle) and  missing transverse energy spectrum (right) for {\bf BM I} and {\bf BM II} after applying all cuts in Table \ref{tab:Signal}. We include the SM backgrounds: $t {\bar t}$, single top, di-bosons, and gauge boson+jets. }}
	\label{fig:Significance27}
\end{figure}

\section{Summary and Conclusions}
\label{sec:conclusion}

We have presented an analysis of the {secluded}  UMSSM, a non-minimal SUSY scenario wherein the gauge symmetry of the MSSM is augmented by a $U(1)^\prime$ group and where a secluded sector is also added in the form  of three additional scalar superfields. Their role is to separate the SUSY-breaking scale from the mass of the $Z^\prime$, the gauge boson introduced by the additional gauge symmetry following its spontaneous breaking, so that the latter can have a value well within the LHC reach irrespectively of the SUSY mass scale. 

Our analysis here has highlighted, in particular,  some novel phenomenological features pertaining to this BSM scenario, which would make it distinguishable from the MSSM or $E_6$ motivated UMSSM scenarios. 
For a start,  the  $Z^\prime$ can be leptophobic without invoking gauge kinetic mixing. Thus one can naturally lower the experimentally imposed limits on its  mass coming from its LHC hadroproduction followed by di-lepton and di-jet decays. In addition, and setting it apart from that of $U(1)^\prime$ scenarios with gauge kinetic mixing, the $Z^\prime$ is also $d$-quark-phobic, allowing one to reduce its mass constraints event further.

Then, we have shown that the model predicts the existence of very light charginos and neutralinos, the lightest of the latter being a singlino-like DM candidate satisfying relic density constraints as well as direct and indirect detection bounds. In fact, alongside this new singlino state, an LSP with mass $M_{{\tilde \chi}_1^0} \lsim 50 $ GeV, our BSM scenario also accommodates a similarly light lightest chargino companion, with  $M_{{\tilde \chi}_1^\pm} \lsim 350$ GeV,  both of which are 
respecting collider constraints. Furthermore, the next-to-LSP and next-to-next-to-LSP are higgsinos and, together with the lightest chargino, they are largely responsible (once appropriately combined with the lightest sleptons in one-loop Feynman diagrams) for obtaining a value for the muon anomalous moment consistent with experimental measurements at 1$\sigma$.

Finally, armed with such specific model setup, we have investigated the prospects of detecting such a light $Z^\prime$ boson  in its SUSY cascade decays via the aforementioned lightest charginos and neutralinos, eventually yielding a di-lepton final state in presence of significant missing transverse energy. The fact that the model is $d$-quark phobic, useful to reduce the mass constraints,  has an adverse effect on the production cross section for $Z^\prime$, rendering it smaller than in the $E_6$ motivated UMSSM. In addition, the $S, T, U$ parameters impose conditions on the $U(1)^\prime$ associated charges, constraining them to be small. The secluded UMSSM is a good model for loosening $Z^\prime$ mass bounds, but no so promising for signal observability.

Requiring the parameter space to satisfy all experimental conditions, including the DM and  $(g-2)_\mu$ ones simultaneously, or just the relic density, we have devised most favourable benchmark points with $M_{Z^\prime}\approx 3.3$ TeV. Relaxing the $(g-2)_\mu$ requirement, our second benchmark allows $M_{Z^\prime}\approx 2.3$ TeV. Of the two benchmarks, the latter one  shows more promise to be observed at the HE-LHC at $3 \sigma$ or better, as proved from a prototypical MC analysis performed, while the former would be observed only assuming small systematic errors. Our analysis should justify dedicated searches with real data from ATLAS and/or CMS. 

In summary, we enumerate the interesting and novel features of our model:
\begin{itemize}
\item The model framework is not new but ours is the only up-to date study of its phenomenology. 
\item The conditions for gauge invariance and anomaly cancellations have appeared before. There  are linear, quadratic and cubic in the $U(1)^\prime$ charges, and also depend on the electric charges and number of generations of the exotic fermions. Solving them is non-trivial; and finding a solution obeying rational numbers for exotic fermion charges requirements, even less so. 
\item Our choice of $U(1)^\prime$ charges is innovative because it renders this to be a  $U(1)^\prime$ model which is leptophobic {\it without} kinetic mixing, or requiring family-non-universality. Our choice of $U(1)^\prime$ charges which effects this is particularly simple and transparent.
\item The model was previously used because it decouples the $Z^\prime$ scale from the SUSY sector:  $Z^\prime$ was always considered to be heavy, while the chargino-neutralino sector could be light.  In our scenario, both charginos and neutralinos can be light, at the same time  $Z^\prime$, since it is leptohobic,  can also be light. A model featuring {\it both} a very light singlino $<50$ GeV (escaping LHC bounds)  {\it and } a light $Z^\prime$, while obeying family universality, is new. 
\item   While a light singlino is possible in secluded models (containing extra singlet fields), here we implement it in the context of a leptophobic scenario.
\item In addition, we have shown that we can distinguish this scenario from $E_6$ motivated $U(1)^\prime$ models with kinetic mixing, because in our scenario, the $Z^\prime$ is $d$-quark phobic. Again,  this is a novel feature in universal $Z^\prime$ models.
\item In this model, we have also investigated, and found out a link between satisfying $(g-2)_\mu$ and relaxing mass constraints on $Z^\prime$.
\item Finally, the model can be tested at the HL-LHC, making it relevant for searches at Run III.
\end{itemize}

\begin{acknowledgments}
Parts of the numerical calculations reported in this paper were performed using High Performance Computing (HPC), managed by Calcul Qu{\'e}bec and Compute Canada, and the IRIDIS High Performance Computing Facility, and associated support services, at the University of Southampton. 
The database entry for ATLAS-SUSY-2019-08 and ATLAS-SUSY-2018-32  were implemented and validated by the SModelS authors upon request, for which we are grateful to Sabine Kraml and Wolfgang Waltenberger. 
  SM is supported in part through the NExT Institute and the STFC consolidated Grant No. ST/L000296/1. The work of MF and \"{O}\"{O} has been partly supported by NSERC through grant number SAP105354, and by  MITACS International Fellowship. The work of YH is supported by The Scientific and Technological Research Council of Turkey (TUBITAK) in the framework of the 2219-International Postdoctoral Research Fellowship Programme, and by Balikesir University Scientific Research Projects with grant No. BAP-2017/198.
   \end{acknowledgments}
\section{Appendix: Anomaly cancellation conditions}
\label{sec:appendix}
Partial anomaly conditions have been explored before in \cite{Kang:2004pp}, and complete expressions exist in \cite{Demir:2010is}. As our choices for $U(1)^\prime$ differ from the usual assignments, we include them here, for completeness.
For the model to be anomaly-free the $U(1)^{\prime}$ charges of fields must
satisfy
\begin{eqnarray}
0&=&3(2Q_{Q}+Q_{U}+Q_{D})+
n_{\Upsilon}(Q_{\Upsilon}+Q_{\overline{\Upsilon}}),
\\
0&=&3(3Q_{Q}+Q_{L})+Q_{H_d}+Q_{H_u},
\\
0&=&3(\frac{1}{6}Q_{Q}+\frac{1}{3}Q_{D}+
\frac{4}{3}Q_{U}+
\frac{1}{2}Q_{L}+Q_{E})
+\frac{1}{2}(Q_{H_d}+Q_{H_u})\nonumber \\
&+&3n_{\Upsilon} Y^2_{\Upsilon} (Q_{\Upsilon}+
Q_{\overline{\Upsilon}})+ n_{\varphi}
Y^2_{\varphi} (Q_{\varphi}+
Q_{\overline{\varphi}} ),
 \\
0&=&3(6Q_{Q}+3Q_{U}+3Q_{D}+2Q_{\ell}+
Q_{e}+Q_{N})
+2Q_{H_d}+2Q_{H_u}\nonumber \\
&+&Q_{S}+Q_{S_1}+Q_{S_2}+Q_{S_3}+
3 n_{\Upsilon}
(Q_{\Upsilon}+Q_{\overline{\Upsilon}})+
n_{\varphi}(Q_{\varphi}+Q_{\overline{\varphi}} ),
\\
0&=&3(Q^{ 2}_{Q}+Q^2_{D}-2Q^{2}_{U}-Q^{2}_{\ell}+Q^{2}_{e})-Q^{2}_{H_d}+Q^{2}_{H_u}+ 3n_{\Upsilon} Y_{\Upsilon}
(Q^{2}_{\Upsilon}- Q^{2}_{\overline{\Upsilon}})\nonumber \\
&+&n_{\varphi} Y_{\varphi}
(Q^{2}_{\varphi}-Q^{2}_{\overline{\varphi}} ),
 \\
0&=&3(6Q^{3}_{Q}+3Q^{3}_{D}+3Q^{3}_{U}+2Q^{3}_{\ell}+Q^{3}_{e}+Q^{3}_{N})+ 2Q^{3}_{H_d}+2Q^{3}_{H_u}+Q^{3}_{S}\nonumber \\
&+&Q^{3}_{S_1}+Q^{3}_{S_2}+Q^{3}_{S_3}+
3n_{\Upsilon}(Q^{3}_{\Upsilon}+Q^{3}_{\overline{\Upsilon}} )+n_{\varphi}(Q^{3}_{\varphi}+Q^{3}_{\overline{\varphi}}),
\end{eqnarray}
which correspond to vanishing of
$U(1)^{\prime}$-$SU(3)_C$-$SU(3)_C$,
$U(1)^{\prime}$-$SU(2)_L$-$SU(2)_L$,
$U(1)^{\prime}$-$U(1)_Y$-$U(1)_Y$,
$U(1)^{\prime}$-graviton-graviton,
$U(1)^{\prime}$-$U(1)^{\prime}$-$U(1)_Y$, and
$U(1)^{\prime}$-$U(1)^{\prime}$-$U(1)^{\prime}$ anomalies,
respectively. All these anomaly cancellation conditions are
satisfied for a particular pattern of charges and parameters. The
$U(1)^\prime$ charges for Higgs fields in the model are chosen as
\begin{eqnarray}
Q_{S_2} = -2Q_{S_1} = -2Q_{S_3}\, , \qquad
Q_{H_u}+Q_{H_d}+Q_{S}=0.
\end {eqnarray}
For the $U(1)$ charge assignments in the model, Eq. \ref{U1ChargesIII}, the exotic fields satisfy the relations:
 \begin{eqnarray}
n_\Upsilon \left[ Q_\Upsilon +Q_{\overline {\Upsilon}} \right ]&=&-27 \alpha \nonumber \\
n_\varphi \left[ Q_\varphi +Q_{\overline {\varphi} }\right]&=&-18 \alpha \nonumber \\
 9Y_\Upsilon^2+2Y^2_\varphi &=& 9 \nonumber \\
 9 Y_\Upsilon \left[ Q_\Upsilon -Q_{\overline {\Upsilon}} \right ]+ 2Y_\varphi \left[ Q_\varphi -Q_{\overline {\varphi}} \right ]&=&33 \alpha \nonumber \\
 3 n_\Upsilon \left[ Q^3_\Upsilon +Q^3_{\overline {\Upsilon} }\right ] +n_\varphi \left[ Q^3_\varphi +Q^3_{\overline {\varphi} }\right ]&=& 6(\delta^3-999\alpha^3)
 \end{eqnarray}
 we 
found that a possible solution to the mixed anomaly constraints allows
$n_{\Upsilon} = 3$ color triplet pairs with hypercharge $Y_{\Upsilon}
= \pm 1/3$, and $n_{\varphi} =  2$ singlet pairs with $Y_{\varphi} = \pm 
2$. This still allows some freedom in the $U(1)^\prime$ charges of $\Upsilon, \overline{\Upsilon}$ and $\varphi, \overline{\varphi}$ as solutions of the last quartic equations.

\bibliography{SecludedMSSM}

\providecommand{\href}[2]{#2}\begingroup\raggedright\begin{thebibliography}{100}

\bibitem{Cvetic:1996mf}
M.~Cvetic and P.~Langacker, \emph{{New gauge bosons from string models}},
  \href{http://dx.doi.org/10.1142/S0217732396001260}{\emph{Mod. Phys. Lett.}
  {\bf A11} (1996) 1247--1262},
  [\href{http://arxiv.org/abs/hep-ph/9602424}{{\tt hep-ph/9602424}}].

\bibitem{Suematsu:1994qm}
D.~Suematsu and Y.~Yamagishi, \emph{{Radiative symmetry breaking in a
  supersymmetric model with an extra U(1)}},
  \href{http://dx.doi.org/10.1142/S0217751X95002096}{\emph{Int. J. Mod. Phys.}
  {\bf A10} (1995) 4521--4536},
  [\href{http://arxiv.org/abs/hep-ph/9411239}{{\tt hep-ph/9411239}}].

\bibitem{Lee:2007fw}
H.-S. Lee, K.~T. Matchev and T.~T. Wang, \emph{{A U(1) -prime solution to the
  $\mu^-$ problem and the proton decay problem in supersymmetry without
  R-parity}}, \href{http://dx.doi.org/10.1103/PhysRevD.77.015016}{\emph{Phys.
  Rev.} {\bf D77} (2008) 015016}, [\href{http://arxiv.org/abs/0709.0763}{{\tt
  0709.0763}}].

\bibitem{Demir:1998dm}
D.~A. Demir, \emph{{Two Higgs doublet models from TeV scale supersymmetric
  extra U(1) models}},
  \href{http://dx.doi.org/10.1103/PhysRevD.59.015002}{\emph{Phys. Rev.} {\bf
  D59} (1999) 015002}, [\href{http://arxiv.org/abs/hep-ph/9809358}{{\tt
  hep-ph/9809358}}].

\bibitem{Demir:2006jj}
D.~A. Demir and Y.~Farzan, \emph{{Correlating mu parameter and right-handed
  neutrino masses in N=1 supergravity}},
  \href{http://dx.doi.org/10.1088/1126-6708/2006/03/010}{\emph{JHEP} {\bf 03}
  (2006) 010}, [\href{http://arxiv.org/abs/hep-ph/0601096}{{\tt
  hep-ph/0601096}}].

\bibitem{Demir:2007dt}
D.~A. Demir, L.~L. Everett and P.~Langacker, \emph{{Dirac Neutrino Masses from
  Generalized Supersymmetry Breaking}},
  \href{http://dx.doi.org/10.1103/PhysRevLett.100.091804}{\emph{Phys. Rev.
  Lett.} {\bf 100} (2008) 091804}, [\href{http://arxiv.org/abs/0712.1341}{{\tt
  0712.1341}}].

\bibitem{Langacker:2008yv}
P.~Langacker, \emph{{The Physics of Heavy $Z^\prime$ Gauge Bosons}},
  \href{http://dx.doi.org/10.1103/RevModPhys.81.1199}{\emph{Rev. Mod. Phys.}
  {\bf 81} (2009) 1199--1228}, [\href{http://arxiv.org/abs/0801.1345}{{\tt
  0801.1345}}].

\bibitem{Erler:2009jh}
J.~Erler, P.~Langacker, S.~Munir and E.~Rojas, \emph{{Improved Constraints on
  Z-prime Bosons from Electroweak Precision Data}},
  \href{http://dx.doi.org/10.1088/1126-6708/2009/08/017}{\emph{JHEP} {\bf 08}
  (2009) 017}, [\href{http://arxiv.org/abs/0906.2435}{{\tt 0906.2435}}].

\bibitem{Anoka:2004vf}
O.~C. Anoka, K.~S. Babu and I.~Gogoladze, \emph{{Constraining Z-prime from
  supersymmetry breaking}},
  \href{http://dx.doi.org/10.1016/j.nuclphysb.2004.03.009}{\emph{Nucl. Phys.}
  {\bf B687} (2004) 3--30}, [\href{http://arxiv.org/abs/hep-ph/0401133}{{\tt
  hep-ph/0401133}}].

\bibitem{Aad:2019fac}
{\scshape ATLAS} collaboration, G.~Aad et~al., \emph{{Search for high-mass
  dilepton resonances using 139 fb$^{-1}$ of $pp$ collision data collected at
  $\sqrt{s}=$13 TeV with the ATLAS detector}},
  \href{http://dx.doi.org/10.1016/j.physletb.2019.07.016}{\emph{Phys. Lett.}
  {\bf B796} (2019) 68--87}, [\href{http://arxiv.org/abs/1903.06248}{{\tt
  1903.06248}}].

\bibitem{Araz:2017wbp}
J.~Y. Araz, G.~Corcella, M.~Frank and B.~Fuks, \emph{{Loopholes in Z$^{′}$
  searches at the LHC: exploring supersymmetric and leptophobic scenarios}},
  \href{http://dx.doi.org/10.1007/JHEP02(2018)092}{\emph{JHEP} {\bf 02} (2018)
  092}, [\href{http://arxiv.org/abs/1711.06302}{{\tt 1711.06302}}].

\bibitem{Sirunyan:2019vgj}
{\scshape CMS} collaboration, A.~M. Sirunyan et~al., \emph{{Search for high
  mass dijet resonances with a new background prediction method in
  proton-proton collisions at $\sqrt{s}=$ 13 TeV}},
  \href{http://arxiv.org/abs/1911.03947}{{\tt 1911.03947}}.

\bibitem{Coleppa:2018fau}
B.~Coleppa, S.~Kumar and A.~Sarkar, \emph{{Fermiophobic gauge boson
  phenomenology in 221 Models}},
  \href{http://dx.doi.org/10.1103/PhysRevD.98.095009}{\emph{Phys. Rev. D} {\bf
  98} (2018) 095009}, [\href{http://arxiv.org/abs/1808.09728}{{\tt
  1808.09728}}].

\bibitem{Rizzo:1998ut}
T.~G. Rizzo, \emph{{Gauge kinetic mixing and leptophobic $Z^\prime$ in E(6) and
  SO(10)}}, \href{http://dx.doi.org/10.1103/PhysRevD.59.015020}{\emph{Phys.
  Rev.} {\bf D59} (1998) 015020},
  [\href{http://arxiv.org/abs/hep-ph/9806397}{{\tt hep-ph/9806397}}].

\bibitem{Babu:1996vt}
K.~S. Babu, C.~F. Kolda and J.~March-Russell, \emph{{Leptophobic U(1) $s$ and
  the R($b$) - R($c$) crisis}},
  \href{http://dx.doi.org/10.1103/PhysRevD.54.4635}{\emph{Phys. Rev.} {\bf D54}
  (1996) 4635--4647}, [\href{http://arxiv.org/abs/hep-ph/9603212}{{\tt
  hep-ph/9603212}}].

\bibitem{Chiang:2014yva}
C.-W. Chiang, T.~Nomura and K.~Yagyu, \emph{{Phenomenology of $E_6$-Inspired
  Leptophobic $Z'$ Boson at the LHC}},
  \href{http://dx.doi.org/10.1007/JHEP05(2014)106}{\emph{JHEP} {\bf 05} (2014)
  106}, [\href{http://arxiv.org/abs/1402.5579}{{\tt 1402.5579}}].

\bibitem{Celis:2015ara}
A.~Celis, J.~Fuentes-Martin, M.~Jung and H.~Serodio, \emph{{Family nonuniversal
  Z′ models with protected flavor-changing interactions}},
  \href{http://dx.doi.org/10.1103/PhysRevD.92.015007}{\emph{Phys. Rev. D} {\bf
  92} (2015) 015007}, [\href{http://arxiv.org/abs/1505.03079}{{\tt
  1505.03079}}].

\bibitem{Allanach:2019mfl}
B.~Allanach, J.~Butterworth and T.~Corbett, \emph{{Collider constraints on
  Z$^{′}$ models for neutral current B-anomalies}},
  \href{http://dx.doi.org/10.1007/JHEP08(2019)106}{\emph{JHEP} {\bf 08} (2019)
  106}, [\href{http://arxiv.org/abs/1904.10954}{{\tt 1904.10954}}].

\bibitem{Alvarado:2019gyh}
J.~Alvarado, C.~E. Diaz and R.~Martinez, \emph{{A $U(1)_X$ extension to the
  MSSM with three families}},  in \emph{{Meeting of the Division of Particles
  and Fields of the American Physical Society}}, 9, 2019.
\newblock \href{http://arxiv.org/abs/1909.02891}{{\tt 1909.02891}}.

\bibitem{Mantilla:2016lui}
S.~Mantilla, R.~Martinez and F.~Ochoa, \emph{{Neutrino and $CP$-even Higgs
  boson masses in a nonuniversal $\mathrm{U}(1)'$ extension}},
  \href{http://dx.doi.org/10.1103/PhysRevD.95.095037}{\emph{Phys. Rev. D} {\bf
  95} (2017) 095037}, [\href{http://arxiv.org/abs/1612.02081}{{\tt
  1612.02081}}].

\bibitem{Tang:2017gkz}
Y.~Tang and Y.-L. Wu, \emph{{Flavor non-universal gauge interactions and
  anomalies in B-meson decays}},
  \href{http://dx.doi.org/10.1088/1674-1137/42/3/033104}{\emph{Chin. Phys. C}
  {\bf 42} (2018) 033104}, [\href{http://arxiv.org/abs/1705.05643}{{\tt
  1705.05643}}].

\bibitem{Kamenik:2017tnu}
J.~F. Kamenik, Y.~Soreq and J.~Zupan, \emph{{Lepton flavor universality
  violation without new sources of quark flavor violation}},
  \href{http://dx.doi.org/10.1103/PhysRevD.97.035002}{\emph{Phys. Rev. D} {\bf
  97} (2018) 035002}, [\href{http://arxiv.org/abs/1704.06005}{{\tt
  1704.06005}}].

\bibitem{Alves:2013tqa}
A.~Alves, S.~Profumo and F.~S. Queiroz, \emph{{The dark $Z^{'}$ portal: direct,
  indirect and collider searches}},
  \href{http://dx.doi.org/10.1007/JHEP04(2014)063}{\emph{JHEP} {\bf 04} (2014)
  063}, [\href{http://arxiv.org/abs/1312.5281}{{\tt 1312.5281}}].

\bibitem{Erler:2002pr}
J.~Erler, P.~Langacker and T.-j. Li, \emph{{The $Z$ - $Z^\prime$ mass hierarchy
  in a supersymmetric model with a secluded U(1) -prime breaking sector}},
  \href{http://dx.doi.org/10.1103/PhysRevD.66.015002}{\emph{Phys. Rev.} {\bf
  D66} (2002) 015002}, [\href{http://arxiv.org/abs/hep-ph/0205001}{{\tt
  hep-ph/0205001}}].

\bibitem{Chiang:2009fs}
C.-W. Chiang and E.~Senaha, \emph{{Electroweak phase transitions in the
  secluded U(1)-prime-extended MSSM}},
  \href{http://dx.doi.org/10.1007/JHEP06(2010)030}{\emph{JHEP} {\bf 06} (2010)
  030}, [\href{http://arxiv.org/abs/0912.5069}{{\tt 0912.5069}}].

\bibitem{Demir:2010is}
D.~A. Demir, M.~Frank, L.~Selbuz and I.~Turan, \emph{{Scalar Neutrinos at the
  LHC}}, \href{http://dx.doi.org/10.1103/PhysRevD.83.095001}{\emph{Phys. Rev.}
  {\bf D83} (2011) 095001}, [\href{http://arxiv.org/abs/1012.5105}{{\tt
  1012.5105}}].

\bibitem{Kang:2007ib}
J.~Kang, P.~Langacker and B.~D. Nelson, \emph{{Theory and Phenomenology of
  Exotic Isosinglet Quarks and Squarks}},
  \href{http://dx.doi.org/10.1103/PhysRevD.77.035003}{\emph{Phys. Rev.} {\bf
  D77} (2008) 035003}, [\href{http://arxiv.org/abs/0708.2701}{{\tt
  0708.2701}}].

\bibitem{Frank:2012ne}
M.~Frank, L.~Selbuz and I.~Turan, \emph{{Neutralino and Chargino Production in
  U(1)' at the LHC}},
  \href{http://dx.doi.org/10.1140/epjc/s10052-013-2656-7}{\emph{Eur. Phys. J.
  C} {\bf 73} (2013) 2656}, [\href{http://arxiv.org/abs/1212.4428}{{\tt
  1212.4428}}].

\bibitem{Langacker:1998tc}
P.~Langacker and J.~Wang, \emph{{U(1)-prime symmetry breaking in supersymmetric
  E(6) models}},
  \href{http://dx.doi.org/10.1103/PhysRevD.58.115010}{\emph{Phys. Rev. D} {\bf
  58} (1998) 115010}, [\href{http://arxiv.org/abs/hep-ph/9804428}{{\tt
  hep-ph/9804428}}].

\bibitem{Kang:2004bz}
J.~Kang and P.~Langacker, \emph{{$Z$ ' discovery limits for supersymmetric E(6)
  models}}, \href{http://dx.doi.org/10.1103/PhysRevD.71.035014}{\emph{Phys.
  Rev.} {\bf D71} (2005) 035014},
  [\href{http://arxiv.org/abs/hep-ph/0412190}{{\tt hep-ph/0412190}}].

\bibitem{Hewett:1988xc}
J.~L. Hewett and T.~G. Rizzo, \emph{{Low-Energy Phenomenology of Superstring
  Inspired E(6) Models}},
  \href{http://dx.doi.org/10.1016/0370-1573(89)90071-9}{\emph{Phys. Rept.} {\bf
  183} (1989) 193}.

\bibitem{Langacker:1980js}
P.~Langacker, \emph{{Grand Unified Theories and Proton Decay}},
  \href{http://dx.doi.org/10.1016/0370-1573(81)90059-4}{\emph{Phys. Rept.} {\bf
  72} (1981) 185}.

\bibitem{Kang:2004pp}
J.~Kang, P.~Langacker, T.-j. Li and T.~Liu, \emph{{Electroweak baryogenesis in
  a supersymmetric U(1)-prime model}},
  \href{http://dx.doi.org/10.1103/PhysRevLett.94.061801}{\emph{Phys. Rev.
  Lett.} {\bf 94} (2005) 061801},
  [\href{http://arxiv.org/abs/hep-ph/0402086}{{\tt hep-ph/0402086}}].

\bibitem{Barger:2004dy}
V.~Barger, P.~Langacker and H.-S. Lee, \emph{{Big bang nucleosynthesis
  constraints on Z' properties}},  in \emph{{11th International Conference on
  Supersymmetry and the Unification of Fundamental Interactions}}, 2, 2004.
\newblock \href{http://arxiv.org/abs/hep-ph/0402048}{{\tt hep-ph/0402048}}.

\bibitem{Kazana:2016goy}
{\scshape ATLAS, CMS} collaboration, M.~Kazana, \emph{{Searches for heavy
  stable charged particles and other exotic signatures with large ionization at
  the LHC }}, {\emph{CMS-CR-2016-003} (2016) 132--137}.

\bibitem{Rosner:1999ub}
J.~L. Rosner, \emph{{Mixing of charge -1/3 quarks and charged leptons with
  exotic fermions in E(6)}},  \href{http://arxiv.org/abs/hep-ph/9907438}{{\tt
  hep-ph/9907438}}.

\bibitem{Frank:2020pui}
M.~Frank, Y.~Hicyilmaz, S.~Moretti and O.~Ozdal, \emph{{$E_6$ Motivated UMSSM
  Confronts Experimental Data}},  \href{http://arxiv.org/abs/2004.01415}{{\tt
  2004.01415}}.

\bibitem{Staub:2008uz}
F.~Staub, \emph{{SARAH}},  \href{http://arxiv.org/abs/0806.0538}{{\tt
  0806.0538}}.

\bibitem{Staub:2010jh}
F.~Staub, \emph{{Automatic Calculation of supersymmetric Renormalization Group
  Equations and Self Energies}},
  \href{http://dx.doi.org/10.1016/j.cpc.2010.11.030}{\emph{Comput. Phys.
  Commun.} {\bf 182} (2011) 808--833},
  [\href{http://arxiv.org/abs/1002.0840}{{\tt 1002.0840}}].

\bibitem{Staub:2015kfa}
F.~Staub, \emph{{Exploring new models in all detail with SARAH}},
  \href{http://dx.doi.org/10.1155/2015/840780}{\emph{Adv. High Energy Phys.}
  {\bf 2015} (2015) 840780}, [\href{http://arxiv.org/abs/1503.04200}{{\tt
  1503.04200}}].

\bibitem{Belyaev:2012qa}
A.~Belyaev, N.~D. Christensen and A.~Pukhov, \emph{{CalcHEP 3.4 for collider
  physics within and beyond the Standard Model}},
  \href{http://dx.doi.org/10.1016/j.cpc.2013.01.014}{\emph{Comput. Phys.
  Commun.} {\bf 184} (2013) 1729--1769},
  [\href{http://arxiv.org/abs/1207.6082}{{\tt 1207.6082}}].

\bibitem{Degrande:2011ua}
C.~Degrande, C.~Duhr, B.~Fuks, D.~Grellscheid, O.~Mattelaer and T.~Reiter,
  \emph{{UFO - The Universal FeynRules Output}},
  \href{http://dx.doi.org/10.1016/j.cpc.2012.01.022}{\emph{Comput. Phys.
  Commun.} {\bf 183} (2012) 1201--1214},
  [\href{http://arxiv.org/abs/1108.2040}{{\tt 1108.2040}}].

\bibitem{Christensen:2009jx}
N.~D. Christensen, P.~de~Aquino, C.~Degrande, C.~Duhr, B.~Fuks, M.~Herquet
  et~al., \emph{{A Comprehensive approach to new physics simulations}},
  \href{http://dx.doi.org/10.1140/epjc/s10052-011-1541-5}{\emph{Eur. Phys. J.
  C} {\bf 71} (2011) 1541}, [\href{http://arxiv.org/abs/0906.2474}{{\tt
  0906.2474}}].

\bibitem{Belanger:2018ccd}
G.~Bélanger, F.~Boudjema, A.~Goudelis, A.~Pukhov and B.~Zaldivar,
  \emph{{micrOMEGAs5.0 : Freeze-in}},
  \href{http://dx.doi.org/10.1016/j.cpc.2018.04.027}{\emph{Comput. Phys.
  Commun.} {\bf 231} (2018) 173--186},
  [\href{http://arxiv.org/abs/1801.03509}{{\tt 1801.03509}}].

\bibitem{Porod:2003um}
W.~Porod, \emph{{SPheno, a program for calculating supersymmetric spectra, SUSY
  particle decays and SUSY particle production at e+ e- colliders}},
  \href{http://dx.doi.org/10.1016/S0010-4655(03)00222-4}{\emph{Comput. Phys.
  Commun.} {\bf 153} (2003) 275--315},
  [\href{http://arxiv.org/abs/hep-ph/0301101}{{\tt hep-ph/0301101}}].

\bibitem{Porod:2011nf}
W.~Porod and F.~Staub, \emph{{SPheno 3.1: Extensions including flavour,
  CP-phases and models beyond the MSSM}},
  \href{http://dx.doi.org/10.1016/j.cpc.2012.05.021}{\emph{Comput. Phys.
  Commun.} {\bf 183} (2012) 2458--2469},
  [\href{http://arxiv.org/abs/1104.1573}{{\tt 1104.1573}}].

\bibitem{Fuks:2007gk}
B.~Fuks, M.~Klasen, F.~Ledroit, Q.~Li and J.~Morel, \emph{{Precision
  predictions for $Z^\prime$ - production at the CERN LHC: QCD matrix elements,
  parton showers, and joint resummation}},
  \href{http://dx.doi.org/10.1016/j.nuclphysb.2008.01.017}{\emph{Nucl. Phys. B}
  {\bf 797} (2008) 322--339}, [\href{http://arxiv.org/abs/0711.0749}{{\tt
  0711.0749}}].

\bibitem{Fuks:2017vtl}
B.~Fuks and R.~Ruiz, \emph{{A comprehensive framework for studying $W'$ and
  $Z'$ bosons at hadron colliders with automated jet veto resummation}},
  \href{http://dx.doi.org/10.1007/JHEP05(2017)032}{\emph{JHEP} {\bf 05} (2017)
  032}, [\href{http://arxiv.org/abs/1701.05263}{{\tt 1701.05263}}].

\bibitem{Alloul:2013bka}
A.~Alloul, N.~D. Christensen, C.~Degrande, C.~Duhr and B.~Fuks,
  \emph{{FeynRules 2.0 - A complete toolbox for tree-level phenomenology}},
  \href{http://dx.doi.org/10.1016/j.cpc.2014.04.012}{\emph{Comput. Phys.
  Commun.} {\bf 185} (2014) 2250--2300},
  [\href{http://arxiv.org/abs/1310.1921}{{\tt 1310.1921}}].

\bibitem{Degrande:2014vpa}
C.~Degrande, \emph{{Automatic evaluation of UV and R2 terms for beyond the
  Standard Model Lagrangians: a proof-of-principle}},
  \href{http://dx.doi.org/10.1016/j.cpc.2015.08.015}{\emph{Comput. Phys.
  Commun.} {\bf 197} (2015) 239--262},
  [\href{http://arxiv.org/abs/1406.3030}{{\tt 1406.3030}}].

\bibitem{Hahn:2000kx}
T.~Hahn, \emph{{Generating Feynman diagrams and amplitudes with FeynArts 3}},
  \href{http://dx.doi.org/10.1016/S0010-4655(01)00290-9}{\emph{Comput. Phys.
  Commun.} {\bf 140} (2001) 418--431},
  [\href{http://arxiv.org/abs/hep-ph/0012260}{{\tt hep-ph/0012260}}].

\bibitem{Alwall:2014hca}
J.~Alwall, R.~Frederix, S.~Frixione, V.~Hirschi, F.~Maltoni, O.~Mattelaer
  et~al., \emph{{The automated computation of tree-level and next-to-leading
  order differential cross sections, and their matching to parton shower
  simulations}}, \href{http://dx.doi.org/10.1007/JHEP07(2014)079}{\emph{JHEP}
  {\bf 07} (2014) 079}, [\href{http://arxiv.org/abs/1405.0301}{{\tt
  1405.0301}}].

\bibitem{Ball:2017nwa}
{\scshape NNPDF} collaboration, R.~D. Ball et~al., \emph{{Parton distributions
  from high-precision collider data}},
  \href{http://dx.doi.org/10.1140/epjc/s10052-017-5199-5}{\emph{Eur. Phys. J.
  C} {\bf 77} (2017) 663}, [\href{http://arxiv.org/abs/1706.00428}{{\tt
  1706.00428}}].

\bibitem{Sirunyan:2018xlo}
{\scshape CMS} collaboration, A.~M. Sirunyan et~al., \emph{{Search for narrow
  and broad dijet resonances in proton-proton collisions at $ \sqrt{s}=13 $ TeV
  and constraints on dark matter mediators and other new particles}},
  \href{http://dx.doi.org/10.1007/JHEP08(2018)130}{\emph{JHEP} {\bf 08} (2018)
  130}, [\href{http://arxiv.org/abs/1806.00843}{{\tt 1806.00843}}].

\bibitem{Bechtle:2013wla}
P.~Bechtle, O.~Brein, S.~Heinemeyer, O.~Stål, T.~Stefaniak, G.~Weiglein
  et~al., \emph{{$\mathsf{HiggsBounds}-4$: Improved Tests of Extended Higgs
  Sectors against Exclusion Bounds from LEP, the Tevatron and the LHC}},
  \href{http://dx.doi.org/10.1140/epjc/s10052-013-2693-2}{\emph{Eur. Phys. J.}
  {\bf C74} (2014) 2693}, [\href{http://arxiv.org/abs/1311.0055}{{\tt
  1311.0055}}].

\bibitem{Bechtle:2013xfa}
P.~Bechtle, S.~Heinemeyer, O.~Stål, T.~Stefaniak and G.~Weiglein,
  \emph{{$HiggsSignals$: Confronting arbitrary Higgs sectors with measurements
  at the Tevatron and the LHC}},
  \href{http://dx.doi.org/10.1140/epjc/s10052-013-2711-4}{\emph{Eur. Phys. J.}
  {\bf C74} (2014) 2711}, [\href{http://arxiv.org/abs/1305.1933}{{\tt
  1305.1933}}].

\bibitem{Buckley:2013jua}
A.~Buckley, \emph{{PySLHA: a Pythonic interface to SUSY Les Houches Accord
  data}}, \href{http://dx.doi.org/10.1140/epjc/s10052-015-3638-8}{\emph{Eur.
  Phys. J. C} {\bf 75} (2015) 467}, [\href{http://arxiv.org/abs/1305.4194}{{\tt
  1305.4194}}].

\bibitem{Skands:2003cj}
P.~Z. Skands et~al., \emph{{SUSY Les Houches accord: Interfacing SUSY spectrum
  calculators, decay packages, and event generators}},
  \href{http://dx.doi.org/10.1088/1126-6708/2004/07/036}{\emph{JHEP} {\bf 07}
  (2004) 036}, [\href{http://arxiv.org/abs/hep-ph/0311123}{{\tt
  hep-ph/0311123}}].

\bibitem{Cacciapaglia:2006pk}
G.~Cacciapaglia, C.~Csaki, G.~Marandella and A.~Strumia, \emph{{The Minimal Set
  of Electroweak Precision Parameters}},
  \href{http://dx.doi.org/10.1103/PhysRevD.74.033011}{\emph{Phys. Rev.} {\bf
  D74} (2006) 033011}, [\href{http://arxiv.org/abs/hep-ph/0604111}{{\tt
  hep-ph/0604111}}].

\bibitem{Altarelli:1990zd}
G.~Altarelli and R.~Barbieri, \emph{{Vacuum polarization effects of new physics
  on electroweak processes}},
  \href{http://dx.doi.org/10.1016/0370-2693(91)91378-9}{\emph{Phys. Lett.} {\bf
  B253} (1991) 161--167}.

\bibitem{Peskin:1990zt}
M.~E. Peskin and T.~Takeuchi, \emph{{A New constraint on a strongly interacting
  Higgs sector}},
  \href{http://dx.doi.org/10.1103/PhysRevLett.65.964}{\emph{Phys. Rev. Lett.}
  {\bf 65} (1990) 964--967}.

\bibitem{Peskin:1991sw}
M.~E. Peskin and T.~Takeuchi, \emph{{Estimation of oblique electroweak
  corrections}}, \href{http://dx.doi.org/10.1103/PhysRevD.46.381}{\emph{Phys.
  Rev.} {\bf D46} (1992) 381--409}.

\bibitem{Maksymyk:1993zm}
I.~Maksymyk, C.~P. Burgess and D.~London, \emph{{Beyond S, T and U}},
  \href{http://dx.doi.org/10.1103/PhysRevD.50.529}{\emph{Phys. Rev.} {\bf D50}
  (1994) 529--535}, [\href{http://arxiv.org/abs/hep-ph/9306267}{{\tt
  hep-ph/9306267}}].

\bibitem{Baak:2014ora}
{\scshape Gfitter Group} collaboration, M.~Baak, J.~Cúth, J.~Haller,
  A.~Hoecker, R.~Kogler, K.~Mönig et~al., \emph{{The global electroweak fit at
  NNLO and prospects for the LHC and ILC}},
  \href{http://dx.doi.org/10.1140/epjc/s10052-014-3046-5}{\emph{Eur. Phys. J.}
  {\bf C74} (2014) 3046}, [\href{http://arxiv.org/abs/1407.3792}{{\tt
  1407.3792}}].

\bibitem{Tanabashi:2018oca}
{\scshape Particle Data Group} collaboration, M.~Tanabashi et~al.,
  \emph{{Review of Particle Physics}},
  \href{http://dx.doi.org/10.1103/PhysRevD.98.030001}{\emph{Phys. Rev.} {\bf
  D98} (2018) 030001}.

\bibitem{Chatrchyan:2012xdj}
{\scshape CMS} collaboration, S.~Chatrchyan et~al., \emph{{Observation of a new
  boson at a mass of 125 GeV with the CMS experiment at the LHC}},
  \href{http://dx.doi.org/10.1016/j.physletb.2012.08.021}{\emph{Phys. Lett.}
  {\bf B716} (2012) 30--61}, [\href{http://arxiv.org/abs/1207.7235}{{\tt
  1207.7235}}].

\bibitem{Aaij:2012nna}
{\scshape LHCb} collaboration, R.~Aaij et~al., \emph{{First Evidence for the
  Decay $B_s^0 \to \mu^+ \mu^-$}},
  \href{http://dx.doi.org/10.1103/PhysRevLett.110.021801}{\emph{Phys. Rev.
  Lett.} {\bf 110} (2013) 021801}, [\href{http://arxiv.org/abs/1211.2674}{{\tt
  1211.2674}}].

\bibitem{Asner:2010qj}
{\scshape Heavy Flavor Averaging Group} collaboration, D.~Asner et~al.,
  \emph{{Averages of $b$-hadron, $c$-hadron, and $\tau$-lepton properties}},
  \href{http://arxiv.org/abs/1010.1589}{{\tt 1010.1589}}.

\bibitem{Amhis:2012bh}
{\scshape Heavy Flavor Averaging Group} collaboration, Y.~Amhis et~al.,
  \emph{{Averages of B-Hadron, C-Hadron, and tau-lepton properties as of early
  2012}},  \href{http://arxiv.org/abs/1207.1158}{{\tt 1207.1158}}.

\bibitem{Bahl:2019hmm}
H.~Bahl, S.~Heinemeyer, W.~Hollik and G.~Weiglein, \emph{{Theoretical
  uncertainties in the MSSM Higgs boson mass calculation}},
  \href{http://dx.doi.org/10.1140/epjc/s10052-020-8079-3}{\emph{Eur. Phys. J.
  C} {\bf 80} (2020) 497}, [\href{http://arxiv.org/abs/1912.04199}{{\tt
  1912.04199}}].

\bibitem{Gogoladze:2011db}
I.~Gogoladze, R.~Khalid, S.~Raza and Q.~Shafi, \emph{{Higgs and Sparticle
  Spectroscopy with Gauge-Yukawa Unification}},
  \href{http://dx.doi.org/10.1007/JHEP06(2011)117}{\emph{JHEP} {\bf 06} (2011)
  117}, [\href{http://arxiv.org/abs/1102.0013}{{\tt 1102.0013}}].

\bibitem{Gogoladze:2011aa}
I.~Gogoladze, Q.~Shafi and C.~S. Un, \emph{{Higgs Boson Mass from t-b-$\tau$
  Yukawa Unification}},
  \href{http://dx.doi.org/10.1007/JHEP08(2012)028}{\emph{JHEP} {\bf 08} (2012)
  028}, [\href{http://arxiv.org/abs/1112.2206}{{\tt 1112.2206}}].

\bibitem{Ajaib:2013zha}
M.~Adeel~Ajaib, I.~Gogoladze, Q.~Shafi and C.~S. Un, \emph{{A Predictive Yukawa
  Unified SO(10) Model: Higgs and Sparticle Masses}},
  \href{http://dx.doi.org/10.1007/JHEP07(2013)139}{\emph{JHEP} {\bf 07} (2013)
  139}, [\href{http://arxiv.org/abs/1303.6964}{{\tt 1303.6964}}].

\bibitem{Un:2016hji}
C.~S. Un and O.~Ozdal, \emph{{Mass Spectrum and Higgs Profile in BLSSM}},
  \href{http://dx.doi.org/10.1103/PhysRevD.93.055024}{\emph{Phys. Rev. D} {\bf
  93} (2016) 055024}, [\href{http://arxiv.org/abs/1601.02494}{{\tt
  1601.02494}}].

\bibitem{ArkaniHamed:2006mb}
N.~Arkani-Hamed, A.~Delgado and G.~Giudice, \emph{{The Well-tempered
  neutralino}},
  \href{http://dx.doi.org/10.1016/j.nuclphysb.2006.02.010}{\emph{Nucl. Phys. B}
  {\bf 741} (2006) 108--130}, [\href{http://arxiv.org/abs/hep-ph/0601041}{{\tt
  hep-ph/0601041}}].

\bibitem{Cao:2015efs}
J.~Cao, Y.~He, L.~Shang, W.~Su and Y.~Zhang, \emph{{Testing the light dark
  matter scenario of the MSSM at the LHC}},
  \href{http://dx.doi.org/10.1007/JHEP03(2016)207}{\emph{JHEP} {\bf 03} (2016)
  207}, [\href{http://arxiv.org/abs/1511.05386}{{\tt 1511.05386}}].

\bibitem{Calibbi:2013poa}
L.~Calibbi, J.~M. Lindert, T.~Ota and Y.~Takanishi, \emph{{Cornering light
  Neutralino Dark Matter at the LHC}},
  \href{http://dx.doi.org/10.1007/JHEP10(2013)132}{\emph{JHEP} {\bf 10} (2013)
  132}, [\href{http://arxiv.org/abs/1307.4119}{{\tt 1307.4119}}].

\bibitem{Ellwanger:2016sur}
U.~Ellwanger, \emph{{Present Status and Future Tests of the Higgsino-Singlino
  Sector in the NMSSM}},
  \href{http://dx.doi.org/10.1007/JHEP02(2017)051}{\emph{JHEP} {\bf 02} (2017)
  051}, [\href{http://arxiv.org/abs/1612.06574}{{\tt 1612.06574}}].

\bibitem{Belanger:2015cra}
G.~B\'elanger, J.~Da~Silva, U.~Laa and A.~Pukhov, \emph{{Probing U(1)
  extensions of the MSSM at the LHC Run I and in dark matter searches}},
  \href{http://dx.doi.org/10.1007/JHEP09(2015)151}{\emph{JHEP} {\bf 09} (2015)
  151}, [\href{http://arxiv.org/abs/1505.06243}{{\tt 1505.06243}}].

\bibitem{Barger:2004bz}
V.~Barger, C.~Kao, P.~Langacker and H.-S. Lee, \emph{{Neutralino relic density
  in a supersymmetric U(1)-prime model}},
  \href{http://dx.doi.org/10.1016/j.physletb.2004.08.070}{\emph{Phys. Lett. B}
  {\bf 600} (2004) 104--115}, [\href{http://arxiv.org/abs/hep-ph/0408120}{{\tt
  hep-ph/0408120}}].

\bibitem{Ade:2013zuv}
{\scshape Planck} collaboration, P.~A.~R. Ade et~al., \emph{{Planck 2013
  results. XVI. Cosmological parameters}},
  \href{http://dx.doi.org/10.1051/0004-6361/201321591}{\emph{Astron.
  Astrophys.} {\bf 571} (2014) A16},
  [\href{http://arxiv.org/abs/1303.5076}{{\tt 1303.5076}}].

\bibitem{Aghanim:2018eyx}
{\scshape Planck} collaboration, N.~Aghanim et~al., \emph{{Planck 2018 results.
  VI. Cosmological parameters}},  \href{http://arxiv.org/abs/1807.06209}{{\tt
  1807.06209}}.

\bibitem{Frank:2017ohg}
M.~Frank and O.~\"Ozdal, \emph{{Exploring the supersymmetric U(1)$_{B-L}
  \times$ U(1)$_{R}$ model with dark matter, muon $g-2$ and $Z^\prime$ mass
  limits}}, \href{http://dx.doi.org/10.1103/PhysRevD.97.015012}{\emph{Phys.
  Rev. D} {\bf 97} (2018) 015012}, [\href{http://arxiv.org/abs/1709.04012}{{\tt
  1709.04012}}].

\bibitem{Araz:2017qcs}
J.~Y. Araz, M.~Frank and B.~Fuks, \emph{{Differentiating $U(1)^\prime$
  supersymmetric models with right sneutrino and neutralino dark matter}},
  \href{http://dx.doi.org/10.1103/PhysRevD.96.015017}{\emph{Phys. Rev. D} {\bf
  96} (2017) 015017}, [\href{http://arxiv.org/abs/1705.01063}{{\tt
  1705.01063}}].

\bibitem{Aprile:2018dbl}
{\scshape XENON} collaboration, E.~Aprile et~al., \emph{Dark matter search
  results from a one ton-year exposure of xenon1t},
  \href{http://dx.doi.org/10.1103/PhysRevLett.121.111302}{\emph{Phys.Rev.Lett.}
  {\bf 121} (2018) 111302}, [\href{http://arxiv.org/abs/1805.12562}{{\tt
  1805.12562}}].

\bibitem{Aalbers:2016jon}
{\scshape DARWIN} collaboration, J.~Aalbers et~al., \emph{{DARWIN: towards the
  ultimate dark matter detector}},
  \href{http://dx.doi.org/10.1088/1475-7516/2016/11/017}{\emph{JCAP} {\bf 1611}
  (2016) 017}, [\href{http://arxiv.org/abs/1606.07001}{{\tt 1606.07001}}].

\bibitem{Ackermann:2015zua}
{\scshape Fermi-LAT} collaboration, M.~Ackermann et~al., \emph{{Searching for
  Dark Matter Annihilation from Milky Way Dwarf Spheroidal Galaxies with Six
  Years of Fermi Large Area Telescope Data}},
  \href{http://dx.doi.org/10.1103/PhysRevLett.115.231301}{\emph{Phys. Rev.
  Lett.} {\bf 115} (2015) 231301}, [\href{http://arxiv.org/abs/1503.02641}{{\tt
  1503.02641}}].

\bibitem{Ahnen:2016qkx}
{\scshape MAGIC, Fermi-LAT} collaboration, M.~L. Ahnen et~al., \emph{{Limits to
  Dark Matter Annihilation Cross-Section from a Combined Analysis of MAGIC and
  Fermi-LAT Observations of Dwarf Satellite Galaxies}},
  \href{http://dx.doi.org/10.1088/1475-7516/2016/02/039}{\emph{JCAP} {\bf 1602}
  (2016) 039}, [\href{http://arxiv.org/abs/1601.06590}{{\tt 1601.06590}}].

\bibitem{Bennett:2006fi}
{\scshape Muon g-2} collaboration, G.~W. Bennett et~al., \emph{{Final Report of
  the Muon E821 Anomalous Magnetic Moment Measurement at BNL}},
  \href{http://dx.doi.org/10.1103/PhysRevD.73.072003}{\emph{Phys. Rev.} {\bf
  D73} (2006) 072003}, [\href{http://arxiv.org/abs/hep-ex/0602035}{{\tt
  hep-ex/0602035}}].

\bibitem{Parker_2018}
R.~H. Parker, C.~Yu, W.~Zhong, B.~Estey and H.~Muller, \emph{Measurement of the
  fine-structure constant as a test of the standard model},
  \href{http://dx.doi.org/10.1126/science.aap7706}{\emph{Science} {\bf 360}
  (Apr, 2018) 191?195}.

\bibitem{Sirunyan:2019ctn}
{\scshape CMS} collaboration, A.~M. Sirunyan et~al., \emph{{Search for
  supersymmetry in proton-proton collisions at 13 TeV in final states with jets
  and missing transverse momentum}},
  \href{http://dx.doi.org/10.1007/JHEP10(2019)244}{\emph{JHEP} {\bf 10} (2019)
  244}, [\href{http://arxiv.org/abs/1908.04722}{{\tt 1908.04722}}].

\bibitem{Sirunyan:2018nwe}
{\scshape CMS} collaboration, A.~M. Sirunyan et~al., \emph{{Search for
  supersymmetric partners of electrons and muons in proton-proton collisions at
  $\sqrt{s}=$ 13 TeV}},
  \href{http://dx.doi.org/10.1016/j.physletb.2019.01.005}{\emph{Phys. Lett. B}
  {\bf 790} (2019) 140--166}, [\href{http://arxiv.org/abs/1806.05264}{{\tt
  1806.05264}}].

\bibitem{Sirunyan:2019mlu}
{\scshape CMS} collaboration, A.~M. Sirunyan et~al., \emph{{Search for
  Supersymmetry with a Compressed Mass Spectrum in Events with a Soft $\tau$
  Lepton, a Highly Energetic Jet, and Large Missing Transverse Momentum in
  Proton-Proton Collisions at $\sqrt{s}=$ TeV}},
  \href{http://dx.doi.org/10.1103/PhysRevLett.124.041803}{\emph{Phys. Rev.
  Lett.} {\bf 124} (2020) 041803}, [\href{http://arxiv.org/abs/1910.01185}{{\tt
  1910.01185}}].

\bibitem{ATLAS:2019ucg}
{\scshape ATLAS} collaboration, T.~A. Collaboration, \emph{{Search for direct
  stau production in events with two hadronic tau leptons in $\sqrt{s}=$ 13 TeV
  pp collisions with the ATLAS detector}}, .

\bibitem{Aad:2019vvf}
{\scshape ATLAS} collaboration, G.~Aad et~al., \emph{{Search for direct
  production of electroweakinos in final states with one lepton, missing
  transverse momentum and a Higgs boson decaying into two $b$-jets in (pp)
  collisions at $\sqrt{s}=13$ TeV with the ATLAS detector}},
  \href{http://arxiv.org/abs/1909.09226}{{\tt 1909.09226}}.

\bibitem{Aad:2019vnb}
{\scshape ATLAS} collaboration, G.~Aad et~al., \emph{{Search for electroweak
  production of charginos and sleptons decaying into final states with two
  leptons and missing transverse momentum in $\sqrt{s}=13$ TeV $pp$ collisions
  using the ATLAS detector}},
  \href{http://dx.doi.org/10.1140/epjc/s10052-019-7594-6}{\emph{Eur. Phys. J.}
  {\bf C80} (2020) 123}, [\href{http://arxiv.org/abs/1908.08215}{{\tt
  1908.08215}}].

\bibitem{Aoyama:2014sxa}
T.~Aoyama, M.~Hayakawa, T.~Kinoshita and M.~Nio, \emph{{Tenth-Order Electron
  Anomalous Magnetic Moment --- Contribution of Diagrams without Closed Lepton
  Loops}}, \href{http://dx.doi.org/10.1103/PhysRevD.91.033006}{\emph{Phys. Rev.
  D} {\bf 91} (2015) 033006}, [\href{http://arxiv.org/abs/1412.8284}{{\tt
  1412.8284}}].

\bibitem{Volkov:2018jhy}
S.~Volkov, \emph{{Numerical calculation of high-order QED contributions to the
  electron anomalous magnetic moment}},
  \href{http://dx.doi.org/10.1103/PhysRevD.98.076018}{\emph{Phys. Rev. D} {\bf
  98} (2018) 076018}, [\href{http://arxiv.org/abs/1807.05281}{{\tt
  1807.05281}}].

\bibitem{Aoyama:2017uqe}
T.~Aoyama, T.~Kinoshita and M.~Nio, \emph{{Revised and Improved Value of the
  QED Tenth-Order Electron Anomalous Magnetic Moment}},
  \href{http://dx.doi.org/10.1103/PhysRevD.97.036001}{\emph{Phys. Rev. D} {\bf
  97} (2018) 036001}, [\href{http://arxiv.org/abs/1712.06060}{{\tt
  1712.06060}}].

\bibitem{Volkov:2017xaq}
S.~Volkov, \emph{{New method of computing the contributions of graphs without
  lepton loops to the electron anomalous magnetic moment in QED}},
  \href{http://dx.doi.org/10.1103/PhysRevD.96.096018}{\emph{Phys. Rev. D} {\bf
  96} (2017) 096018}, [\href{http://arxiv.org/abs/1705.05800}{{\tt
  1705.05800}}].

\bibitem{Aoyama:2019ryr}
T.~Aoyama, T.~Kinoshita and M.~Nio, \emph{{Theory of the Anomalous Magnetic
  Moment of the Electron}},
  \href{http://dx.doi.org/10.3390/atoms7010028}{\emph{Atoms} {\bf 7} (2019)
  28}.

\bibitem{Chun:2019oix}
E.~J. Chun, J.~Kim and T.~Mondal, \emph{{Electron EDM and Muon anomalous
  magnetic moment in Two-Higgs-Doublet Models}},
  \href{http://dx.doi.org/10.1007/JHEP12(2019)068}{\emph{JHEP} {\bf 12} (2019)
  068}, [\href{http://arxiv.org/abs/1906.00612}{{\tt 1906.00612}}].

\bibitem{DeConto:2016ith}
G.~De~Conto and V.~Pleitez, \emph{{Electron and muon anomalous magnetic dipole
  moment in a 3--3--1 model}},
  \href{http://dx.doi.org/10.1007/JHEP05(2017)104}{\emph{JHEP} {\bf 05} (2017)
  104}, [\href{http://arxiv.org/abs/1603.09691}{{\tt 1603.09691}}].

\bibitem{Badziak:2019gaf}
M.~Badziak and K.~Sakurai, \emph{{Explanation of electron and muon g $-$ 2
  anomalies in the MSSM}},
  \href{http://dx.doi.org/10.1007/JHEP10(2019)024}{\emph{JHEP} {\bf 10} (2019)
  024}, [\href{http://arxiv.org/abs/1908.03607}{{\tt 1908.03607}}].

\bibitem{Ambrogi:2017neo}
F.~Ambrogi, S.~Kraml, S.~Kulkarni, U.~Laa, A.~Lessa, V.~Magerl et~al.,
  \emph{{SModelS v1.1 user manual: Improving simplified model constraints with
  efficiency maps}},
  \href{http://dx.doi.org/10.1016/j.cpc.2018.02.007}{\emph{Comput. Phys.
  Commun.} {\bf 227} (2018) 72--98},
  [\href{http://arxiv.org/abs/1701.06586}{{\tt 1701.06586}}].

\bibitem{Ambrogi:2018ujg}
F.~Ambrogi et~al., \emph{{SModelS v1.2: long-lived particles, combination of
  signal regions, and other novelties}},
  \href{http://dx.doi.org/10.1016/j.cpc.2019.07.013}{\emph{Comput. Phys.
  Commun.} {\bf 251} (2020) 106848},
  [\href{http://arxiv.org/abs/1811.10624}{{\tt 1811.10624}}].

\bibitem{Dutta:2018ioj}
J.~Dutta, S.~Kraml, A.~Lessa and W.~Waltenberger, \emph{{SModelS extension with
  the CMS supersymmetry search results from Run 2}},
  \href{http://dx.doi.org/10.31526/LHEP.1.2018.02}{\emph{LHEP} {\bf 1} (2018)
  5--12}, [\href{http://arxiv.org/abs/1803.02204}{{\tt 1803.02204}}].

\bibitem{Khosa:2020zar}
C.~K. Khosa, S.~Kraml, A.~Lessa, P.~Neuhuber and W.~Waltenberger,
  \emph{{SModelS database update v1.2.3}},
  \href{http://arxiv.org/abs/2005.00555}{{\tt 2005.00555}}.

\bibitem{Zyla:2020zbs}
{\scshape Particle Data Group} collaboration, P.~Zyla et~al., \emph{{Review of
  Particle Physics}}, \href{http://dx.doi.org/10.1093/ptep/ptaa104}{\emph{PTEP}
  {\bf 2020} (2020) 083C01}.

\bibitem{Aaboud:2016zdn}
{\scshape ATLAS} collaboration, M.~Aaboud et~al., \emph{{Search for squarks and
  gluinos in final states with jets and missing transverse momentum at
  $\sqrt{s} =$ 13 TeV with the ATLAS detector}},
  \href{http://dx.doi.org/10.1140/epjc/s10052-016-4184-8}{\emph{Eur. Phys. J.
  C} {\bf 76} (2016) 392}, [\href{http://arxiv.org/abs/1605.03814}{{\tt
  1605.03814}}].

\bibitem{Aaboud:2017vwy}
{\scshape ATLAS} collaboration, M.~Aaboud et~al., \emph{{Search for squarks and
  gluinos in final states with jets and missing transverse momentum using 36
  fb$^{-1}$ of $\sqrt{s}=13$ TeV pp collision data with the ATLAS detector}},
  \href{http://dx.doi.org/10.1103/PhysRevD.97.112001}{\emph{Phys. Rev. D} {\bf
  97} (2018) 112001}, [\href{http://arxiv.org/abs/1712.02332}{{\tt
  1712.02332}}].

\bibitem{Sirunyan:2017cwe}
{\scshape CMS} collaboration, A.~M. Sirunyan et~al., \emph{{Search for
  supersymmetry in multijet events with missing transverse momentum in
  proton-proton collisions at 13 TeV}},
  \href{http://dx.doi.org/10.1103/PhysRevD.96.032003}{\emph{Phys. Rev. D} {\bf
  96} (2017) 032003}, [\href{http://arxiv.org/abs/1704.07781}{{\tt
  1704.07781}}].

\bibitem{ATLAS:2019vcq}
{\scshape ATLAS} collaboration, \emph{{Search for squarks and gluinos in final
  states with jets and missing transverse momentum using 139 fb$^{-1}$ of
  $\sqrt{s}$ =13 TeV $pp$ collision data with the ATLAS detector}}, .

\bibitem{Sjostrand:2014zea}
T.~Sjöstrand, S.~Ask, J.~R. Christiansen, R.~Corke, N.~Desai, P.~Ilten et~al.,
  \emph{{An Introduction to PYTHIA 8.2}},
  \href{http://dx.doi.org/10.1016/j.cpc.2015.01.024}{\emph{Comput. Phys.
  Commun.} {\bf 191} (2015) 159--177},
  [\href{http://arxiv.org/abs/1410.3012}{{\tt 1410.3012}}].

\bibitem{deFavereau:2013fsa}
{\scshape DELPHES 3} collaboration, J.~de~Favereau, C.~Delaere, P.~Demin,
  A.~Giammanco, V.~Lemaître, A.~Mertens et~al., \emph{{DELPHES 3, A modular
  framework for fast simulation of a generic collider experiment}},
  \href{http://dx.doi.org/10.1007/JHEP02(2014)057}{\emph{JHEP} {\bf 02} (2014)
  057}, [\href{http://arxiv.org/abs/1307.6346}{{\tt 1307.6346}}].

\bibitem{Anderson:2013kxz}
J.~Anderson et~al., \emph{{Snowmass Energy Frontier Simulations}},
  \href{http://arxiv.org/abs/1309.1057}{{\tt 1309.1057}}.

\bibitem{Avetisyan:2013onh}
A.~Avetisyan et~al., \emph{{Methods and Results for Standard Model Event
  Generation at $\sqrt{s}$ = 14 TeV, 33 TeV and 100 TeV Proton Colliders (A
  Snowmass Whitepaper)}},  in \emph{{Community Summer Study 2013}: {Snowmass on
  the Mississippi}}, 8, 2013.
\newblock \href{http://arxiv.org/abs/1308.1636}{{\tt 1308.1636}}.

\bibitem{Cacciari:2008gp}
M.~Cacciari, G.~P. Salam and G.~Soyez, \emph{{The anti-$k_t$ jet clustering
  algorithm}},
  \href{http://dx.doi.org/10.1088/1126-6708/2008/04/063}{\emph{JHEP} {\bf 04}
  (2008) 063}, [\href{http://arxiv.org/abs/0802.1189}{{\tt 0802.1189}}].

\bibitem{Cacciari:2011ma}
M.~Cacciari, G.~P. Salam and G.~Soyez, \emph{{FastJet User Manual}},
  \href{http://dx.doi.org/10.1140/epjc/s10052-012-1896-2}{\emph{Eur. Phys. J.
  C} {\bf 72} (2012) 1896}, [\href{http://arxiv.org/abs/1111.6097}{{\tt
  1111.6097}}].

\bibitem{Conte:2012fm}
E.~Conte, B.~Fuks and G.~Serret, \emph{{MadAnalysis 5, A User-Friendly
  Framework for Collider Phenomenology}},
  \href{http://dx.doi.org/10.1016/j.cpc.2012.09.009}{\emph{Comput. Phys.
  Commun.} {\bf 184} (2013) 222--256},
  [\href{http://arxiv.org/abs/1206.1599}{{\tt 1206.1599}}].

\end{thebibliography}\endgroup

\end{document}